\documentclass[12pt,twoside]{article}
\usepackage{amsmath, amssymb, amsthm, mathrsfs, amsfonts, bbm ,graphicx,color,citesort,stmaryrd,textpos,upgreek,protosem,ifpdf} 

\usepackage[pdftex,bookmarks]{hyperref}

\usepackage[text={6.9in,9in},centering,includehead,includefoot,headheight=14.5pt]{geometry}
\usepackage[square,comma,numbers,sort&compress]{natbib}
\usepackage{hypernat}

\usepackage[version=3,arrows=pgf-filled]{mhchem}

\usepackage{fancyhdr}
\renewcommand{\subsectionmark}[1]{}
\newlength\Li \newlength\Lii
\newcommand{\vdim}{\mbox{{dim}}}

\date{}

\makeatletter
\def\@seccntformat#1{}
\makeatother
\numberwithin{equation}{section}
\renewcommand{\numberline}[1]{}

\makeatletter
\def\blfootnote{\xdef\@thefnmark{}\@footnotetext}
\makeatother

\title{Modeling Chemical Reactors I: Quiescent Reactors} 
\author{\\ C.E.~Michoski \textsuperscript{\htmladdnormallink{\dag}{https://webspace.utexas.edu/michoski/Michoski.html}} \textsuperscript{\htmladdnormallink{\textproto{\AAsade}}{https://webspace.utexas.edu/michoski/Michoski.html}}  \\[-2mm] \emph{\footnotesize{ Department of Chemistry and Biochemistry, Computational Sciences and Engineering Mathematics,}} \\ [-2mm]\emph{\footnotesize{University of Texas at Austin, Austin, TX, 78712}} \\ \\ \ J.A.~Evans  \textsuperscript{\htmladdnormallink{$\star$}{http://users.ices.utexas.edu/~evans/Site/Welcome.html}} \\[-2mm] \emph{\footnotesize{Computational Sciences and Engineering Mathematics, University of Texas at Austin, Austin, TX, 78712}} \\ \\  \ P.G.~Schmitz \textsuperscript{\htmladdnormallink{\ddag}{http://www.ma.utexas.edu/users/pschmitz/}}  \\[-2mm] \emph{\footnotesize{Department of Mathematics, University of Texas at Austin, Austin, TX, 78712}}}

\begin{document}
\maketitle
\begin{abstract}   
We introduce a fully generalized quiescent chemical reactor system in arbitrary space $\vdim =1,2$ or $3$, with $n\in\mathbb{N}$ chemical constituents $\alpha_{i}$, where the character of the numerical solution is strongly determined by the relative scaling between the local reactivity of species $\alpha_{i}$ and the local functional diffusivity $\mathscr{D}_{ij}(\alpha)$ of the reaction mixture.  We develop an operator time-splitting predictor multi-corrector RK--LDG scheme, and utilize $hp$-adaptivity relying only on the entropy $\mathscr{S}_{\mathfrak{R}}$ of the reactive system $\mathfrak{R}$.  This condition preserves these bounded nonlinear entropy functionals as a necessarily enforced stability condition on the coupled system.  We apply this scheme to a number of application problems in chemical kinetics; including a difficult classical problem arising in nonequilibrium thermodynamics known as the Belousov-Zhabotinskii reaction where we utilize a concentration-dependent diffusivity tensor $\mathscr{D}_{ij}(\alpha)$, in addition to solving a simple equilibrium problem in order to evaluate the numerical error behavior.\end{abstract}\blfootnote{\textdagger {\it michoski@cm.utexas.edu}, \textproto{\AAsade}\scriptsize{Corresponding author}}\blfootnote{$\star$ {\it evans@ices.utexas.edu}, \ \ddag {\it pschmitz@math.utexas.edu}}\tableofcontents

\section{\texorpdfstring{\protect\centering $\S 1$ Introduction}{\S 1 Introduction}} 

Chemical reactors are of fundamental importance in a large array of scientific fields, spanning applications in chemistry and chemical engineering \cite{Hirschfelder,Gardiner,SK}, mechanical and aerospace engineering \cite{Snytnikov}, atmospheric and oceanic sciences \cite{Suh}, astronomy and plasma physics \cite{Ferris,YRMAIT}; as well as generally in any numbers of biologically related fields (\emph{viz.} \cite{Pompano} for example).  More fundamentally it is the basic prevalence of these dynamic reactive chemical systems in nature that makes the ability to effectively model them so essential.  

From a theoretical point of view, much of the underlying theory for reactor systems may be found in \cite{Hirschfelder,erdi,Chapman}, where generally reactor systems may be derived using kinetic theory by way of a Chapman--Enskog or Hilbert type perturbative expansion, which immediately raises a set of important concerns that are far beyond the present scope of this paper (see for example \cite{TruMun} for an example of the formal complications that may arise in rigorous treatments).  Here we rather restrict ourselves to the study of a set of simplifications leading to a generalized system of reaction-diffusion equations, that may be referred to collectively as \emph{quiescent reactors}.  

The foundational theory provides that \emph{quiescent reactor systems} may be derived explicitly from fluid particle systems (\emph{i.e.} the Boltzmann equation), where the \emph{reaction diffusion multivariate master equation} \cite{Smoller,Vanden,Isaa} serves as the rigorous justification underlying the model.  From the point of view of the experimental sciences, a quiescent reactor may be defined as a chemical reactor system where no explicit stirring is either present or plays a significant role in the dynamic behavior of the medium.  It must be noted however, that often in these quiescent experimental systems, as seen in \cite{Pang}, heat gradients may be utilized for chemical catalysis.  Because of the complicated formal coupling between state variables, in particular those of density $\rho=\rho(t,\boldsymbol{x})$, temperature $\vartheta=\vartheta(t,\boldsymbol{x})$ and internal energy $\mathscr{E}=\mathscr{E}(t,\boldsymbol{x})$, all of which may have the effect of imparting local velocity gradients on fluids elements, we restrict ourselves in this paper to isothermal systems, and define a quiescent reactor as one which, up to the transport properties of the system, is diffusion dominated in the sense of the Fickian regime (which we expand upon below); or, more precisely, may be modeled up to an implicit stochasticity as a system of reaction-diffusion equations \cite{Atzberger,Isaac2}.  

More clearly, from the point of view of the simplified mathematics of the system, such a restriction to the quiescent regime may be presented as any flow system obeying (\ref{quiescent}) which satisfies the following approximate bound: \begin{equation}\label{bound1}\nabla_{x}\cdot(\alpha_{i}\boldsymbol{u})\lesssim \mathscr{D}_{i}'|\nabla_{x}\alpha_{i}|^{2} + \mathscr{D}_{i}\Delta_{x}\alpha_{i},\end{equation} where $\boldsymbol{u}=\boldsymbol{u}(t,\boldsymbol{x})$ is the flow velocity of the system, and where when the formal inequality holds (\emph{i.e.} for $\leq$ in \ref{bound1}), the quiescent approximation is particularly strong.   Formally we can say that if the concentration and velocity gradients are comparable $\nabla_{x}u_{j}\sim\nabla_{x}\alpha_{i}$ for each constituent $i$ and each component $j$, or even more strongly whenever $\nabla_{x}u_{j}\lesssim\nabla_{x}\alpha_{i}$, then if the velocity components are bounded from above by $u_{j}\lesssim \alpha_{i}\mathscr{D}'_{i}\nabla_{x}\ln \alpha_{i}-\alpha_{i}$, then (\ref{bound1}) is satisfied, and strictly satisfied when the bounds are precise (\emph{i.e.} $\lesssim \implies \leq$).  In the case of, for example, the Chapman-Enskog expansion of $\mathscr{D}_{ij}$ as developed in \textsection{4}, this merely suggests that for a bounded concentration gradient $|\nabla_{x}\alpha_{i}| \leq C$ the diffusive gradient $\mathscr{D}_{i}'$ is controlled from below such that $\mathscr{D}_{i}' \geq \kappa\alpha_{i}$ for $\kappa=\kappa(\alpha_{j\neq i},C)$ having only functional dependencies on the fractional weighting of the other constituents of the fluid.  By contrast, when the diffusion coefficient is taken to be constant such that  $\mathscr{D}_{i}'=0$, it follows that in local areas of appreciable concentration, \emph{i.e.} $\alpha_{i}\gg 0$, the advection must scale with diffusive collisions, and similarly in areas of measurable velocities it is the concentration gradient which must scale with the collisional motions.

From the point of view of the physics and chemistry of the system, a diffusion dominated flow regime is merely one in which the random collisional molecular motion of the fluid dominates the advective flow characteristic.  Such systems are frequently used as approximate models to restrict to systems that implicitly contain substantially more complicated dynamics (\emph{e.g.} such as in combustion models \cite{Williams}). 

In fact, it is remarkable the number of complicated and important physical phenomena that are understood merely by way of modeling coupled reaction-diffusion equations.  For example, the spatially distributed FitzHugh-Nagumo model is a reaction-diffusion system of primary importance in tracking the formation, propagation and recovery of action potentials in biological and artificial neural networks \cite{Dikansky}.  In fact the Nagumo formulation \cite{PesYur} (of which the FitzHugh-Nagumo model may viewed as a special case) for single component reaction-diffusion models comprises the core of the underlying mathematics responsible for the chemical basis of morphogenesis in biological processes (\emph{e.g.} the Kolmogorov-Petrovsky-Piskunov (KPP) Equation) \cite{PesYur,Turing}.  Fisher's equation is also a biologically relevant reaction-diffusion system, and is used for modeling the propagation of genetic variation over sample populations \cite{QiuSloan}, while in plasma physics the modeling of multicomponent reactive hot plasmas are often reduced down to systems of coupled reaction-diffusion equations \cite{WiLaz}.   Additionally, reaction-diffusion models are of central importance in the study of phase-field models \cite{Takezawa,AllenCahn} and nonequilibrium thermodynamics (we provide a detailed discussion of the latter in \textsection{4.3}), just to mention a very sparse few. 

One important and emergent feature of coupled reaction-diffusion systems is the nuance that arises in understanding that a \emph{diffusion dominated} regime is not necessarily a \emph{diffusion limited} regime --- \emph{e.g.} in the sense of the standard parlance of analytic chemistry \cite{Vijayendran}.  That is, the diffusion rates (or diffusion ``velocity'') will limit the reaction front in a reaction whose kinetics occur on shorter timescales than the particles diffuse (which is a \emph{diffusion limited} process), but in a reaction with timescales that are appreciably longer than the timescales of the diffusion rates of the systems components, the chemical reaction rate can becomes the limiting process (\emph{i.e.} a \emph{reaction limited} process), and the diffusion regime switches from a \emph{diffusion limited} process to a fully \emph{diffusion dominated} process.  In contrast, when the reaction rates occur much faster than the diffusion rates, and when the domain is for example homogenized, then we see a fully \emph{reaction dominated process}.  Notice that different parts of the domain may be characterizes by different regimes.

It should further be noted, as discussed for example in \cite{Vanden} and \cite{lutsko}, that frequently one encounters a full decoupling between the reactive and elastic regimes when the reactive time scales are much slower than those of the dissipative time scales.  This does not, however, account for the popular engineering trend towards multiscale applications \cite{seriesMult,Ingram}, where substantial differences in reaction and diffusing timescales may be present and yet still coupled through a standard reactor regime.

In fact, it is precisely this difference in relative timescales between the diffusion processes of the system and the reaction processes of the system which makes developing a general numerical scheme difficult.  One generally finds, when solving a parabolic system, that the stability condition on the timestep $\Delta t \leq C \Delta x^{2}$ makes formulating an explicit solution unattractive, and implicit methods are favoured.  However, in generalized quiescent reactor systems the timescales of the reactive components of the system may vary wildly in the effective scaling (\emph{viz.} reactions rates on the order of $10^{-15}$ to $10^{15}$ in standard units), while the diffusion scaling may demonstrate substantially less variation.  Because of this, the time-stepping limitations in quiescent reactor models may be either strongly reaction limited, or strongly diffusion limited, or both --- in the sense of oscillating between the regimes, or being split across the regimes.  That is, in reactor systems, where many different reactions may be occurring simultaneously, the depletion of a certain constituent $\alpha_{i}$ at time $t^{n}$ may cause a local in time transition from a \emph{reaction dominated} time-stepping regime to a \emph{diffusion dominated} or \emph{diffusion limited} time-stepping regime, and \emph{vice verse} (in the sense of the operator-splitting regimes of \cite{Spotisse}).  Or, as may occur, parts of the domain may be dominated by constituents which are inert with respect to each other, are uncoupled, and which consequently operate with respect to fundamentally different regimes (\emph{i.e.} \emph{diffusion} versus \emph{reaction limited} regimes).

It must be additionally noted here that this distinction is in many ways a simplification of what can be a very subtle interplay between the reactive and diffusive modes present in complicated reactive mixtures.  For example, it is well known that Fick's law of diffusion is in some cases ill-suited for describing the behavior of some hysteretic mixtures, or diffusion regimes with memory (\emph{e.g.} such as the electrochemistry induced near an active electrolytic cell \cite{Aoki,Einstein,Fedotov}).  In these cases, the local propagation speed caused by the gradient of the concentration forces the Fick's component of the diffusion to obey a telegraph equation, which may often reduce to an integro--differential equation over all time $[0,T]$ coupled to a mass transport equation in the reactive components.  These complications arise in systems that demonstrate large variations in concentrations over short time frames, though a large class of reactions demonstrate even more complicated and subtle behavior that might require the inclusion of quantum effects, such as in \cite{MESV2}.  As a general rule we will not directly address these complications below, as we uniformly make the assumption that the reaction--diffusion system of equations employed is an appropriate approximate model for the system in question. 

Nevertheless, it is because of both the time-stepping nuance mentioned above, as well as the fact that some systems require maximal resolution of highly localized fluctuations in the concentration in order to be well-suited to the particular model system, that we choose to model our quiescent reactor systems by way of an explicit LDG numerical scheme.  We also note that this particular numerical scheme has the advantage of being relatively easily generalized to advection dominated compressible regimes, in which case capturing numerical shock profiles becomes a concern, and is often more easily dealt with in the explicit formulations.
  
More specifically, we introduce a generalized approach to modeling quiescent reactor systems, the theory of which is largely inspired by Ref.~\cite{GV1,MCV1,erdi,Williams,Hirschfelder}.  In \textsection{2} we provide a formulation of the model problem, then develop the temporal discretization and numerical method for performing a predictor multi-corrector over the chemical modes of the system.  We proceed by showing a fairly standard discontinuous Galerkin spatial discretization, and discuss in some detail the iterative methods used along with the temporal mode splitting.  

Let us also note here that a number of very nice numerical approaches to reaction-diffusion systems already exist in the literature.  In addition to the very nice operator splitting methods in the temporal space that employ the Strang method formalism \cite{DM,Miller} and its \emph{entropic structure},  Petrov-Galerkin (SUPG) residual methods have been applied \cite{Gale}, fully adaptive finite volume (multiresolution) methods have been proposed \cite{Rous}, in addition to compact implicit integration factor (cIFF,IFF,cIFF2,ETD) methods over adaptive spatial meshes \cite{LiuNie}, and particle trajectory based methods \cite{Bergdorf}, in addition to the stochastic methods dealing with substantially more complicated molecular scale data \cite{Ferm,Atzberger}.  Moreover, exponential convergence results have been shown for $hp$-adaptive reaction-diffusion systems \cite{Melenk,Xeno,Xeno2} where boundary layer data must be retained in order to achieve full convergence.  In this context we introduce the first --- to our knowledge --- spatially dimension independent $hp$-adaptive operator splitting SSP RKDG predictor multi-corrector scheme for fully generalized reaction--diffusion systems of equations with functionally dependent parameters (\emph{e.g.} $\mathscr{D}(\alpha)$).  

In \textsection{3} we derive the exact entropy relation satisfied by the system, which is borrowed and extended from the regularity analysis of \cite{GV1}, then applying this entropy functional in order to develop an $hp$-adaptive scheme that is fully entropy-consistent --- which is to say entropy-preserving and bounded ---  relying only on the global $\varrho\mathscr{S}^{k+1}_{\mathfrak{R}}$ and local  $\rho\mathscr{S}_{\mathfrak{R},\Omega_{e_{i}}}^{k+1}$ entropy densities, as well as the local change in the density of the entropic jump  $\mathscr{J}_{\mathfrak{R},\Omega_{e_{i}}}^{k+1}$. 

Finally, in \textsection{4} we present some example applications (that were developed in part using a C++ finite element library \cite{dealii}). We address the complication arising from reactive/diffusion dominated/limited regimes explicitly, where we provide four example applications, one which is strongly \emph{reaction dominated} in some areas and \emph{diffusion limited} in others (a set of fast hypergolic combustion reactions), one which is strongly \emph{diffusion dominated} in some areas and \emph{reaction limited} in others (a set of gas-phase alkyl halide atomic transition metal reactions), and one that oscillates between all four regimes (autocatalysis in excitable media across oscillating reactions).  Finally we utilize an equilibrium system in order to demonstrate the standard and expected error convergence results for the method.

\section{\texorpdfstring{\protect\centering $\S 2$ Formulation}{\S 2 Formulation}}

\subsection{\S 2.1 Governing equations}

Consider the stationary reaction-diffusion system, which in chemistry and chemical engineering contexts generalizes our notion of a quiescent reactor, comprised of $i=1,\ldots,n$, species in $N=1,2,$ or $3$ spatial dimensions, satisfying the system of equations: \begin{equation} \begin{aligned}\label{quiescent} &\partial_{t}\alpha_{i}-\nabla_{x}\cdot(\mathscr{D}_{i}(\alpha)\nabla_{x}\alpha_{i}) - \mathscr{A}_{i}(\alpha) =0, \\ & \mathscr{A}_{i}(\alpha)=\sum_{r\in\mathfrak{R}}(\nu_{i,r}^{b}-\nu^{f}_{i,r})\left(k_{f,r}\prod_{j=1}^{n}\alpha_{j}^{\nu_{j,r}^{f}}-k_{b,r}\prod_{j=1}^{n}\alpha_{j}^{\nu_{j,r}^{b}}\right),\end{aligned}\end{equation} with initial-boundary data given by  \begin{equation}\label{robin}\alpha_{i}(t=0)=\alpha_{i,0},\quad \mathrm{and}\quad a_{i}\alpha_{i,b} +  \nabla_{x}\alpha_{i,b}\left( b_{i}\cdot \boldsymbol{n} + c_{i}\cdot \boldsymbol{\tau}\right) - g = 0 \quad \mathrm{on} \ \ \partial\Omega,\end{equation} taking arbitrary functions $a_{i}=a_{i}(t,\boldsymbol{x}_{b}), b_{i}=b_{i}(t,\boldsymbol{x}_{b})$, $c_{i}=c_{i}(t,\boldsymbol{x}_{b})$ and $g_{i} = g_{i}(t,\boldsymbol{x}_{b})$ restricted to the boundary, where $\boldsymbol{n}$ is the unit outward normal and $\boldsymbol{\tau}$ the unit tangent vector at the boundary $\partial\Omega$.  Here $\alpha_{i}$ is the concentration of the $i$-th chemical constituent, the $\mathscr{D}_{i}(\alpha)$ are the inter-species diffusion coefficients which form an $N\times N\times n$ tensor that characterizes the directional dependence on the concentration and its gradient, while $\nu_{i,r}^{f}\in\mathbb{N}$ and $\nu_{i,r}^{b}\in\mathbb{N}$ are the forward and backward stoichiometric coefficients of elementary reaction $r\in\mathbb{N}$ (if $r$ is not elementary then $\nu_{i,r}^{f},\nu_{i,r}^{b}\in\mathbb{Q}$), and $k_{f,r},k_{b,r}\in\mathbb{R}$ are the respective forward and backward reactions rates of reaction $r$.

More precisely, we are interested in systems comprised of $n$ distinct chemical species $\mathfrak{M}_{i}$ (recalling that each constituent's corresponding concentration is given by $\alpha_{i}$) indexed by $r\in\mathfrak{R}$ elementary chemical reactions, for $\mathfrak{R}\subset\mathbb{N}$ where \begin{equation}\label{react}\sum_{j\in\mathscr{R}^{r}} \nu_{j,r}^{f} \mathfrak{M}_{j} \ \ce{<=>T[{$\ \ k_{f,r} \ $\ }][{$\ \ k_{b,r} \ \ $}] } \sum_{k\in\mathscr{P}^{r}}\nu_{k,r}^{b} \mathfrak{M}_{k},\quad \mathrm{for} \ r\in\mathfrak{R},\end{equation} and where $k_{f,r}$ characterizes the forward rate of reaction $r$, and $k_{b,r}$ the backward rate of reaction $r$.  The indexing sets $\mathscr{R}^{r}$ and $\mathscr{P}^{r}$ are the reactant and product sets $\mathscr{R}^{r},\mathscr{P}^{r}\subset\mathbb{N}$ for reaction $r$, respectively.  The $\nu_{j,r}$'s are the stoichiometric coefficients $\nu_{j,r} \in \mathbb{Q}^{+}$ of the products and reagents, and for elementary (or fully reduced form) chemical reactions are positive integers $\nu_{j,r}\in\mathbb{Z}^{+}$ when $i\in\mathscr{R}^{r}$ or $j\in\mathscr{P}^{r}$, since in elementary reactions atoms may only react as absolute entities (which is to say in whole number quantities).   Furthermore, all chemical constituents of the flow are either reactants or products, where inert species may be viewed as the product of a unimolecular reaction denoted $r_{I}$, where $\nu_{r_{I},j}^{f}=\nu_{k,r_{I}}^{b}$ for all $j\in\mathscr{P}^{r_{I}}$ and $k\in\mathscr{R}^{r_{I}}$, such that we may view $\mathscr{P}^{r_{I}}\subset\mathscr{P}^{r}$ and $\mathscr{R}^{r_{I}}\subset\mathscr{P}^{r}$.

Let us proceed by introducing the following $n$ dimensional state vectors: \[\begin{aligned}\boldsymbol{\alpha} = (\alpha_{1},\ldots,\alpha_{n})^{T}, \quad & \mathscr{D}=(\mathscr{D}_{1}(\alpha),\ldots,\mathscr{D}_{n}(\alpha))^{T}, \quad \mathscr{A}(\boldsymbol{\alpha}) = (\mathscr{A}_{1}(\alpha),\ldots,\mathscr{A}_{n}(\alpha))^{T},\end{aligned}\]  and additionally defining the ``auxiliary variable'' $\boldsymbol{\sigma}$, such that using $\mathscr{A} =\mathscr{A}(\boldsymbol{\alpha})$ we may recast (\ref{quiescent}) as the coupled system, \begin{equation} \begin{aligned}\label{quiescentaux} \boldsymbol{\alpha}_{t}-&\nabla_{x}\cdot(\mathscr{D}\boldsymbol{\sigma}) - \mathscr{A} = 0, \quad \mathrm{and}\quad  \boldsymbol{\sigma} - \nabla_{x}\boldsymbol{\alpha} = 0,\end{aligned}\end{equation} where we denote the spatial gradient as, $\nabla_{x}\boldsymbol{\alpha} = \sum_{i=1}^{N}\partial_{x_{i}}\boldsymbol{\alpha}.$

Then we solve (\ref{quiescentaux}) by employing a predictor multi-corrector method coupled to an RKDG scheme which is solved over a reaction mode time-splitting.  Let us first describe the solution vaguely in terms of three basic steps.  In the first step, we use a predictor multi-corrector method to solve (\ref{quiescentaux}), where we exploit the fact that a partially decoupled version of the reaction source term may be solved analytically in order to generate a prediction of the concentrations ($\alpha_{i}$'s) at each timestep.  This predictor is then corrected by way of a fixed point iteration.   In the second step, we solve the components of (\ref{quiescent}) in the usual DG formulation, by integrating against test functions in space and determining local approximations for each of those terms, respectively, were we use an arbitrary order time integrator.  Finally, the third step simply requires determining the ``fast'' and ``slow'' modes with respect to the reactivity of the system, such that for some smallest positive $n\in\mathbb{N}$ we may iterate our solution until $n\Delta t_{f} =\Delta t_{s}$, where we then proceed with the same procedure over all the reacting modes (\emph{i.e.} for each $\alpha_{i}\in \boldsymbol{\alpha}$).  The following sections are devoted to deriving this methodology.

\subsection{\S 2.2 The predictor multi-corrector}

First notice that the reaction term $\mathscr{A}^{n}=\mathscr{A}^{n}(\boldsymbol{\alpha})$ may be viewed as the source of a proliferating set of nontrivial numerical difficulties.  That is, not only is it well known that $\mathscr{A}^{n}$ may cause numerical instabilities due to the varying ``fast'' and ``slow'' timescales discussed above, but due to the presence of nonlinearities arising from the stoichiometric coefficients $\nu_{i,r}$,  the $\mathscr{A}^{n}(\boldsymbol{\alpha})$ term is responsible for generating a coupled system of $n$ highly nonlinear first order ordinary differential equations.  Thus, in order to formulate a computationally realistic numerical method for solving our system (\ref{quiescentt}) over some modest (yet realistic) number of constituents $n$, we find it necessary to employ the following linearization.

Let us first denote the vector $\mathscr{A}_{i}(\beta_{i},\boldsymbol{\alpha})$ as \[\mathscr{A}_{i}(\beta_{i},\boldsymbol{\alpha})=\sum_{r\in\mathfrak{R}}(\nu_{j,r}^{b}-\nu^{f}_{j,r})\left(k_{f,r}\beta_{i}^{\nu_{i,r}^{f}}\prod_{j=1,j\neq i}^{n}\alpha_{j}^{\nu_{j,r}^{f}}-k_{b,r}\beta_{i}^{\nu_{i,r}^{b}}\prod_{j=1,j\ne i}^{n}\alpha_{j}^{\nu_{j,r}^{b}}\right),\] where $\boldsymbol{\alpha}$ is treated as constant for all $j\neq i$.  In other words, we wish to think of  $\mathscr{A}_{i}(\beta_{i},\boldsymbol{\alpha})$ as the mass action vector such that all but $\beta_{i}$ is treated as temporally inert.  

Then using this notation, we proceed by considering the system of equations (\ref{quiescentaux}) and discretizing in time, such that at time $t^{n+1}$ we are interested in solving the semi-implicit system of equations: \begin{equation}\begin{aligned}\label{quiescentt} \left(\frac{\boldsymbol{\alpha}^{n+1} - \boldsymbol{\alpha}^{n}}{\Delta t}\right) & = \nabla_{x}\cdot(\mathscr{D}^{n}\nabla_{x}\boldsymbol{\alpha}^{n}) + \mathring{\mathscr{A}}(\hat{\boldsymbol{\alpha}}^{n+1},\boldsymbol{\alpha}^{n}), \quad\mathrm{and}\quad \boldsymbol{\sigma}^{n} = \nabla_{x}\boldsymbol{\alpha}^{n}.\end{aligned}\end{equation} 

The predictor term $\mathring{\mathscr{A}}(\hat{\boldsymbol{\alpha}}^{n+1},\boldsymbol{\alpha}^{n})$ is predicated on the notion that we may ``predict'' the approximate value of the coupled reaction rate of each species $\mathfrak{M}_{i}$ at time $t^{n+1}$ by decoupling the $n$-th order system of first order ordinary differential equations containing the nonlinear $\nu_{i,r}^{f}$ and $\nu_{i,r}^{b}$ factors (as seen in (\ref{quiescent})) by simply using an analytic rate law derived with respect to the value at the previous timestep $t^{n}$.  

That is, for the $i$-th molecular constituent $\mathfrak{M}_{i}$ we predict its concentration $\alpha_{i}$ at time $t^{n+1}$, in either the reactant or product well (\emph{i.e.} in the reaction coordinate representation), by analytically solving the following first order ordinary differential equation, \begin{equation}\label{fir}\partial_{t}\tilde{\alpha}_{i}= \mathscr{A}_{i}(\tilde{\alpha}_{i},\boldsymbol{\alpha}).\end{equation}  We solve the integrated rate law in time over a discrete timestep $\Delta t = t^{n+1}-t^{n}$, such that we obtain an analytic form for each $\tilde{\alpha}_{i}^{n+1}$ in terms of the solution at the previous timestep $t^{n}$ treated as constants (again as denoted $\boldsymbol{\alpha}$ in the term $\mathscr{A}(\alpha_{i},\boldsymbol{\alpha})$ for all species of index $j\neq i$).  We refer to these solutions (of (\ref{fir})) as the ``predicted values'' of $\alpha_{i}^{n+1}$ and denote them either componentwise by $\tilde{\alpha}_{i}^{n+1}$ or in vector form by $\tilde{\boldsymbol{\alpha}}^{n+1}$.

Now, of course, the term $\mathring{\mathscr{A}}(\hat{\boldsymbol{\alpha}}^{n+1},\boldsymbol{\alpha}^{n})$ abstractly represents the rate of change of the concentrations $\alpha_{i}$'s at time $t^{n+1}$, while the vector $\hat{\boldsymbol{\alpha}}^{n+1}$ will be an averaged form of the predicted total concentration at time $t^{n+1}$.  Thus, in order to find the rate of change predictor in the residual representation $\mathring{\mathscr{A}}(\hat{\boldsymbol{\alpha}}^{n+1},\boldsymbol{\alpha}^{n})$ we simply find the formal difference, \begin{equation}\label{diffalpha} \mathring{\mathscr{A}}(\hat{\boldsymbol{\alpha}}^{n+1},\boldsymbol{\alpha}^{n})= \left(\frac{\hat{\boldsymbol{\alpha}}^{n+1}-\boldsymbol{\alpha}^{n}}{\Delta t}\right).\end{equation}   

Next we implement a fixed point iteration corrector over $i=1,\ldots,\ell$ iterates in order to provide convergence in the solution $\boldsymbol{\alpha}^{n+1}$.  That is, for fixed timestep $\Delta t$ we denote the $i$-th corrector of (\ref{quiescentt}), implementing our predictor to find a corrector over each iterate $i=0,\ldots,\ell$, via: 
\begin{equation}\begin{aligned}\label{quiescentpred} & (\boldsymbol{\alpha}^{n+1})^{i+1} = \boldsymbol{\alpha}^{n} + \Delta t\left(\nabla_{x}\cdot(\mathscr{D}^{n}\boldsymbol{\sigma}^{n})+ \mathring{\mathscr{A}}((\hat{\boldsymbol{\alpha}}^{n+1})^{i+1},\boldsymbol{\alpha}^{n})\right), \quad (\boldsymbol{\alpha}^{n+1})^{0}= (\hat{\boldsymbol{\alpha}}^{n+1})^{0},  \\ & \qquad \qquad\qquad\quad (\hat{\boldsymbol{\alpha}}^{n+1})^{i+1} = \left(\frac{(\tilde{\boldsymbol{\alpha}}^{n+1})^{i+1}+(\tilde{\boldsymbol{\alpha}}^{n+1})^{i}}{2}\right), \quad \boldsymbol{\sigma}^{n} = \nabla_{x}\boldsymbol{\alpha}^{n}.\end{aligned}\end{equation}  It is important here to recall that the $(\tilde{\boldsymbol{\alpha}}^{n+1})$'s are explicitly determined by the derived analytic solutions to (\ref{fir}), which depend on the specific reaction system.  It is also worth noting that (\ref{fir}) is solved iteratively in the sense that $\partial_{t}\tilde{\alpha}_{i}^{i+1}= \mathscr{A}_{i}(\tilde{\alpha}_{i}^{i+1},(\boldsymbol{\alpha}^{n+1})^{i})$ is integrated over $\Delta t$ to form what is the analytic rate law of the new predictor at the $(i+1)$-st iterate of timestep $t^{n+1}$.  Note here that we have chosen a splitting between the explicit diffusion terms and the semi-implicit reaction terms, which we have found (by trial and error) to be the correct splitting to maximize the robustness of our method, with respect to both fast and complicated multistable reaction regimes (see \textsection{4} for examples).

The endpoint of the iteration $\ell$ is chosen in tandem with the following bound on the ``relative change'' of the iterated corrector componentwise in $j$, (\emph{i.e.} the component $(\alpha_{j}^{n+1})^{\ell}$): \begin{equation}\label{converge}\left(\frac{\|(\alpha_{j}^{n+1})^{\ell} -  (\alpha_{j}^{n+1})^{\ell-1} \|_{L^{\infty}(\Omega)}}{\| (\alpha_{j}^{n+1})^{\ell-1}\|_{L^{\infty}(\Omega)}}\right)\leq C,\end{equation} for a judicially chosen constant $C$ (\emph{e.g.} see \textsection{4} where $C\in\{10^{-6},10^{-14}\}$).  In fact, we set a slightly stronger condition than (\ref{converge}), after spatial discretization in the discontinuous Galerkin setting, as we provide componentwise convergence in the above sense with respect to each quadrature point.  Note that we must converge componentwise in each chemical species $\mathfrak{M}_{j}$, since the relative orders of concentrations over $\mathfrak{R}$ may substantially vary, such that while the rate limiting products may readily converge, the coupled ancillary products may oscillate wildly, and \emph{vice versa}. In order to circumvent this pathology, we demand convergence componentwise globally in space for each timestep.


\subsection{\S 2.3 Spatial discretization}

Now let use discretize in space.  Take an open $\Omega\subset\mathbb{R}$ with boundary $\partial\Omega=\Gamma$, given $T>0$ such that $\mathcal{Q}_{T}=((0,T)\times\Omega)$.  Let $\mathscr{T}_{h}$ denote the partition of the closure of the polygonal triangulation of $\Omega$, which we denote $\Omega_{h}$, into a finite number of polygonal elements denoted $\Omega_{e}$, such that  $\mathscr{T}_{h}= \{\Omega_{e_1},\Omega_{e_2}, \ldots,\Omega_{e_{ne}}\}$, for $ne\in\mathbb{N}$ the number of elements in $\Omega_{h}$.   In this work we define the mesh diameter $h$ to satisfy $h = \min_{ij}(d_{ij})$ for the distance function $d_{ij}= d(\boldsymbol{x}_{i},\boldsymbol{x}_{j})$ and elementwise edge vertices $\boldsymbol{x}_{i},\boldsymbol{x}_{j}\in\partial\Omega_{e}$ when the mesh is structured and regular.  For unstructured meshes we mean the average value of $h$ over the mesh.

Now, let $\Gamma_{ij}$ denote the edge shared by two neighboring elements $\Omega_{e_{i}}$ and $\Omega_{e_{j}}$, and for $i\in I\subset\mathbb{Z}^{+}=\{1,2,\ldots\}$ define the indexing set $r(i)=\{j \in I : \Omega_{e_{j}}$ is a neighbor of $\Omega_{e_{i}}\}$.  Let us denote all boundary edges of $\Omega_{e_{i}}$ contained in $\partial\Omega_{h}$ by $S_{j}$ and letting $I_{B}\subset \mathbb{Z}^{-}=\{-1,-2,\ldots\}$ define $s(i)=\{j\in I_{B}:S_{j}$ is an edge of $\Omega_{e_{i}}\}$ such that $\Gamma_{ij}=S_{j}$ for $\Omega_{e_{i}}\in \Omega_{h}$ when $S_{j}\in\partial\Omega_{e_{i}}$, $j\in I_{B}$.  Then for $\Xi_{i}=r(i)\cup s(i)$, we have \[\partial\Omega_{e_{i}}=\bigcup_{j\in \Xi(i)}\Gamma_{ij},\quad\mathrm{and}\quad \partial\Omega_{e_{i}}\cap\partial\Omega_{h} = \bigcup_{j\in s(i)}\Gamma_{ij}.\] 

We are interested in obtaining an approximate solution to $\boldsymbol{U}$ at time $t$ on the finite dimensional space of discontinuous piecewise polynomial functions over $\Omega$ restricted to $\mathscr{T}_{h}$, given as \[S_{h}^{d}(\Omega_{h},\mathscr{T}_{h})=\{v:v_{|\Omega_{e_{i}}}\in \mathscr{P}^{d}(\Omega_{e_{i}}) \ \ \forall\Omega_{e_{i}}\in\mathscr{T}_{h}\}\] for $\mathscr{P}^{d}(\Omega_{e_{i}})$ the space of degree $\leq d$ polynomials over $\Omega_{e_{i}}$.      

Choosing a set of degree $d$ polynomial basis functions $N_{l}\in\mathscr{P}^{d}(\mathcal{G}_{i})$ for $ l=0,\ldots, p$ we can denote the state vector at time $t$ over $\Omega_{h}$, by \[\boldsymbol{\alpha}_{h}(t,\boldsymbol{x})=\sum_{l=0}^{d}\boldsymbol{\alpha}_{l}^{i}(t)N^{i}_{l}(\boldsymbol{x}),\quad  \forall \boldsymbol{x}\in\Omega_{e_{i}},\] where the $N^{i}_{l}$'s are the finite element shape functions in the DG setting, and the $\boldsymbol{\alpha}_{l}^{i}$'s correspond to the nodal unknowns.   The finite dimensional test functions  $\boldsymbol{\varphi}_{h}, \boldsymbol{\varsigma}_{h}\in W^{2,2}(\Omega_{h},\mathscr{T}_{h})$ are characterized by \[\begin{aligned}\boldsymbol{\varphi}_{h}(\boldsymbol{x})=\sum_{l=0}^{d}\boldsymbol{\varphi}_{l}^{i}N_{l}^{i}(\boldsymbol{x})  \quad\mathrm{and}\quad \boldsymbol{\varsigma}_{h}(\boldsymbol{x})=\sum_{l=0}^{d}\boldsymbol{\varsigma}_{l}^{i}N_{l}^{i}(\boldsymbol{x}) \quad \forall \boldsymbol{x}\in\mathcal{G}_{i},\end{aligned}\] where $\boldsymbol{\varphi}_{\ell}^{i}$ and $\boldsymbol{\varsigma}_{\ell}^{i}$ are the nodal values of the test functions in each $\Omega_{e_{i}}$, and with the broken Sobolev space over the partition $\mathscr{T}_{h}$ defined by \[W^{k,2}(\Omega_{h},\mathscr{T}_{h})=\{v : v_{|\Omega_{e_{i}}}\in W^{k,2}(\Omega_{e_{i}}) \ \ \forall\Omega_{e_{i}}\in\mathscr{T}_{h}\}.\]

We thus multiply (\ref{quiescentpred}) by the test functions $\boldsymbol{\varphi}_{h}$ and $\boldsymbol{\varsigma}_{h}$ and then integrate locally over elements $\Omega_{e_{i}}$ in space, defining global scalar products, $(\boldsymbol{a}_{h}^{n},\boldsymbol{b}_{h})_{\Omega_{\mathcal{G}}} = \sum_{\Omega_{e_{_{i}}}\in\mathscr{T}_{h}}\int_{\Omega_{e_{i}}}\boldsymbol{a}_{h}^{n}:\boldsymbol{b}_{h} dx$, such that we obtain the system: \begin{equation}\begin{aligned} \label{dgsteady} \frac{1}{\Delta t}&\left((\boldsymbol{\alpha}^{n+1})^{i+1} - \boldsymbol{\alpha}^{n}, \boldsymbol{\varphi}_{h}\right)_{\Omega_{\mathcal{G}}} = \left( \nabla_{x}\cdot(\mathscr{D}^{n}\boldsymbol{\sigma}^{n}),\boldsymbol{\varphi}_{h}\right)_{\Omega_{\mathcal{G}}} + \left(\mathring{\mathscr{A}}((\hat{\boldsymbol{\alpha}}^{n+1})^{i+1},\boldsymbol{\alpha}^{n}), \boldsymbol{\varphi}_{h} \right)_{\Omega_{\mathcal{G}}}, \\ & \qquad\qquad\qquad\qquad\left((\boldsymbol{\alpha}^{n+1})^{0},\boldsymbol{\varphi}_{h}\right)_{\Omega_{\mathcal{G}}} =  \left((\hat{\boldsymbol{\alpha}}^{n+1})^{0}, \boldsymbol{\varphi}_{h}\right)_{\Omega_{\mathcal{G}}},\\ & \qquad\qquad \left((\hat{\boldsymbol{\alpha}}^{n+1})^{i+1},\boldsymbol{\varphi}_{h}\right)_{\Omega_{\mathcal{G}}} = \left(\frac{(\tilde{\boldsymbol{\alpha}}^{n+1})^{i+1}+(\tilde{\boldsymbol{\alpha}}^{n+1})^{i}}{2},\boldsymbol{\varphi}_{h}\right)_{\Omega_{\mathcal{G}}}, \\ & \qquad\qquad\qquad\qquad\left(\boldsymbol{\sigma}^{n},\boldsymbol{\varsigma}_{h}\right)_{\Omega_{\mathcal{G}}} - \left( \nabla_{x} \boldsymbol{\alpha}^{n},\boldsymbol{\varsigma}_{h}\right)_{\Omega_{\mathcal{G}}} = 0.\end{aligned}\end{equation}  
 
We proceed by approximating each term of (\ref{dgsteady}) in the usual DG sense.  That is, we approximate the first term on the left in the first equation in (\ref{dgsteady}) by, \begin{equation}
\begin{aligned}\label{timeterm}
\frac{1}{\Delta t}\left((\boldsymbol{\alpha}_{h}^{n+1})^{i+1} - \boldsymbol{\alpha}_{h}^{n}, \boldsymbol{\varphi}_{h}\right)_{\Omega_{\mathcal{G}}} \approx  \frac{1}{\Delta t}\left((\boldsymbol{\alpha}^{n+1})^{i+1} - \boldsymbol{\alpha}^{n}, \boldsymbol{\varphi}_{h}\right)_{\Omega_{\mathcal{G}}}.
\end{aligned}
\end{equation}

Now, let $\boldsymbol{n}_{ij}$ be the unit outward normal to $\partial\Omega_{e_{i}}$ on $\Gamma_{ij}$, and let $\varphi_{|\Gamma_{ij}}$ and  $\varphi_{|\Gamma_{ji}}$ denote the values of $\varphi$ on $\Gamma_{ij}$ considered from the interior and the exterior of $\Omega_{e_{i}}$, respectively. Then the second term of the first equation in (\ref{dgsteady}), after an integration by parts, yields, \begin{equation}\label{diffusion} \left( \nabla_{x}\cdot(\mathscr{D}^{n}\boldsymbol{\sigma}^{n}),\boldsymbol{\varphi}_{h}\right)_{\Omega_{\mathcal{G}}} =  \sum_{\Omega_{e_{i}}\in\mathscr{T}_{h}}\int_{\Omega_{e_{i}}}\nabla_{x}\cdot (\boldsymbol{\varphi}_{h}\mathscr{D}^{n}\boldsymbol{\sigma}^{n}) dx -   \left(\mathscr{D}^{n}\boldsymbol{\sigma}^{n},\nabla_{x}\boldsymbol{\varphi}_{h}\right)_{\Omega_{\mathcal{G}}},\end{equation} such that we approximate the first term on the right in (\ref{diffusion}) using a generalized viscous flux $\hat{\mathscr{G}}$ (see Ref.~\cite{ABCM}) across the boundary, where upon setting $\mathscr{G}_{i}=\mathscr{G}_{i}(\boldsymbol{\sigma}_{h}^{n},\boldsymbol{\alpha}_{h}^{n},\boldsymbol{\varphi}_{h})$, we approximate
\begin{equation}
\begin{aligned}
\label{viscousreact1}
\mathscr{G}_{i} & =  \sum_{j\in S(i)}\int_{\Gamma_{ij}}\hat{\mathscr{G}}(\boldsymbol{\sigma}_{h}^{n}|_{\Gamma_{ij}},\boldsymbol{\sigma}_{h}^{n}|_{\Gamma_{ji}}, \boldsymbol{\alpha}_{h}^{n}|_{\Gamma_{ij}}, \boldsymbol{\alpha}_{h}^{n}|_{\Gamma_{ji}}, \boldsymbol{n}_{ij})\cdot\boldsymbol{\varphi}_{h}|_{\Gamma_{ij}} d\Xi \\ & \approx   \sum_{j\in S(i)}\int_{\Gamma_{ij}}\sum_{s=1}^{N}(\mathscr{D}_{h}^{n}\boldsymbol{\sigma}^{n})_{s}\cdot (n_{ij})_{s}\boldsymbol{\varphi}_{h}|_{\Gamma_{ij}}d\Xi,
\end{aligned}
\end{equation} while the second term in (\ref{diffusion}) is approximated by:
\begin{equation}\label{seconddif}\mathscr{H} = \mathscr{H}(\boldsymbol{\sigma}_{h}^{n},\boldsymbol{\alpha}_{h}^{n},\boldsymbol{\varphi}_{h})= \left(\mathscr{D}\boldsymbol{\sigma}^{n}_{h},\boldsymbol{\varphi}^{h}_{x}\right)_{\Omega_{\mathcal{G}}} \approx \left(\mathscr{D}\nabla_{x}\boldsymbol{\alpha}^{n},\boldsymbol{\varphi}^{h}_{x}\right)_{\Omega_{\mathcal{G}}}.\end{equation} 

The reaction term in (\ref{dgsteady}), which is dealt with using the predictor multi-corrector, is projecting into the basis in the obvious way, \begin{equation}\label{adis} \left(\mathring{\mathscr{A}}((\hat{\boldsymbol{\alpha}}_{h}^{n+1})^{i+1},\boldsymbol{\alpha}_{h}^{n}), \boldsymbol{\varphi}_{h} \right)_{\Omega_{\mathcal{G}}}\approx \left(\mathring{\mathscr{A}}((\hat{\boldsymbol{\alpha}}^{n+1})^{i+1},\boldsymbol{\alpha}^{n}), \boldsymbol{\varphi}_{h} \right)_{\Omega_{\mathcal{G}}},\end{equation} where the second and third equations in (\ref{dgsteady}) are approximated componentwise in the usual sense, simply setting: \begin{equation}\begin{aligned}\label{2and3} \left((\hat{\boldsymbol{\alpha}}_{h}^{n+1})^{0},\boldsymbol{\varphi}_{h}\right)_{\Omega_{\mathcal{G}}} & \approx \left((\hat{\boldsymbol{\alpha}}^{n+1})^{0},\boldsymbol{\varphi}_{h}\right)_{\Omega_{\mathcal{G}}}, \\ \left(\frac{(\tilde{\boldsymbol{\alpha}}_{h}^{n+1})^{i}+(\tilde{\boldsymbol{\alpha}}_{h}^{n+1})^{i-1}}{2},\boldsymbol{\varphi}_{h}\right)_{\Omega_{\mathcal{G}}}&\approx  \left(\frac{(\tilde{\boldsymbol{\alpha}}^{n+1})^{i}+(\tilde{\boldsymbol{\alpha}}^{n+1})^{i-1}}{2},\boldsymbol{\varphi}_{h}\right)_{\Omega_{\mathcal{G}}}.\end{aligned}\end{equation}

Finally, for the fourth equation in (\ref{dgsteady}) a numerical flux is chosen which satisfies: \begin{equation} \begin{aligned}\label{penaltyreact1}  & \mathscr{L}_{i} = \mathscr{L}_{i}(\check{\boldsymbol{\alpha}},\boldsymbol{\sigma}_{h}^{n},\boldsymbol{\alpha}_{h}^{n},\boldsymbol{\varsigma}_{h},\boldsymbol{\varsigma}_{x}^{h})  = \left(\boldsymbol{\sigma}_{h}^{n},\boldsymbol{\varsigma}_{h}\right)_{\Omega_{e_{i}}}  + \left(\boldsymbol{\alpha}_{h}^{n},\boldsymbol{\varsigma}^{h}_{x}\right)_{\Omega_{e_{i}}} \\ & \qquad\qquad\qquad\qquad\qquad\qquad - \sum_{j\in S(i)}\int_{\Gamma_{ij}}\check{\boldsymbol{\alpha}}(\boldsymbol{\alpha}^{n}_{h}|_{\Gamma_{ij}},\boldsymbol{\alpha}^{n}_{h}|_{\Gamma_{ji}},\boldsymbol{\varsigma}_{h}|_{\Gamma_{ij}},\boldsymbol{n}_{ij}) d\Xi, \\ & \quad \mathrm{where} \ \ \sum_{j\in S(i)}\int_{\Gamma_{ij}}\check{\boldsymbol{\alpha}}(\boldsymbol{\alpha}^{n}_{h}|_{\Gamma_{ij}},\boldsymbol{\alpha}^{n}_{h}|_{\Gamma_{ji}},\boldsymbol{\varsigma}_{h}|_{\Gamma_{ij}},\boldsymbol{n}_{ij}) d\Xi \approx  \sum_{j\in S(i)}\int_{\Gamma_{ij}}\sum_{s=1}^{N}(\boldsymbol{\alpha}^{n})_{s}\cdot (n_{ij})_{s} \boldsymbol{\varsigma}_{h}|_{\Gamma_{ij}}d\Xi.\end{aligned}\end{equation}

\subsection{\S 2.4 Formulation of the problem}

Combining (\ref{timeterm}), and (\ref{viscousreact1})--(\ref{penaltyreact1}) while setting $\mathscr{X} = \sum_{\mathcal{G}_{i}\in\mathscr{T}_{h}}\mathscr{X}_{i}$, we formulate our approximate solution to (\ref{quiescent}) via (\ref{quiescentpred}) which by construction may be stated as: for each $n\geq 0$, $C\in[0,1]$ and $\ell>0$, find the pair $(\boldsymbol{\alpha}_{h}^{n},\boldsymbol{\sigma}_{h}^{n})$ such that:\begin{center}\underline{The Predictor Multi-corrector DG Solution}\end{center}
\begin{equation}
\begin{aligned}
\label{aproxreact}
& a) \ \ \ \boldsymbol{\alpha}_{h}\in C^{1}([0,T);S_{h}^{d}), \ \boldsymbol{\sigma}_{h}\in S_{h}^{d}, \\
& b)  \  \left((\boldsymbol{\alpha}_{h}^{n+1})^{i+1} - \boldsymbol{\alpha}^{n}_{h}, \boldsymbol{\varphi}_{h}\right)_{\Omega_{\mathcal{G}}} = \Delta t\left(\mathscr{G} + \mathscr{H}\right) + \left(\mathring{\mathscr{A}}((\hat{\boldsymbol{\alpha}}_{h}^{n+1})^{i+1},\boldsymbol{\alpha}_{h}^{n}), \boldsymbol{\varphi}_{h} \right)_{\Omega_{\mathcal{G}}},\\
& c) \ \left((\boldsymbol{\alpha}_{h}^{n+1})^{0},\boldsymbol{\varphi}_{h}\right)_{\Omega_{\mathcal{G}}} =  \left((\hat{\boldsymbol{\alpha}}_{h}^{n+1})^{0}, \boldsymbol{\varphi}_{h}\right)_{\Omega_{\mathcal{G}}}, \\
& d) \ \left((\hat{\boldsymbol{\alpha}}^{n+1})_{h}^{i+1},\boldsymbol{\varphi}_{h}\right)_{\Omega_{\mathcal{G}}} = \left(\frac{(\tilde{\boldsymbol{\alpha}}^{n+1})^{i+1}+(\tilde{\boldsymbol{\alpha}}^{n+1})^{i}}{2}, \boldsymbol{\varphi}_{h}\right)_{\Omega_{\mathcal{G}}}, \\
& e) \ \left(\frac{\|(\alpha_{j,h}^{i,n+1})^{\ell}\varphi_{h}^{i} -  (\alpha_{j,h}^{i,n+1})^{\ell-1}\varphi_{h}^{i} \|_{L^{\infty}(\Omega)}}{\| (\alpha_{j,h}^{i,n+1})^{\ell-1}\varphi_{h}^{i}\|_{L^{\infty}(\Omega)}}\right)\leq C,\quad \forall(\boldsymbol{x},j)\in (\Omega_{e_{i}},\mathfrak{R}),
\\
& f) \ \ \  \mathscr{L}(\check{\boldsymbol{\alpha}},\boldsymbol{\sigma}_{h}^{n},\boldsymbol{\alpha}_{h}^{n},\boldsymbol{\varsigma}_{h},\boldsymbol{\varsigma}_{x}^{h}) = 0, 
\\
& g) \ \ \ \boldsymbol{\alpha}_{h}(0)=\Pi_{h}\boldsymbol{\alpha}_{0}, \ \nabla_{x}\boldsymbol{\alpha}_{h}(0)=\tilde{\Pi}_{h}\boldsymbol{\alpha}_{0}.
\end{aligned}
\end{equation}  Here, $\Pi_{h}$ is a projection operator onto the space of discontinuous piecewise polynomials $S_{h}^{p}$, and where below we always utilize a standard $L^{2}$--projection, given for a function $\boldsymbol{f}_{0}\in L^{2}(\Omega_{e_{i}})$ such that our approximate projection $\boldsymbol{f}_{0,h}\in L^{2}(\Omega_{e_{i}})$ is obtained by solving, $\int_{\Omega_{e_{i}}}\boldsymbol{f}_{0,h}\boldsymbol{v}_{h} dx = \int_{\Omega_{e_{i}}}\boldsymbol{f}_{0}\boldsymbol{v}_{h} dx.$   

The gradient projection $\tilde{\Pi}_{h}$ merely approximates the initial gradients using numerical difference quotients, for example below we frequently employ the approximate fourth order scheme: \[ f' \approx \left(\frac{f(x-2h)-8 f(x-h)+8f(x+h)-f(x+2h)}{12h}\right).\]

Now, it follows that (\ref{aproxreact}) is a solution for any reaction scheme of arbitrary order satisfying the \emph{law of mass action} (\ref{quiescent}), whether or not the reactions in $\mathscr{A}_{i}(\alpha)$ are elementary, and regardless of the reaction order (\emph{e.g.} mixed reaction orders of arbitrary type that change order during the course of the reaction, and fractional order reactions, \emph{etc}.) 

We conclude using a standard time discretization for (\ref{aproxreact}), where we employ a family of SSP (strong stability preserving) Runge-Kutta schemes as discussed in \cite{Ruuth,SO}.  That is, we may abstractly represent our ODE in (\ref{aproxreact}) by $\frac{d}{dt}\boldsymbol{\alpha} = \mathcal{L}(\boldsymbol{\alpha})$, such that previously the first order forward Euler time discretization was assumed (equation b in (\ref{aproxreact})).  However, to generalize to a $\gamma$ stage SSP Runge-Kutta method of order $\wp$ (which we denote $\mathrm{SSP}(\gamma,\wp)$), we simply augment the second equation in (\ref{aproxreact}) in the diffusion terms by: \begin{equation}\begin{aligned}\label{SSPRK} & b_{i}) \ \ \ \boldsymbol{\alpha}_{h}^{(0)}=\boldsymbol{\alpha}_{h}^{n}, \\ & b_{ii}) \ \ \left((\boldsymbol{\alpha}_{h}^{(j)})^{i+2},\boldsymbol{\varphi}_{h}) \right)_{\Omega_{\mathcal{G}}}= \sum_{k = 0}^{j-1}\left(\lambda_{jk}\boldsymbol{\alpha}_{h}^{k}+\Delta t\tilde{\lambda}_{jk}\mathcal{L}^{k},\boldsymbol{\varphi}_{h} \right)_{\Omega_{\mathcal{G}}} \quad\mathrm{for} \ j=1,\ldots,\gamma \\ & b_{iii}) \ \ \ \boldsymbol{\alpha}_{h}^{n+1} =\boldsymbol{\alpha}_{h}^{\gamma},\end{aligned}\end{equation} where $\mathcal{L}^{k} = \mathcal{L}(\boldsymbol{\alpha}_{h}^{k})$ can be viewed as the abstract diffusion operator, and the solution at the $n$--th timestep is given as $\boldsymbol{\alpha}_{h}^{n}=\boldsymbol{\alpha}_{h,|t=t^{n}}$ and at the $n$--th plus first timestep by $\boldsymbol{\alpha}_{h}^{n+1}=\boldsymbol{\alpha}_{h,|t=t^{n+1}}$, where $\Delta t =t^{n+1}-t^{n}$.  The order $\wp$ of the method is fully determined by the choice of the coefficients $\lambda_{jk}$ and $\tilde{\lambda}_{jk}$ in the Butcher tableau.  More clearly, the abstract operator $\mathcal{L}^{k}$ does not alter the mass action term containing $\mathring{\mathscr{A}}((\hat{\boldsymbol{\alpha}}_{h}^{n+1})^{i+1},\boldsymbol{\alpha}_{h}^{n})$ in (\ref{aproxreact}).  That is, as already explained in \textsection{2.1} the predictor multi-corrector scheme provides an approximate solution chosen with respect to a distinct temporal integrator which is strongly dependent on the exact solution of a partially decoupled system of ODEs; and is thus taken outside the Runge-Kutta loop. 

It remains to identify the ``fast'' $\Delta t_{f}$ and ``slow'' $\Delta t_{s}$ modes of the system with respect to the form of the equations (\ref{quiescentaux} often occurring on substantially different time scales, as illustrated in \cite{DM}.  However, the fast modes and the slow modes separate not only with respect to the reaction coordinate (as determined by $\mathscr{A}$), but also separate with respect to the interspecies diffusion as determined by Fick's law, which may also be decomposed into fast and slow moving modes.  To account for this complication we simply decompose the concentration over ``fast'' $\boldsymbol{\alpha}_{f}$ and ``slow'' $\boldsymbol{\alpha}_{s}$ variables, using $\boldsymbol{\alpha}=(\boldsymbol{\alpha}_{f},\boldsymbol{\alpha}_{s})^{T}$.

Such an operator splitting into fast and slow modes has been thoroughly presented in \cite{DM}, and is known to lead to a fully well-posed system of reaction-diffusion equations, providing the existence of an \emph{entropic structure} and a partial equilibrium manifold.  The \emph{entropic structure} discussed in \cite{DM} is quite strong however, requiring for example the existence of a $C^{\infty}$ monotone entropy functional.  Clearly in the context of our approximate variational solution (\ref{aproxreact}) this constraint is too restrictive (esp.  with respect to a discontinuous polynomial basis and with an eye towards generalizing in order to easily extend our formalism to the reactive multicomponent Navier--Stokes regimes, for example).

Thus in order to stabilize our method we introduce an exact entropic restriction as outlined in \textsection{3} below, based on the analytic well-posedness results of A.~Vasseur, T.~Goudon and C.~Caputo \cite{GV1,MCV1}, which depends strongly on an explicit analytic entropy functional $\mathscr{S}_{\mathfrak{R}}$.  We enforce entropy consistency on our solution, which provides for the usual monotonicity constraint on the systems entropy, but we further expand this constraint and then utilize the entropy consistent scheme as a foundation for a dynamic $hp$-adaptive strategy as fully derived in \textsection{3}. 

We further note that the notion of ``fast'' and ``slow'' modes here is made to highlight a qualitative choice, where the physics of the system may, of course, be substantially more complicated.  That is, for simplicity in our derivation, we have assumed that the rate laws split into no more than two distinct sets, of ``fast'' and ``slow;'' while there may of course be $k$ arbitrary such sets representing $k$ grouped rates each of a quantitatively different order of magnitude.  While in some physical systems it is essential to neglect the chemical kinetics of reactions occurring on substantially different timescales (\emph{e.g.} neutrino production rates in atmospheric chemistry, \emph{etc}.), in many settings (such as in environmental science, for example) it is important to include reactions occurring in a number of different phases (\emph{i.e.} ice, water, water vapour, \emph{etc.}), which can have a large array of different timescales for their coupled rates laws.  More explicitly, in standard units, common chemical reaction rates can differ in a particular setting and choice of units up to some twenty orders of magnitude.

Thus the solution obtained from (\ref{aproxreact}) trivially lends itself to these split time discretizations ($\Delta t_{f},\cdots,\Delta t_{s})$, and thus we numerically integrate over the ``faster'' variables (\emph{i.e.} the $\alpha_{i}\in\boldsymbol{\alpha}_{f}$) in $\Delta t_{f}$.  That is, for some smallest positive $n\in\mathbb{N}$ we recurse our solution until $n\Delta t_{f} =\Delta t_{s}$ (thus slightly restricting the permissible choice of $\Delta t_{f}$), where we proceed with the same procedure over all the reacting modes (\emph{i.e.} $\forall \alpha_{i}\in \boldsymbol{\alpha}$).  In this way we easily acquire the approximate solution over the time split modes.

\section{\texorpdfstring{\protect\centering $\S 3$ Entropy enriched $hp$-adaptivity}{\S 3 Entropy enriched \emph{hp}-adaptivity}}

We are now concerned with an entropy based $hp$-adaptive methodology which does not depend on the powerful, though often computationally prohibitive, adjoint problem formulation which relies upon the calculation of \emph{a posteriori} error estimates and minimization techniques (\emph{e.g.} see \cite{Dedner}).  That is, our method is potentially cheaper computationally, while still maintaining a rigorous foundation based on satisfying \emph{a priori} entropy bounds.

\subsection{\S 3.1 Bounded entropy in quiescent reactors}

We use a formal entropy inequality discovered in \cite{GV1} in order to develop a stability analysis of our approximate solutions.  

Let us derive the entropy of the system $\mathscr{S}_{\mathfrak{R}}$ over each reaction $r\in\mathfrak{R}$ in the quiescent reactor.  First define the mass action term reactionwise, such that we have \begin{equation}\mathscr{Q}_{i,r}(\alpha) = (\nu_{i,r}^{b}-\nu^{f}_{i,r})\left(k_{f,r}\prod_{j=1}^{n}\alpha_{j}^{\nu_{j,r}^{f}}-k_{b,r}\prod_{j=1}^{n}\alpha_{j}^{\nu_{j,r}^{b}}\right).\end{equation} Now notice that $\partial_{t}\alpha_{i}=\alpha_{i}\partial_{t}(\ln\alpha_{i})$ such that in addition to multiplying (\ref{quiescent}) by $\ln\alpha_{i}$ and summing, after integration in $x$ we obtain the relation: \[\frac{d}{dt}\sum_{i=0}^{n}\int_{\Omega}\alpha_{i}\ln\alpha_{i} dx - \int_{\Omega}\sum_{i=0}^{n}\ln\alpha_{i}\nabla_{x}\cdot (\mathscr{D}_{i}(\alpha)\nabla_{x}\alpha_{i})dx - \int_{\Omega}\sum_{i=0}^{n}\mathscr{Q}_{i,r}(\ln\alpha_{i}+1)dx = 0.\]  Now we use the observation employed in \cite{GV1}, which notes that regardless of the form of the $\alpha_{i}$'s, in a system of elementary reactions there always exist reaction dependent constants $(b_{1},\ldots,b_{n})\in\mathbb{N}^{n}$ for $b_{i}\neq 0$ such that the stoichiometry satisfies the following linear relation: \begin{equation}\label{key}\sum_{i=1}^{n}b_{i}\nu_{i,r}^{b}=\sum_{i=1}^{n}b_{i}\nu_{i,r}^{f},\end{equation} and thus yielding formally upon integration of (\ref{quiescent}) that $\frac{d}{dt}\sum_{i=1}^{n}\int_{\Omega} b_{i}\alpha_{i} dx=0$.

Now, we rescale the last term by a constant ($\ln K_{eq,r} = \ln (k_{f,r}/k_{b,r})$) which corresponds (as we show below) to the standard Gibb's free energy of the reaction $\Delta_{r}G_{\vartheta}$, such that integrating by parts and passing to the weak form we obtain the inequality: \begin{equation}\begin{aligned}\label{preentropy}\frac{d}{dt}\sum_{i=0}^{n}\int_{\Omega}\alpha_{i}&(\ln\alpha_{i}+b_{i})dx + \sum_{i=0}^{n}\int_{\Omega}\alpha_{i}^{-1}\mathscr{D}_{i}(\alpha)\nabla_{x}\alpha_{i}\cdot\nabla_{x}\alpha_{i}dx  \\ & - \sum_{i=0}^{n}\int_{\Omega}\mathscr{Q}_{i,r}\ln\left(\alpha_{i}K_{eq,r}^{1/n(\nu_{i,r}^{b}-\nu^{f}_{i,r})}\right)dx \leq 0,\end{aligned}\end{equation} which we rewrite as: \[\begin{aligned} \sum_{i=1}^{n}\frac{d}{dt}\int_{\Omega}\alpha_{i}(\ln\alpha_{i}+b_{i})dx + \sum_{i=1}^{n}\int_{\Omega}\alpha_{i}\mathscr{D}_{i}(\alpha)\nabla_{x}\alpha_{i}\cdot\nabla_{x}\alpha_{i}dx + \sum_{i=1}^{n}\mathfrak{D}(\alpha) \leq 0.\end{aligned}\] 

That is, the reaction term $\sum_{i}\mathfrak{D}(\alpha) = \Delta_{r}G_{\vartheta}$ corresponds to the isothermal Gibbs free energy of the reaction, since \begin{equation}\begin{aligned}\label{reactionGibbs}\sum_{i=1}^{n}\mathfrak{D}(\alpha) & = - \sum_{i=1}^{n} \mathscr{Q}_{i,r}\ln \left(\alpha_{i}K_{eq,r}^{1/n(\nu_{i,r}^{b}-\nu^{f}_{i,r})}\right) \\  & = - \sum_{i=1}^{n}\left( \frac{1}{\nu_{i,r}^{b}-\nu^{f}_{i,r}}\right) \mathscr{Q}_{i,r}\left(\ln\alpha_{i}^{\nu_{i,r}^{b}-\nu^{f}_{i,r}} + n^{-1}\ln K_{eq,r}\right)  \\ & =  \ \xi\ln K_{eq,r} +\left(k_{f,r}\prod_{i=1}^{n}\alpha_{i}^{\nu_{i,r}^{f}}-k_{b,r}\prod_{i=1}^{n}\alpha_{i}^{\nu_{i,r}^{b}}\right) \ln Q_{r}(\alpha) \\ & =  \ \xi\ln Q_{r}(\alpha)+ \Delta G^{\ominus},\end{aligned}\end{equation} with reaction quotient $Q_{r}(\alpha)$ given by: \begin{equation}\label{reactionquotient} Q_{r}(\alpha) =\left(\prod_{i=1}^{n}\alpha_{i}^{\nu_{i,r}^{b}}\bigg/\prod_{i=1}^{n}\alpha_{i}^{\nu_{i,r}^{f}}\right),\end{equation} and where $\xi=\xi(t,\boldsymbol{x})$ is simply a rate-scaled prefactor coefficient.  Thus, for spontaneous reactions at constant temperature $\vartheta$ (the only interesting case for the quiescent reactor systems of the form (\ref{quiescent})), it follows that $\Delta_{r}G_{\vartheta} \leq 0$.  More precisely notice that we may rewrite (\ref{reactionGibbs}) as: \begin{equation}\begin{aligned}\label{reactionGibbs2}\sum_{i=1}^{n}\mathfrak{D}(\alpha) & = - \sum_{i=1}^{n} \mathscr{Q}_{i,r}\ln \left(\alpha_{i}K_{eq,r}^{1/n(\nu_{i,r}^{b}-\nu^{f}_{i,r})}\right) \\  & = -k_{b,r} \left(K_{eq,r}\prod_{i=1}^{n}\alpha_{i}^{\nu_{i,r}^{f}}-\prod_{i=1}^{n}\alpha_{i}^{\nu_{i,r}^{b}}\right)\ln \left(K_{eq,r}\prod_{i=1}^{n}\alpha_{i}^{\nu_{i,r}^{b}}\bigg/\prod_{i=1}^{n}\alpha_{i}^{\nu_{i,r}^{f}}\right) \\ & \leq 0,\end{aligned}\end{equation} since it is clear that we have a term of the form $(A-B)(\ln A - \ln B)$ such that the product is always positive.

As a consequence we obtain the scalar entropy $\mathscr{S}_{\mathfrak{R},\infty}=\mathscr{S}_{\mathfrak{R},\infty}(\boldsymbol{\alpha})$ over the reaction space $\mathfrak{R}$.  That is, given bounded initial reaction state density $P_{0|r\in\mathfrak{R}}$ satisfying \[P_{0|\forall r\in\mathfrak{R}}=\sum_{i=1}^{n}\int_{\Omega}\alpha_{i}^{0}(\ln\alpha_{i}^{0}+b_{i})dx < \infty,\] where $\alpha_{i}^{0}=\alpha_{i|t=0}$ is the initial condition, then summing over all reactions $r\in\mathfrak{R}$ we obtain the following inequality on the system for any fixed $n$ number of constituents over an unbounded domain: \begin{equation}\begin{aligned}\label{entropy} \mathscr{S}_{\mathfrak{R},\infty} = & \sup_{0\leq t\leq T} \bigg\{\sum_{i=1}^{n}\int_{\Omega}\alpha_{i}(\ln\alpha_{i}+b_{i})dx +\sum_{r\in\mathfrak{R}}\sum_{i=1}^{n}\int_{0}^{t}\int_{\Omega}\mathfrak{D}(\alpha)dxds  \\ & \qquad\qquad\qquad + \sum_{i=1}^{n}\int_{0}^{t}\int_{\Omega}\alpha_{i}^{-1}\mathscr{D}_{i}(\alpha)\nabla_{x}\alpha_{i}\cdot\nabla_{x}\alpha_{i}dxds \bigg\} \leq P_{0|\forall r\in\mathfrak{R}} .\end{aligned}\end{equation}  where the first term corresponds to the entropy contribution from the density of states, the second  to the contribution from chemical energy production in the reactor, and the third to the entropy contribution due to the random motion of the system.

\subsection{\S 3.2  Consistent entropy and \emph{p}-enrichment} 

The entropy relation derived above may be used to generate a local smoothness estimator on the solution of each cell's interior.  Moreover, the entropy $\mathscr{S}_{\mathfrak{R}}$ is a particularly attractive functional due to the fact that first, it is globally monotonic and convex (or concave up to the sign convention), and also that it is a functional which approximates the local internal energy of the entire solution space in a totally coupled sense.  In this way, the entropy functional $\mathscr{S}_{\mathfrak{R}}$ provides for an easy test of whether the full approximate solution (\ref{aproxreact}) is entropy consistent in a fully coupled sense, and then, if it is, where inside of $\Omega$ the entropy is most variable.  In the setting presented in \textsection{3.1}, we have assumed a noncompact domain, and thus $\mathscr{S}_{\mathfrak{R}}$ applies globally in the numerical setting to periodic boundaries, or those employing fully transmissive (or radiative) boundary conditions (\emph{e.g.} no forcings on the BCs up to the differential order of the numerical solution).

That is, in order to derive the global discrete total entropy $\mathscr{S}_{\mathfrak{R}}^{k+1}$ at any particular timestep $t^{k+1}$, we may simply integrate in time such that for any discrete $t^{\ell}\in (0,t^{k+1}]$ we have: \begin{equation}\begin{aligned}\label{discreteentropy} \mathscr{S}^{k+1}_{\mathfrak{R},\infty} = &\sup_{0\leq t^{\ell}\leq t^{k+1}} \bigg\{\sum_{i=1}^{n}\int_{\Omega_{\mathcal{G}}}\alpha^{\ell}_{i}(\ln\alpha^{\ell}_{i}+b_{i})dx  \\ & \qquad +\sum_{i=1}^{n}\int_{0}^{t^{k+1}}\int_{\Omega_{\mathcal{G}}}\alpha_{i}^{-1}\mathscr{D}_{i}(\alpha)\nabla_{x}\alpha_{i}\cdot\nabla_{x}\alpha_{i}dxds \\ & \qquad\qquad + \sum_{r\in\mathfrak{R}}\sum_{i=1}^{n}\int_{0}^{t^{k+1}}\int_{\Omega_{\mathcal{G}}}\mathfrak{D}^{s}(\alpha)dxds\bigg\} \leq P_{0|\forall r\in\mathfrak{R}},\end{aligned}\end{equation} where as above $\alpha_{i}^{\ell}=\alpha_{i|t=t^{\ell}}$.

Generally, we solve (\ref{discreteentropy}) when $\mathscr{D}_{i}(\alpha)$ is any (possibly nontrivial) matrix (\emph{e.g.} see \textsection{4}), such that in the numerical setting we must compute the following approximation to (\ref{entropy}) via: \begin{equation}\begin{aligned}\label{discreteentropy2} \mathscr{S}^{k+1}_{\mathfrak{R},\infty} = & \sup_{0\leq t^{\ell}\leq t^{k+1}} \left(\sum_{i=1}^{n}\int_{\Omega_{\mathcal{G}}}\alpha^{\ell}_{i}(\ln\alpha^{\ell}_{i}+b_{i})dx\right)  \\ & \qquad +\sum_{i=1}^{n}\int_{0}^{t^{k+1}}\int_{\Omega_{\mathcal{G}}}\mathbbm{1}_{\{\alpha_{i}\geq L\}}\left(\frac{\mathscr{D}_{i}(\alpha)}{\alpha_{i}^{s}}\right)\nabla_{x}\alpha_{i}^{s}\cdot\nabla_{x}\alpha^{s}_{i} dxds  \\ & \qquad\qquad  +  \sum_{r\in\mathfrak{R}}\sum_{i=1}^{n}\int_{0}^{t^{k+1}}\int_{\Omega_{\mathcal{G}}}\mathfrak{D}^{s}(\alpha)dxds \leq P_{0|\forall r\in\mathfrak{R}},\end{aligned}\end{equation} given a small positive constant $L\in\mathbb{R}^{+}$ where $\mathbbm{1}_{\{\alpha_{i}\geq L\}}$ is the indicator function over the set containing $\alpha_{i}\geq L$. 

Similarly, to find an approximation of the discrete local in $(t,\boldsymbol{x})$ entropy $\mathscr{S}^{k+1}_{\mathfrak{R},\Omega_{e_{i}}}$ we simply integrate over an element $\Omega_{e_{i}}$ restricted to $t^{k+1}$, such that we obtain: \begin{equation}\begin{aligned}\label{discreteentropy4} \mathscr{S}^{k+1}_{\mathfrak{R},\Omega_{e_{i}}} & =\sum_{i=1}^{n}\int_{\Omega_{e_{i}}}\alpha^{k+1}_{i}(\ln\alpha^{k+1}_{i}+b_{i})dx   + \sum_{r\in\mathfrak{R}}\sum_{i=1}^{n}\int_{t^{k}}^{t^{k+1}}\int_{\Omega_{e_{i}}}\mathfrak{D}^{s}(\alpha)dxds \\ & \qquad +\sum_{i=1}^{n}\int_{t^{k}}^{t^{k+1}}\int_{\Omega_{e_{i}}}\mathbbm{1}_{\{\alpha_{i}\geq L\}}\left(\frac{\mathscr{D}_{i}(\alpha)}{\alpha_{i}^{s}}\right)\nabla_{x}\alpha_{i}^{s}\cdot\nabla_{x}\alpha_{i}^{s}dxds,\end{aligned}\end{equation} such that $\alpha_{i}^{k+1}=\alpha_{i|t=t^{k+1}}$, again with  $L\in\mathbb{R}^{+}$ a small positive constant.  Then we proceed by defining the local in time change in entropy density over $\mathrm{int}(\Omega_{e_{i}})$ as satisfying: \begin{equation}\label{entchange}\Delta\rho\mathscr{S}^{k+1}_{\mathfrak{R},\Omega_{e_{i}}}=   \rho\left(\mathscr{S}^{k+1}_{\mathfrak{R},\Omega_{e_{i}}}-\mathscr{S}^{k}_{\mathfrak{R},\Omega_{e_{i}}}\right),\end{equation} where the cell density is taken as $\rho = |\Omega_{e_{i}}|^{-1}$.

We use equation (\ref{entchange}) as an approximate measure of the variation in the local \emph{internal energy} with respect to a fixed volume elements  (at timestep $t^{k+1}$) interior, $\mathrm{int}(\Omega_{e_{i}})$ at constant temperature $\vartheta$.  More explicitly, we use (\ref{entchange}) as a local smoothness estimator over the interior of $\Omega_{e_{i}}$ in order to develop a $p$-enrichment functional $\mathfrak{E}_{i}^{k+1}= \mathfrak{E}_{i}^{k+1}(\mathscr{P}^{s}(\Omega_{e_{i}}^{k+1}))$ which estimates the local internal energy of the element as an approximate measure of the smoothness of the solution, such that for $\mathscr{P}^{p_{\max}}(\Omega_{e_{i}})$ the maximum polynomial order allowed on any $\Omega_{e_{i}}$, and  $\mathscr{P}^{s}(\Omega_{e_{i}}^{k})$ the present polynomial order, we define: \begin{equation}\label{smoothest}\mathfrak{E}_{i}^{k+1} = \left\{\begin{matrix}  \mathscr{P}^{s+1}(\Omega_{e_{i}}^{k+1}) & \mathrm{if} \ \left( \big|\Delta\rho\mathscr{S}^{k+1}_{\mathfrak{R},\Omega_{e_{i}}} -  \Delta\varrho\mathscr{S}^{k+1}_{\mathfrak{R},\Omega_{e_{i}}}\big| < \mu_{s+1} \right)\land (s+1\leq p_{\max}), \\  \mathscr{P}^{s-1}(\Omega_{e_{i}}^{k+1}) & \mathrm{if} \ \left( \big|\Delta\rho\mathscr{S}^{k+1}_{\mathfrak{R},\Omega_{e_{i}}} -  \Delta\varrho\mathscr{S}^{k+1}_{\mathfrak{R},\Omega_{e_{i}}}\big| \geq \mu_{s-1} \right) \land (s-1 \geq p_{\min}),\end{matrix}\right.\end{equation} where the change in the average global entropy density $\Delta\varrho\mathscr{S}^{k+1}_{\mathfrak{R}}$ at $t^{k+1}$ is given by \[ \Delta\varrho\mathscr{S}^{k+1}_{\mathfrak{R}} = \varrho\left(\mathscr{S}^{k+1}_{\mathfrak{R}}-\mathscr{S}^{k}_{\mathfrak{R}}\right).\]  

The global entropy $\mathscr{S}^{k+1}_{\mathfrak{R}}$ at timestep $k+1$ is defined fully in \textsection{3.3}, and the global density is simply taken as $\varrho = |\Omega_{\mathcal{G}}|^{-1}$.  The adjustable parameter $\mu_{s}=\mu(\iota_{s})$ is a composite of the range of the entropy change at time $k+1$ and a weight $\iota_{s}\in(0,1)$.  That is the function may be written $\mu_{s+1}=\iota_{s+1}\delta$ over the midpoint of the range $\delta=\delta(\rho,\mathscr{S}^{k}_{\mathfrak{R},\Omega_{e_{i}}},\mathscr{S}^{k+1}_{\mathfrak{R},\Omega_{e_{i}}})$ of the change in entropy density \[\delta=\max_{i}\Delta\rho\mathscr{S}^{k+1}_{\mathfrak{R},\Omega_{e_{i}}}-\min_{i}\Delta\rho\mathscr{S}^{k+1}_{\mathfrak{R},\Omega_{e_{i}}}.\]  Clearly (\ref{smoothest}) has the effect of using the variation in the entropy change locally to weight some fraction of the cells for $p$-enrichment and the rest for $p$-coarsening, depending only on how far their relative local change in entropy lies from the median of the range. 

In contrast to alternative choices for $p$-enrichment, this scheme provides for a cogent physical interpretation that serves as buttress for the enrichment strategy.  Namely, we see that in areas in which the relative disorder (\emph{i.e.} the relative entropy) of a cell exceeds a specified allowed variation within the cell itself, then we coarsen our solution, thus avoiding CFL instabilities, \emph{etc}.  Likewise in areas of relative order (or stable smoother regions) we readily enrich our solution.  

We have also tested enrichment strategies based on slightly more abstract principles.  For example, one may simply choose a fraction of elements with respect to the magnitude of their relative change in entropy density, or with respect to $|\Delta\rho\mathscr{S}^{k+1}_{\mathfrak{R},\Omega_{e_{i}}} -  \Delta\varrho\mathscr{S}^{k+1}_{\mathfrak{R},\Omega_{e_{i}}}|$ by cell.  Likewise, when using a hierarchical basis one may use the scheme described in \cite{Michoski} to measure the perturbative variation in the higher terms with respect to the $L^q$--norm.  Many of these alternative strategies can lead to stable schemes that effectively ``sense'' relative energy fluctuations with respect the $\Delta t$.  However, it should be noted as a word of caution that (\ref{smoothest}) is particularly well-suited for naturally avoiding the observed phenomenon of \emph{bunching} in the local variational space.  This \emph{bunching} of the solution often leads to a \emph{flickering} of enrichment/coarsening of a substantial number of elements taking values close to the ``center'' of the chosen discriminating parameter (\emph{e.g.}  $|\Delta\rho\mathscr{S}^{k+1}_{\mathfrak{R},\Omega_{e_{i}}} -  \Delta\varrho\mathscr{S}^{k+1}_{\mathfrak{R},\Omega_{e_{I}}}|$ in (\ref{smoothest})).   This behavior is a potential source of debilitating inefficiency in the scheme, and can be difficult to isolate without recourse to an entropy formalism, or similar.

\subsection{\S 3.3 The entropic jump and \emph{hp}-adaptivity} 

As already discussed, the global entropy formulation from \textsection{3.1} is predicated on noncompactness of the space $\Omega$, while in general we are interested in more complicated boundary formulations; in particular any boundary condition satisfying (\ref{robin}).  In this more general setting we see that equation (\ref{entropy}) by way of the divergence theorem becomes:  \begin{equation}\begin{aligned}\label{entropybound} \mathscr{S}_{\mathfrak{R}} & = \sup_{0\leq t\leq T} \bigg\{\sum_{i=1}^{n}\int_{\Omega}\alpha_{i}(\ln\alpha_{i}+b_{i})dx + \sum_{r\in\mathfrak{R}}\sum_{i=1}^{n}\int_{0}^{t}\int_{\Omega}\mathfrak{D}(\alpha)dxds  \\ & \quad\quad\sum_{i=1}^{n}\int_{0}^{t}\int_{\Omega}\alpha_{i}^{-1}\mathscr{D}_{i}(\alpha)\nabla_{x}\alpha_{i}\cdot\nabla_{x}\alpha_{i}dxds  \bigg\} \\ & \leq   P_{0|\forall r\in\mathfrak{R}} + \sum_{i=1}^{n}\int_{0}^{t}\int_{\partial\Omega}(\ln \alpha_{i}+b_{i})\mathscr{D}_{i}(\alpha)\nabla_{x}\alpha_{i}\cdot\boldsymbol{n}dSds .\end{aligned}\end{equation}  Thus, as before, the  discrete approximation to (\ref{entropybound}) simply yields:  \begin{equation}\begin{aligned}\label{discreteentropybound} & \mathscr{S}^{k+1}_{\mathfrak{R}} = \sup_{0\leq t^{\ell}\leq t^{k+1}} \left(\sum_{i=1}^{n}\int_{\Omega_{\mathcal{G}}}\alpha^{\ell}_{i}(\ln\alpha^{\ell}_{i}+b_{i})dx\right)  \\ & \qquad +\sum_{i=1}^{n}\int_{0}^{t^{k+1}}\int_{\Omega_{\mathcal{G}}}\mathbbm{1}_{\{\alpha_{i}\geq L\}}\left(\frac{\mathscr{D}_{i}(\alpha)}{\alpha_{i}^{s}}\right)\nabla_{x}\alpha_{i}^{s}\cdot\nabla_{x}\alpha^{s}_{i} dxds   \\ & \qquad\qquad  + \sum_{r\in\mathfrak{R}}\sum_{i=1}^{n}\int_{0}^{t^{k+1}}\int_{\Omega_{\mathcal{G}}}\mathfrak{D}^{s}(\alpha)dxds \leq P_{0|\forall r\in\mathfrak{R}} \\ & \qquad +  \sum_{i=1}^{n}\int_{0}^{t^{k+1}}\int_{\partial\Omega_{\mathcal{G}}}\mathbbm{1}_{\{\alpha_{i}\geq L\}}(\ln \alpha^{s}_{i}+b_{i})\mathscr{D}_{i}(\alpha)\nabla_{x}\alpha^{s}_{i}\cdot\boldsymbol{n}dSds.\end{aligned}\end{equation}  Further,  notice that for the appropriate choice of boundary conditions both $\mathscr{S}_{\mathfrak{R}}= \mathscr{S}_{\mathfrak{R},\infty}$ and  $\mathscr{S}_{\mathfrak{R}}^{k+1}= \mathscr{S}^{k+1}_{\mathfrak{R},\infty}$ 

Now, in the local approximation it is clear enough how to reformulate (\ref{discreteentropybound}) over cells such that we obtain a local approximation to the entropy in the neighborhood of the cell.  However, for the case of $h$-adaptivity we are more directly concerned with the local jump in entropy across the neighboring cells, since it is these jumps which serve as a proper diagnostic probe for stable $hp$-adaptivity (\emph{e.g.} see \cite{Dem,Dem2,Bangerth,Kanschat}).  Thus we define the local entropic jump $\mathscr{J}^{k+1}_{\mathfrak{R},\Omega_{e_{i}}}$ at time $t^{k+1}$ by \begin{equation}\mathscr{J}^{k+1}_{\mathfrak{R},\Omega_{e_{i}}} = \sum_{i=1}^{n}\int_{0}^{t^{k+1}}\int_{\partial\Omega_{e_{i}}}\mathbbm{1}_{\{\alpha_{i}\geq L\}}(\ln \alpha^{s}_{i}+b_{i})\mathscr{D}_{i}(\alpha)\nabla_{x}\alpha^{s}_{i}\cdot\boldsymbol{n}dSds,\end{equation} such that the density of the change in the entropic jump $\rho\Delta\mathscr{J}^{k+1}_{\mathfrak{R},\Omega_{e_{i}}}$ is given to satisfy \begin{equation}\rho\Delta\mathscr{J}^{k+1}_{\mathfrak{R},\Omega_{e_{i}}}= \rho\left(\mathscr{J}^{k+1}_{\mathfrak{R},\Omega_{e_{i}}}-\mathscr{J}^{k}_{\mathfrak{R},\Omega_{e_{I}}}\right).\end{equation}  

We proceed by estimating the approximate flux of the internal energy of the system by constructing the $h$-adaptivity functional $\mathfrak{A}=\mathfrak{A}(\mathscr{T}_{h'}(\Omega_{e_{i}}^{k+1}))$ where the mesh triangulation $\mathscr{T}_{h}$ at time $t^{k}$ given by $h=h(t^{k},\boldsymbol{x})$ is refined to level $h'=h(t^{k+1},\boldsymbol{x})$ --- that is, we isotropically refine to $h/2$ in each spatial dimension --- over cell $\Omega_{e_{i}}$ at time $t^{k+1}$.  Similarly we may unrefine $\mathfrak{A}_{i}^{k+1}=\mathfrak{A}(\mathscr{T}_{h_{0}}(\Omega_{e_{i}}^{k+1}))$ to level $h_{0}=h(t^{k+1},\boldsymbol{x})$ --- that is, we isotropically coarsen to $2h$ in each spatial dimension.  

For example, in dimension $N=2$ the refinement would take a quadrilateral parent cell  $\Omega_{e_{i}}$ and split it into four child cells $\mathcal{C}_{j}$, while a coarsening would take four child cells denoted $\mathcal{C}_{j}$ and merge them into a single parent element $\Omega_{e_{i}}$.  Thus depending on the evaluation of $\mathfrak{A}$, we obtain the full $h$-adaptivity functional:  \begin{equation}\label{sharpest}\mathfrak{A}_{i}^{k+1} = \left\{\begin{matrix}  \mathscr{T}_{h'}(\Omega_{e_{i}}^{k+1} ) & \mathrm{if} \ \left( \big|\rho\Delta\mathscr{J}^{k+1}_{\mathfrak{R},\Omega_{e_{i}}} -\varrho\Delta\mathscr{J}^{k+1}_{\mathfrak{R}}\big|\geq \eta_{h'}\right)\land (s+1\leq h_{\max}), \\  \mathscr{T}_{h_{0}}(\Omega_{e_{i}}^{k+1} ) & \mathrm{if} \ \left(\big| \rho\Delta\mathscr{J}^{k+1}_{\mathfrak{R},\Omega_{e_{i}}} - \varrho \bar{\Delta}\mathscr{J}^{k+1}_{\mathfrak{R}}\big|<\eta_{h_{0}}\right) \land (s-1 \geq h_{\min}) \ \forall \mathcal{C}_{j},\end{matrix}\right.\end{equation} where $h_{\max},h_{\min}$ correspond to the maximum and minimum refinement levels, respectively.  Here again, the density of the global change in the entropic jump is given such that: \[\varrho\Delta\mathscr{J}^{k+1}_{\mathfrak{R}} = \varrho\left(\mathscr{J}^{k+1}_{\mathfrak{R}}-\mathscr{J}^{k}_{\mathfrak{R}}\right),\] where $\varrho$ is the same as in \textsection{3.2}.  Also as in \textsection{3.2}, the adjustable parameter $\eta_{h}=\eta(\upsilon_{h})$ is again defined over the range of the change in the entropic jump $\psi=\psi(\rho,\mathscr{J}^{k}_{\mathfrak{R},\Omega_{e_{i}}},\mathscr{J}^{k+1}_{\mathfrak{R},\Omega_{e_{i}}})$, and is given by $\eta_{h}=\upsilon_{h}\psi$ such that \[ \psi =\max_{i}\rho\Delta\mathscr{J}^{k+1}_{\mathfrak{R},\Omega_{e_{i}}}-\min_{i}\rho\mathscr{J}^{k+1}_{\mathfrak{R},\Omega_{e_{i}}},\] and $\upsilon_{h}\in(0,1)$. 

It is further interesting to note that the local change in the entropic jump $\Delta\mathscr{J}^{k+1}_{\mathfrak{R},\Omega_{e_{i}}}$ is independent of the reaction entropy at time level $t^{k}$, and ends up depending only on the reaction coupling from the earlier timesteps as well as on the present states diffusivity.  That being said, it is clear that just as (\ref{smoothest}) in \textsection{3.2} effectively $p$-enriches the solution based on the local physics of the system, here, we find that (\ref{sharpest}) has the effect of flagging elements with a high relative change in their entropic jumps for $h$-refinement, and those with low relative change in their entropic jumps for $h$-coarsening.  That is, in areas where the entropy is changing dramatically across the elements boundary, we refine.  However, when coarsening, we are presented with the additional constraint denoted: $\forall\mathcal{C}_{j}$.   That is, by $\forall\mathcal{C}_{j}$ we simply mean that in order to actually coarsen a parent element $\Omega_{e_{i}}$ comprised of $j$ children elements $\cup_{j}\mathcal{C}_{j}=\Omega_{e_{i}}$, each child $\mathcal{C}_{j}$ must be independently flagged for coarsening.  In other words, all children of an isotropically refined element $\Omega_{e_{i}}$ must contain a coarsen flag at time level $k+1$ in order for the parent cell to ultimately be refined at time level $k+1$.  For more details on this isotropic refinement strategy we direct the reader to \cite{Bangerth}.

Finally, we couple the  $h$-adaptivity functional $\mathfrak{A}_{i}^{k+1}$ to the $p$-enrichment functional $\mathfrak{E}_{i}^{k+1}$ such that $h$-adaptivity is always preferentially chosen over $p$-enrichment.  That is, clearly the cell localized entropy $\mathscr{S}^{k+1}_{\mathfrak{R},\Omega_{e_{i}}}$ and its corresponding entropic jump $\mathscr{J}^{k+1}_{\mathfrak{R},\Omega_{e_{i}}}$ are strongly coupled by virtue of (\ref{quiescent}), but in order to avoid numerical instabilities caused by erroneously $p$-enriching relatively inert cells experiencing high entropic fluxes entering through its neighbors, we evaluate the simple kinetic switch functional $\mathfrak{K}^{k+1}_{i} = \mathfrak{K}^{k+1}_{i} ( \mathfrak{A}_{i}^{k+1},\mathfrak{E}_{i}^{k+1} )$ determined by evaluating: \begin{equation}\mathfrak{K}_{i}^{k+1} = \left\{\begin{matrix} \mathfrak{A}_{i}^{k+1} \land  \mathfrak{E}_{i}^{k+1} & \mathrm{if} \  \mathscr{T}_{h'}(\Omega_{e_{i}}^{k+1} ) \land  \mathscr{P}^{s}(\Omega_{e_{i}}^{k+1}) \land \mathscr{S}_{\mathfrak{R}}^{k+1}, \\   \mathfrak{A}_{i}^{k+1} & \mathrm{if} \  \mathscr{T}_{h'}(\Omega_{e_{i}}^{k+1} ) \land  \mathscr{P}^{s+1}(\Omega_{e_{i}}^{k+1}) \land \mathscr{S}_{\mathfrak{R}}^{k+1}, \\  \mathfrak{A}_{i}^{k+1}  \land  \mathfrak{E}_{i}^{k+1} & \mathrm{if} \  \mathscr{T}_{h_{0}}(\Omega_{e_{i}}^{k+1} ) \land  \mathscr{P}^{s}(\Omega_{e_{i}}^{k+1}) \land \mathscr{S}_{\mathfrak{R}}^{k+1},  \\  \mathfrak{A}_{i}^{k+1}  \land  \mathfrak{E}_{i}^{k+1}  & \mathrm{if} \ \mathscr{T}_{h_{0}}(\Omega_{e_{i}}^{k+1} ) \land  \mathscr{P}^{s+1}(\Omega_{e_{i}}^{k+1})  \land \mathscr{S}_{\mathfrak{R}}^{k+1}, \\ 0 & \mathrm{otherwise},\end{matrix}\right.\end{equation} whereby we are able to stabilize these spurious quiescent instabilities, and yet still maintain the entropy consistency of the scheme.

\section{\texorpdfstring{\protect\centering $\S 4$ Example Applications}{\S 4 Example Applications}}

We address several example applications below, and note that all examples in this sections were given reflecting wall boundary conditions, which is just to say that scalar boundary values are determined by their values on the interior at the boundary $a_{i}\alpha_{i,b}|_{\partial\Omega_{e_{i}}}= \alpha_{i}|_{\partial\Omega_{e_{j}}}$, and the gradients are reflected with respect to the normal direction $b_{i}\boldsymbol{\sigma}_{i,b}|_{\partial\Omega_{e_{i}}}= - \boldsymbol{\sigma}_{i}|_{\partial\Omega_{e_{j}}}$, where $c_{i}=0$. 

\subsection{\S 4.1 Reaction dominated hypergolic kinetics in 1D}

We choose as a one dimensional example the nuanced problem of the \emph{reaction dominated} --- as well as \emph{diffusion limited} --- regime in hypergolic kinetics, corresponding to when the rates of the reactions $r$ in $\mathfrak{R}$ occur on substantially smaller timescales than the corresponding diffusivity $\mathscr{D}_{i}$ of the system.  In the proper context (such as in a stable subdomain  $\Omega_{0}\Subset\tilde{\Omega}\subset\Omega$ of a homogenized combustion chamber \cite{AbdElSalam,Merzhanov}, or in the propagation of deflagration flame fronts \cite{CafMel}) such a combustion system may be approximated by the reaction-diffusion equations, and to a low order approximation --- where the compressibility of the fluid may be neglected --- the quiescent reactor regime may be utilized to model the resulting gas phase dynamics of the chamber. 

Along these lines let us consider the second order hypergolic ignition reaction, given by  \[\nu_{1}^{f}\alpha_{1}+\nu_{2}^{f}\alpha_{3} \ \ce{->T[{$ \ \ k_{f} \ \ $}] } \ \nu_{3}^{b}\alpha_{2}+\nu_{4}^{b}\alpha_{4},\]  comprised of a reaction between monomethylhydrazine $\alpha_{1}=\mathrm{CH}_{3}(\mathrm{NH})\mathrm{NH}_{2}$ and nitrogen dioxide $\alpha_{3}=\mathrm{NO}_{2}$, to yield the exhaust radical $\alpha_{2}=\mathrm{CH}_{3}\dot{\mathrm{N}}\mathrm{NH}_{2}$ and nitrous acid $\alpha_{4}=\mathrm{HNO}_{2}$ such that the stoichiometry satisfies: \begin{equation}\label{ignite}\mathrm{CH}_{3}(\mathrm{NH})\mathrm{NH}_{2} + \mathrm{NO}_{2} \  \ce{->T[{$ \ \ k_{f} \ \ $}] } \ \mathrm{CH}_{3}\dot{\mathrm{N}}\mathrm{NH}_{2} + \mathrm{HNO}_{2},\end{equation} with a forward reaction rate (see Ref.~\cite{CCP}) of $k_{f} = 2.2\times 10^{11} e^{-5900/\mathrm{R}\vartheta }\mathrm{cm}^{3}(\mathrm{mol}\cdot \mathrm{s})^{-1}$, where $\vartheta$ is the constant temperature subdomain $\Omega_{0}$ in which the reaction occurs and $\mathrm{R}$ is the ideal gas constant.

Being far from equilibrium, we can readily neglect the back reaction, such that for $\alpha_{1},\alpha_{2},\alpha_{3}$ and $\alpha_{4}$ we see that (\ref{fir}) leads to the coupled system: \begin{equation}\begin{aligned}\label{chemkin} \partial_{t}\alpha_{1}= -k_{f}\alpha_{1}\alpha_{3}, \quad & \partial_{t}\alpha_{2} = k_{f}\alpha_{1}\alpha_{3}, \quad  \partial_{t}\alpha_{3} = -k_{f}\alpha_{3}\alpha_{1}, \quad \partial_{t}\alpha_{4} = k_{f}\alpha_{1}\alpha_{3},\end{aligned}\end{equation} which upon integration over a discrete timestep $\Delta t =t^{n+1}-t^{n}$, taking $\alpha_{i}^{n+1}=\alpha_{i}(t^{n+1})$, yields: \begin{equation}\begin{aligned}\label{simple} & \alpha_{1}^{n+1} = \alpha_{1}^{n}e^{-k_{f}\Delta t \alpha_{3}^{n}}, \quad \alpha_{2}^{n+1} = k_{f}\alpha_{1}^{n}\alpha_{3}^{n}\Delta t+\alpha_{2}^{n},  \\   & \alpha_{3}^{n+1} = \alpha_{3}^{n}e^{-k_{f}\Delta t \alpha_{1}^{n}}, \quad \alpha_{4}^{n+1} = k_{f}\alpha_{1}^{n}\alpha_{3}^{n}\Delta t+\alpha_{4}^{n}.\end{aligned}\end{equation}

\begin{table}[!t]
\centering
\begin{tabular}{|c | c | c |}
\hline
\bf{Index}&  \bf{Species} & \bf{Mass Diffusion$/\mathrm{cm}^{2}\cdot\mathrm{s}^{-1}$}\rule{0pt}{3ex} \rule[-2ex]{0pt}{0pt} \\ 
\hline\hline
$\alpha_{1}$ & $\mathrm{CH}_{3}\mathrm{N}_{2}\mathrm{H}_{3}$ & $\sim 5.9\times 10^{-6} \ (1)$ \rule{0pt}{3ex} \rule[-2ex]{0pt}{0pt}  \\
\hline
$\alpha_{2}$ & $\mathrm{CH}_{2}\mathrm{N}_{2}\mathrm{H}_{2}$ & $\sim 6.0\times 10^{-6}$ \rule{0pt}{3ex} \rule[-2ex]{0pt}{0pt} \\ 
\hline
$\alpha_{3}$ & $\mathrm{NO}_{2}$   & $\sim 9.0\times 10^{-2} \ (2)$  \rule{0pt}{3ex} \rule[-2ex]{0pt}{0pt} \\ 
\hline
$\alpha_{4}$ & $\mathrm{HNO}_{2}$  & $\sim 1.2\times 10^{-2} \ (3)$   \rule{0pt}{3ex} \rule[-2ex]{0pt}{0pt} \\ 
\hline
\end{tabular}
\label{table:diffconst}
\caption{All approximate values taken at STP since at constant pressure $\mathscr{D}_{i}$ scales sublinearly with $\vartheta$ (\emph{e.g.} see \textsection{4.3}).  (1) was measured via chronoamperometry as shown in Ref.~\cite{NRO}, (2) was determined via the single component Chapman-Enskog experimental fits in Ref.~\cite{Massman}, and (3) was calculated using diffusion denuders in Ref.~\cite{Benner}.  The remaining coefficient was adapted using relative magnitude arguments from simple collisional theory \cite{Houston} (viz. $\mathscr{D}_{i}\propto \sigma^{-2}$ the molecular cross sectional radius and \textsection{4.3}).  }
\end{table}

\begin{figure}[!t]
\centering
\includegraphics[width=8cm]{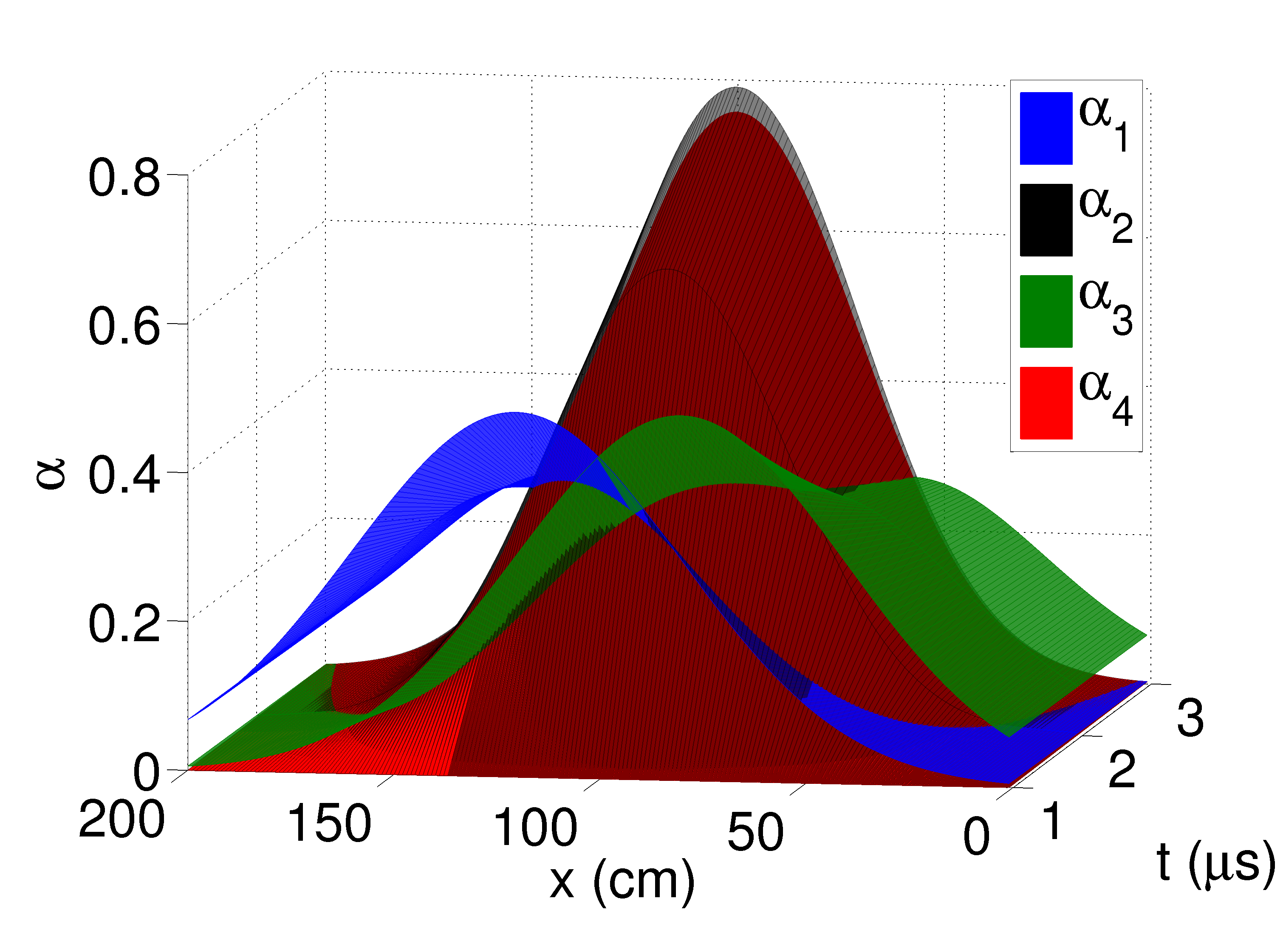} \\ \includegraphics[width=8cm]{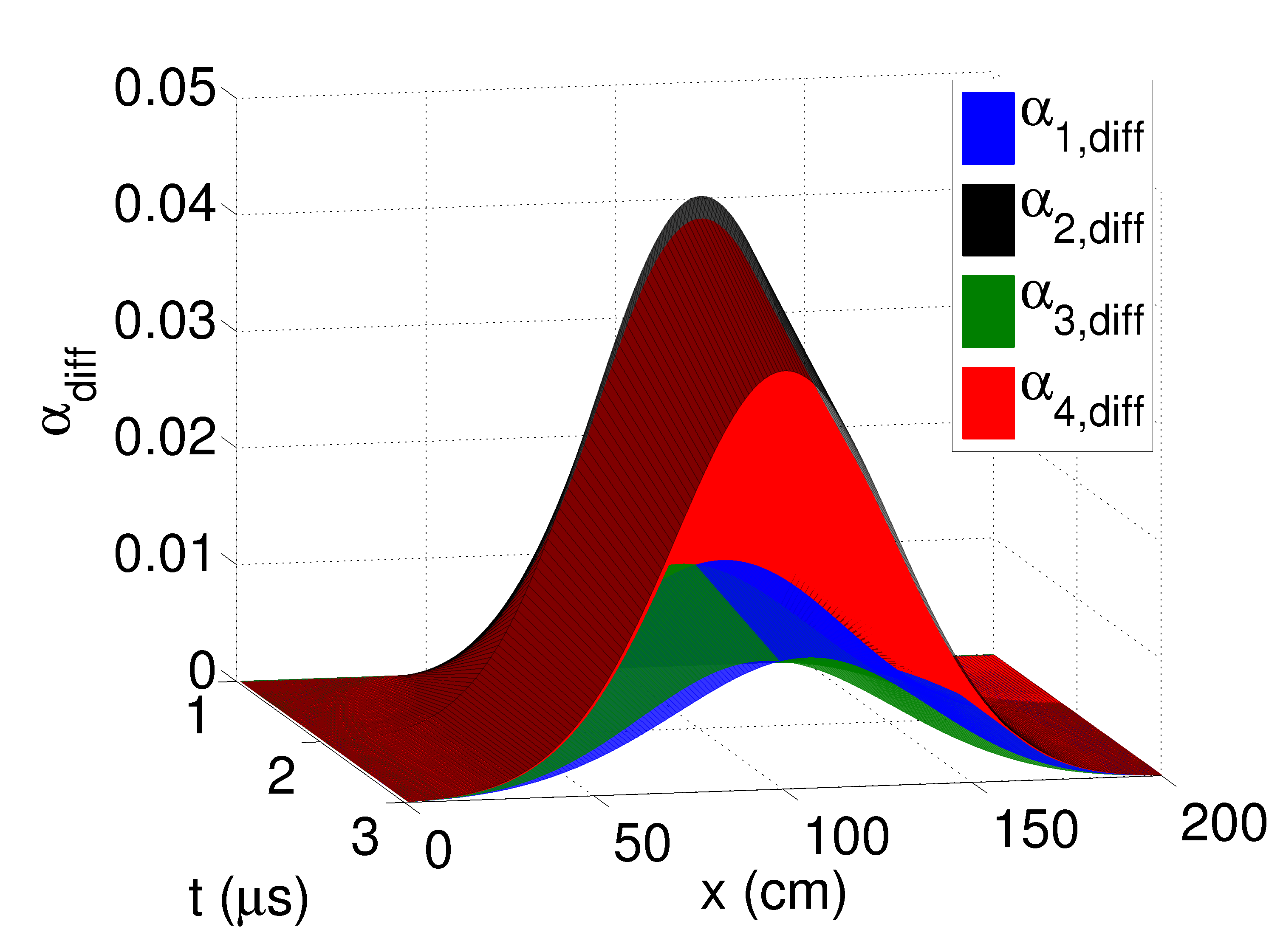} \ \includegraphics[width=8cm]{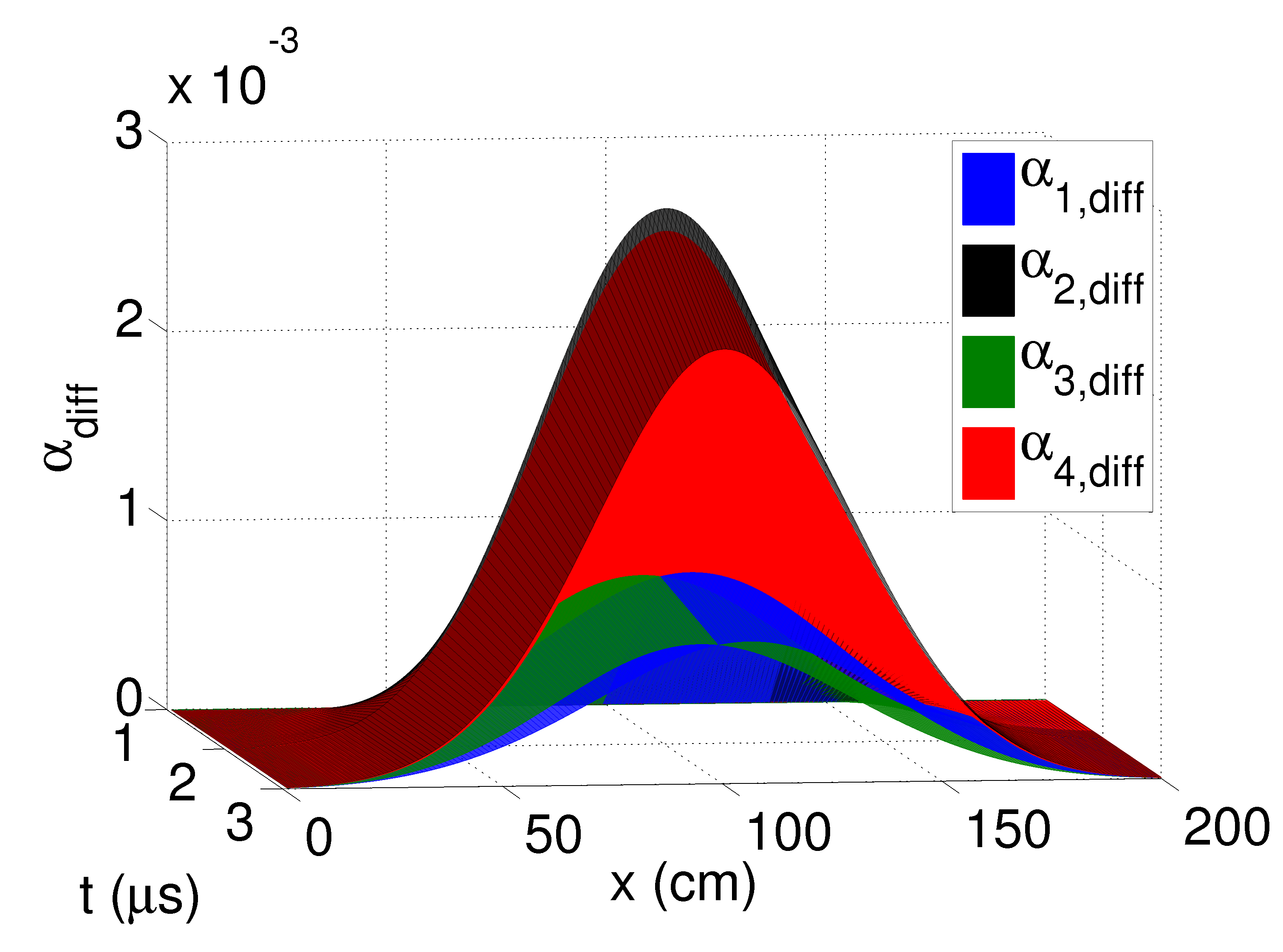} \caption{The top graph shows the solution for $h=0.5$, $\Delta t_{s}= 1\upmu\mathrm{s}$, and $\Delta t_{f} = \Delta t_{s}/50$.  The bottom left shows the absolute difference map between the $\alpha$'s solved using $\Delta t_{f} = \Delta t_{s}/50$ and those solved using $\Delta t_{f} = \Delta t_{s}$, while the bottom right shows the absolute difference map between the $\alpha$'s of  $\Delta t_{f} = \Delta t_{s}/50$ and $\Delta t_{f} = \Delta t_{s}/10$.}
\label{fig:hypergolic}  
\end{figure}

In the case of combustion reactions a mass transfer correction factor $h_{m}\in\mathbb{R}$ is often included, and is in fact necessary in order to stave off the effects of catalytic volume expansions, where again the implicit assumption is that there exists a local subdomain $\Omega_{0}$ of relatively homogeneous reactivity, over which the rate constant $k_{f}$ may be effectively averaged (\emph{e.g.} see Ref.~\cite{Williams}). This correction-based formulation leads to a stable form of the mass action, such that simply replacing $k_{f}$ in (\ref{simple}) by the addend $(k_{f}+h_{m})$ provides the vector $\tilde{\boldsymbol{\alpha}}^{n+1}$ from which we may easily compute $\mathring{\mathscr{A}}(\hat{\boldsymbol{\alpha}}^{n+1},\boldsymbol{\alpha}^{n})$ from (\ref{diffalpha}).

It remains to identify the mass diffusion coefficients $\mathscr{D}_{i}(\alpha,\vartheta)$.  The most straightforward way of choosing  $\mathscr{D}_{i}(\alpha,\vartheta)$ is by simply setting them equal to empirically determined constants, in which case one may make the additional assumption of pure diffusion, such that interspecies diffusion  $\mathscr{D}_{ij}(\alpha,\vartheta)$ --- for a counter example see \textsection{4.3} --- is neglected.  In this case the $\mathscr{D}_{i}$ reduces to a diagonal matrix with positive constant entries, $\mathscr{D}_{i} \in \mathbb{R}^{+}$, and setting the transmissive boundary conditions $\boldsymbol{U}_{h}^{n}|_{\mathcal{K}_{ji}}=\boldsymbol{U}_{h}^{n}|_{\mathcal{K}_{ij}}$, we may proceed to solve (\ref{quiescent}) for (\ref{ignite}).  

The results are presented in Figure (\ref{fig:hypergolic}) using the diffusion constants from Table \ref{table:diffconst}, where the initial conditions are given by: \[\begin{aligned}\alpha_{1,0}= 0.49 & \exp\left(-\frac{(x-60)^{2}}{800}\right), \  \alpha_{3,0}= 0.49 \exp\left(-\frac{(x-40)^{2}}{800}\right), \ \alpha_{2,0} = \alpha_{4,0} = 1\times 10^{-5}.\end{aligned}\]  We use a Bassi-Rebay form of the LDG based viscous flux (see \cite{MESV,ABCM}), and a mass transfer correction factor of $h_{m}=-2.19\times 10^{10} e^{-5900/\mathrm{R}\vartheta }\mathrm{cm}^{3}(\mathrm{mol}\cdot \mathrm{s})^{-1}$ in order to rescale $k_{f}$ to an effective value of $(k_{f}+h_{m})=1\times 10^{4} e^{-5900/\mathrm{R}\vartheta }\mathrm{cm}^{3}(\mathrm{mol}\cdot \mathrm{s})^{-1}$.  Then our slow (diffusion) modes are chosen such that $\Delta t_{s} = 1\upmu \mathrm{s}$ and our fast (reaction) modes to satisfy $\Delta t_{f} = \epsilon\Delta t_{s}$ for $\epsilon \in \big\{\frac{1}{50},\frac{1}{10},1\big\}$.  The convergence bound from (\ref{converge}) is taken as $C=1\times 10^{-8}$ over both fast and slow modes, and the average number of convergence steps is $\sim 25$ where $\ell=50$.

As is clear from figure \ref{fig:hypergolic}, running the reaction with the fast modes has a substantial impact on the solution, but only up to a point.  That is, once a certain timestep is reached with full convergence, little is gained (in a relative sense) by iterating with respect to additional temporal refinements.  Here, as is clear, the diffusion is negligible on the timescale of the simulation, and has nearly no effect on the solution over 2 $\upmu\mathrm{s}$.  

We also note that while the addition of both the mass transfer correction $h_{m}$ and the fast reaction time modes $\Delta t_{f}$ makes solving such problems viable in the reaction-diffusion setting of (\ref{quiescent}), most such problems --- certainly most highly reactive gas phase or combustion reactions --- require the addition of advective fluxes in order to accurately approximate the relevant dynamics of the system, which are determined by complicated couplings between the mass, momentum and energy conservation equations (\emph{e.g.} the Euler and Navier-Stokes equations), not to mention to substantial importance that turbulence plays in these types of combustion regimes.  We shall discuss these systems in more detail in the sequel to this paper. 


\subsection{\S 4.2 Diffusion dominated alkyl halide gas mixture}

Consider the following \emph{diffusion dominated} and \emph{reaction limited} gas phase reaction of fluoromethane $\mathrm{CH}_{3}\mathrm{F}$, with free gold cation $\mathrm{Au}^{+}$ as discussed in \cite{CH3FAu2,CH3FAu1} and whose primary reaction channel follows the termolecular addition reaction in the presence of $\mathrm{He}$ gas:  \begin{equation}\label{Auplus} \mathrm{CH}_{3}\mathrm{F} + \mathrm{Au}^{+} \  \ce{<=>>T[{$ \ \ k_{f_{1}} \ \ $}][{$ \ \ k_{b_{1}} \ \ $}] } \ \mathrm{Au}^{+}\mathrm{CH}_{3}\mathrm{F}  \end{equation} where the helium is an important third body in the gas phase reaction pathway, and is further coupled to the bimolecular elimination reaction (neglecting the hydrogen halide formation): \begin{equation}\label{Auplus2nd} \mathrm{CH}_{3}\mathrm{F} + \mathrm{Au}^{+} \ \ce{<=>>T[{$ \ \ k_{f_{2}} \ \ $}][{$ \ \ k_{b_{2}} \ \ $}] } \ \mathrm{AuCH}_{2}^{+}\end{equation}

\begin{figure}[!t]
\centering
\ \includegraphics[width=8.5cm]{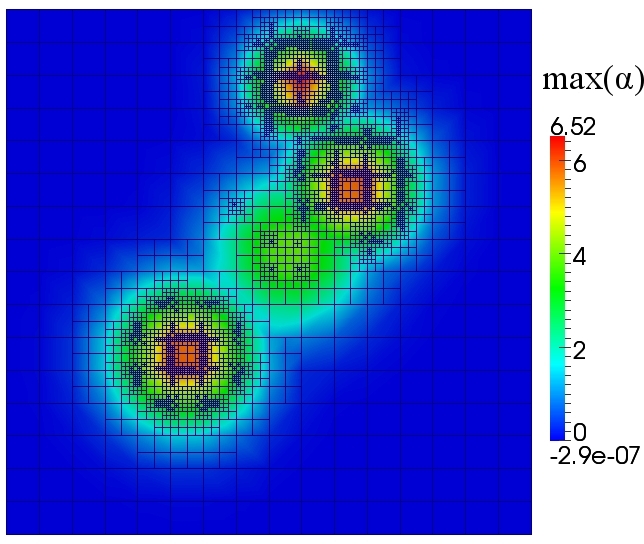} \includegraphics[width=8.5cm]{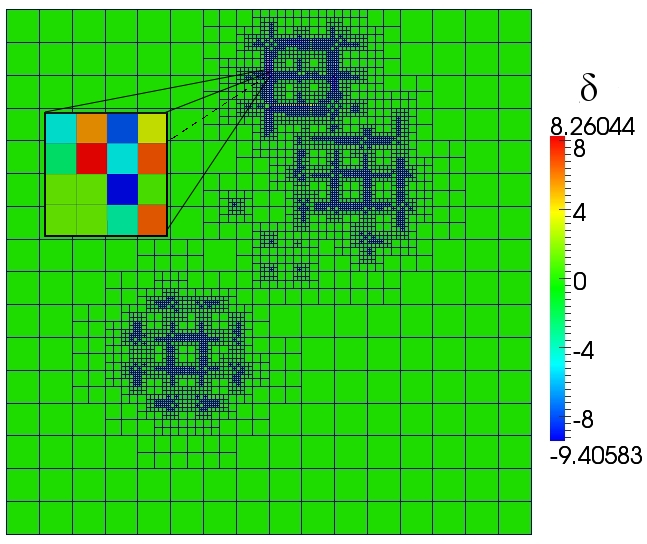} \\ \ \includegraphics[width=8.5cm]{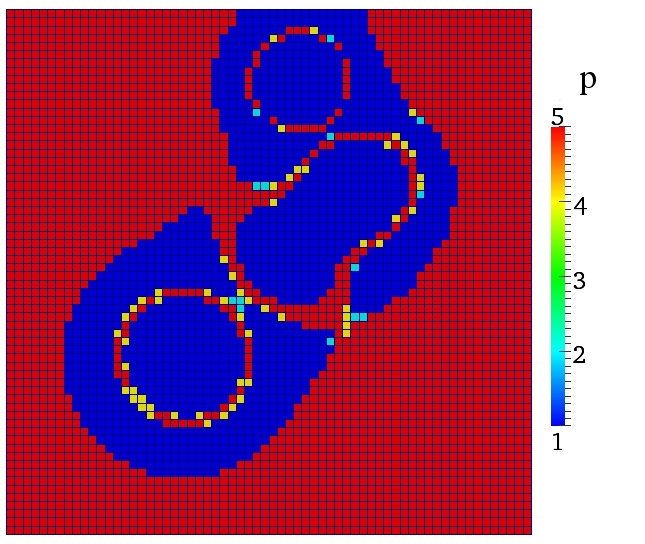} \includegraphics[width=8.5cm]{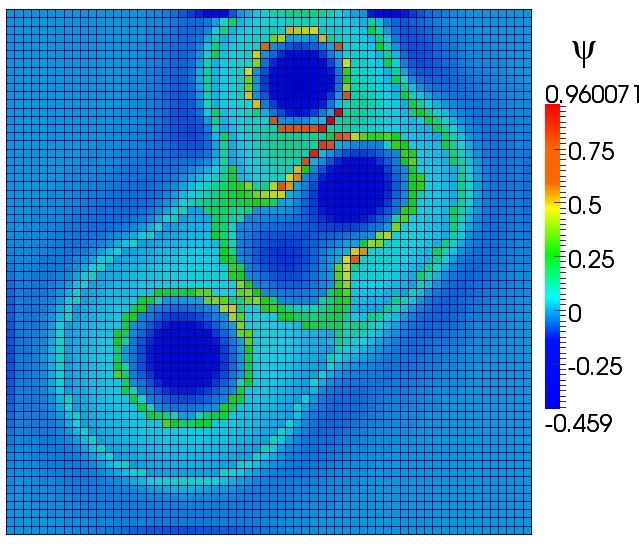}\caption{On the top we show the $N=2$ solution using five levels of $h$ refinement given $p$ fixed at linears, where the contour is coloured by $\max_{x}\boldsymbol{\alpha}$ on the left and by the $\delta$ from \textsection{3.2} on the right, with $\nu_{h}=0.04$ and $h\in\{1/16,1/32,1/64,1/128,1/256\}$ .  At the bottom we have 3 levels of $p$ at fixed $h=1/64$ on the left, and on the right the corresponding values of $\psi$ from \textsection{3.3}, where $\iota_{s}=0.04$.  Each graph is shown at $T=1.3$ minutes.}
\label{fig:Auplus}  
\end{figure}

The forward reaction rates are taken from \cite{CH3FAu2}, and are given to satisfy $k_{f_{1}} \approx 8.9\times 10^{-12} \mathrm{cm}^{3}/\mathrm{mol}\cdot\mathrm{s}$, and $k_{f_{2}} \approx 3.4\times10^{-12}\mathrm{cm}^{3}/\mathrm{mol}\cdot\mathrm{s}$.  The backward rates are simply derived from the equilibrium constant $K_{eq}$ approximation, using the standard isothermal Gibb's free energy change for the reactions $\Delta G^{\ominus}_{\vartheta}=-R\vartheta\ln K_{eq}$, as measured in \cite{CH3FAu2} are given as $K_{eq} = 2.63$ near standard state ($p=1 \ \mathrm{atm}, \vartheta=295\pm 2 \mathrm{K}$), such that we can simply assume that $k_{b_{1}}= k_{f_{1}}/K_{eq}$ and $k_{b_{2}} = k_{f_{2}}/K_{eq}$. 

Then adopting the standard --- though here prolix --- notation for $\alpha_{1}=[\mathrm{CH}_{3}\mathrm{F}]$,   $\alpha_{2}=[\mathrm{Au}^{+}]$, $\alpha_{3}=[\mathrm{Au}^{+}\mathrm{CH}_{3}\mathrm{F}]$, and $\alpha_{4} = [\mathrm{AuCH}_{2}^{+}]$, we obtain:  \begin{equation}\begin{aligned}\label{chemkin2}&\partial_{t}[\mathrm{CH}_{3}\mathrm{F}]= k_{b_{1}}[\mathrm{Au}^{+}\mathrm{CH}_{3}\mathrm{F}] + k_{b_{2}}[ \mathrm{AuCH}_{2}^{+}] -(k_{f_{1}}+k_{f_{2}})[\mathrm{CH}_{3}\mathrm{F}][\mathrm{Au}^{+}], \\ & \ \partial_{t}[\mathrm{Au}^{+}]=  k_{b_{1}}[\mathrm{Au}^{+}\mathrm{CH}_{3}\mathrm{F}] + k_{b_{2}}[ \mathrm{AuCH}_{2}^{+}] -(k_{f_{1}}+k_{f_{2}})[\mathrm{CH}_{3}\mathrm{F}][\mathrm{Au}^{+}], \\ &  \quad \ \  \partial_{t}[\mathrm{Au}^{+}\mathrm{CH}_{3}\mathrm{F} ] = (k_{f_{1}}+k_{f_{2}})[\mathrm{CH}_{3}\mathrm{F}][\mathrm{Au}^{+}] -k_{b_{1}}[\mathrm{Au}^{+}\mathrm{CH}_{3}\mathrm{F}], \\ & \quad\qquad \partial_{t}[\mathrm{AuCH}_{2}^{+}] = (k_{f_{1}}+k_{f_{2}})[\mathrm{CH}_{3}\mathrm{F}][\mathrm{Au}^{+}] -k_{b_{2}}[ \mathrm{AuCH}_{2}^{+}],\end{aligned}\end{equation}  where $[\mathrm{He}]$ is the inert reactive bath.  Now, as discussed in \textsection{2}, it is not difficult to explicitly solve this system of first order ordinary differential equations in terms of the previous timestep.  That is, for an ODE in $\varpi$ of the form $\varpi' = C_{1}\varpi+C_{2}$ (as is each of our constituents in \ref{chemkin2}) we have the general solution over $\Delta t$, \begin{equation}\label{firord}\varpi(t^{n+1}) = \exp^{\int_{\Delta t} C_{1}dt}\left(\varpi(t^{n})+\frac{C_{2}}{C_{1}}\right) - \frac{C_{2}}{C_{1}}.\end{equation}  

Thus letting $f(t^{n})=f^{n}$ for our coupled system (\ref{chemkin2}), and solving each first order ordinary differential equation, we arrive with \begin{equation}\begin{aligned}\label{chemkin3}[\mathrm{CH}_{3}\mathrm{F}]^{n+1}& = \left(\frac{k_{b_{1}}[\mathrm{Au}^{+}\mathrm{CH}_{3}\mathrm{F}]^{n} + k_{b_{2}}[ \mathrm{AuCH}_{2}^{+}]^{n}}{(k_{f_{1}}+k_{f_{2}})[\mathrm{Au}^{+}]^{n}} \right)  \\ &  \ + e^{-(k_{f_{1}}+k_{f_{2}})[\mathrm{Au}^{+}]^{n}\Delta t}\left([\mathrm{CH}_{3}\mathrm{F}]^{n} - \frac{k_{b_{1}}[\mathrm{Au}^{+}\mathrm{CH}_{3}\mathrm{F}]^{n} + k_{b_{2}}[ \mathrm{AuCH}_{2}^{+}]^{n}}{(k_{f_{1}}+k_{f_{2}})[\mathrm{Au}^{+}]^{n}} \right), \\ [\mathrm{Au}^{+}]^{n+1} & =   \left(\frac{k_{b_{1}}[\mathrm{Au}^{+}\mathrm{CH}_{3}\mathrm{F}]^{n} + k_{b_{2}}[ \mathrm{AuCH}_{2}^{+}]^{n}}{(k_{f_{1}}+k_{f_{2}})[\mathrm{CH}_{3}\mathrm{F}]^{n}} \right)  \\ &  \ + e^{- (k_{f_{1}}+k_{f_{2}})[\mathrm{CH}_{3}\mathrm{F}]^{n}\Delta t}\left([\mathrm{Au}^{+}]^{n} - \frac{k_{b_{1}}[\mathrm{Au}^{+}\mathrm{CH}_{3}\mathrm{F}]^{n} + k_{b_{2}}[ \mathrm{AuCH}_{2}^{+}]^{n}}{(k_{f_{1}}+k_{f_{2}})[\mathrm{CH}_{3}\mathrm{F}]^{n}} \right), \\  [\mathrm{Au}^{+}\mathrm{CH}_{3}\mathrm{F}]^{n+1} & =  \frac{1}{k_{b_{1}}} \left((k_{f_{1}}+k_{f_{2}})[\mathrm{CH}_{3}\mathrm{F}]^{n}[\mathrm{Au}^{+}]^{n}\right) \\ & \ + e^{-k_{b_{1}}\Delta t}\left([\mathrm{Au}^{+}\mathrm{CH}_{3}\mathrm{F}]^{n} - \frac{1}{k_{b_{1}}} \left((k_{f_{1}}+k_{f_{2}})[\mathrm{CH}_{3}\mathrm{F}]^{n}[\mathrm{Au}^{+}]^{n}\right)\right), \\  [\mathrm{AuCH}_{2}^{+}]^{n+1} & =  \frac{1}{k_{b_{2}}} \left((k_{f_{1}}+k_{f_{2}})[\mathrm{CH}_{3}\mathrm{F}]^{n}[\mathrm{Au}^{+}]^{n}\right) \\ & \ + e^{-k_{b_{2}}\Delta t}\left([\mathrm{Au}\mathrm{CH}_{2}^{+}]^{n} - \frac{1}{k_{b_{2}}} \left((k_{f_{1}}+k_{f_{2}})[\mathrm{CH}_{3}\mathrm{F}]^{n}[\mathrm{Au}^{+}]^{n}\right)\right).\end{aligned}\end{equation}  Now these comprise the analytic predictor solutions $\tilde{\alpha}_{i}^{n+1}$ from (\ref{fir}), and we are without difficulty able to construct (\ref{diffalpha}) in the mass action operator $\mathring{\mathscr{A}}$.

Note that our initial conditions are taken to satisfy Gaussian distributions in $N=2$ dimensional space: \[\begin{aligned} \big[\mathrm{CH}_{3}\mathrm{F}\big]_{0} = 6e^{-((x-1.6)^{2}+(y-1.6)^{2}) / 1.5}, & \ \ \big[\mathrm{Au}^{+}\big]_{0} =  6e^{-((x-0.6)^{2}+(y-3.6)^{2})}, \\ \big[\mathrm{Au}^{+}\mathrm{CH}_{3}\mathrm{F}\big]_{0} = 4e^{-((x-0.4)^{2}+(y-0.4)^{2}) / 2}, & \ \ \big[\mathrm{AuCH}_{2}^{+}\big]_{0} =  6e^{-((x+1.6)^{2}+(y+1.6)^{2}) / 1.75},\end{aligned}\] where for simplicity here, we take as a first order approximation that the diffusivity is constant with respect to the helium bath $\mathscr{D} \sim 5\times 10^{-5}$ m$^{2}\cdot$s$^{-1}$ (for example see the CRC handbook of chemistry and physics online edition, and compare helium methane/sulfur hexafluoride mixtures, \emph{etc.}).  We show some numerical results in Figure \ref{fig:Auplus}, where $\Delta t = 10$ s, the initial $h$ level locally is $h=1/16$ and the highest level of refinement corresponds to $h=1/256$ locally.  Similarly the initial $p$ level is $p=1$ locally, and the highest level is $p=3$.  The multi-corrector is set with no fast timesteps here, and with the convergence bound of $C=10^{-12}$ taking $\ell =10$, where in most steps $\alpha_{1}$ is the slowest to converge, and usually does not achieve the $C$ bound before reaching $\ell$ --- often achieving a $C$-tolerance corresponding to $\sim 10^{-9}$.

\begin{figure}[!t]
\centering
\quad \includegraphics[width=11cm]{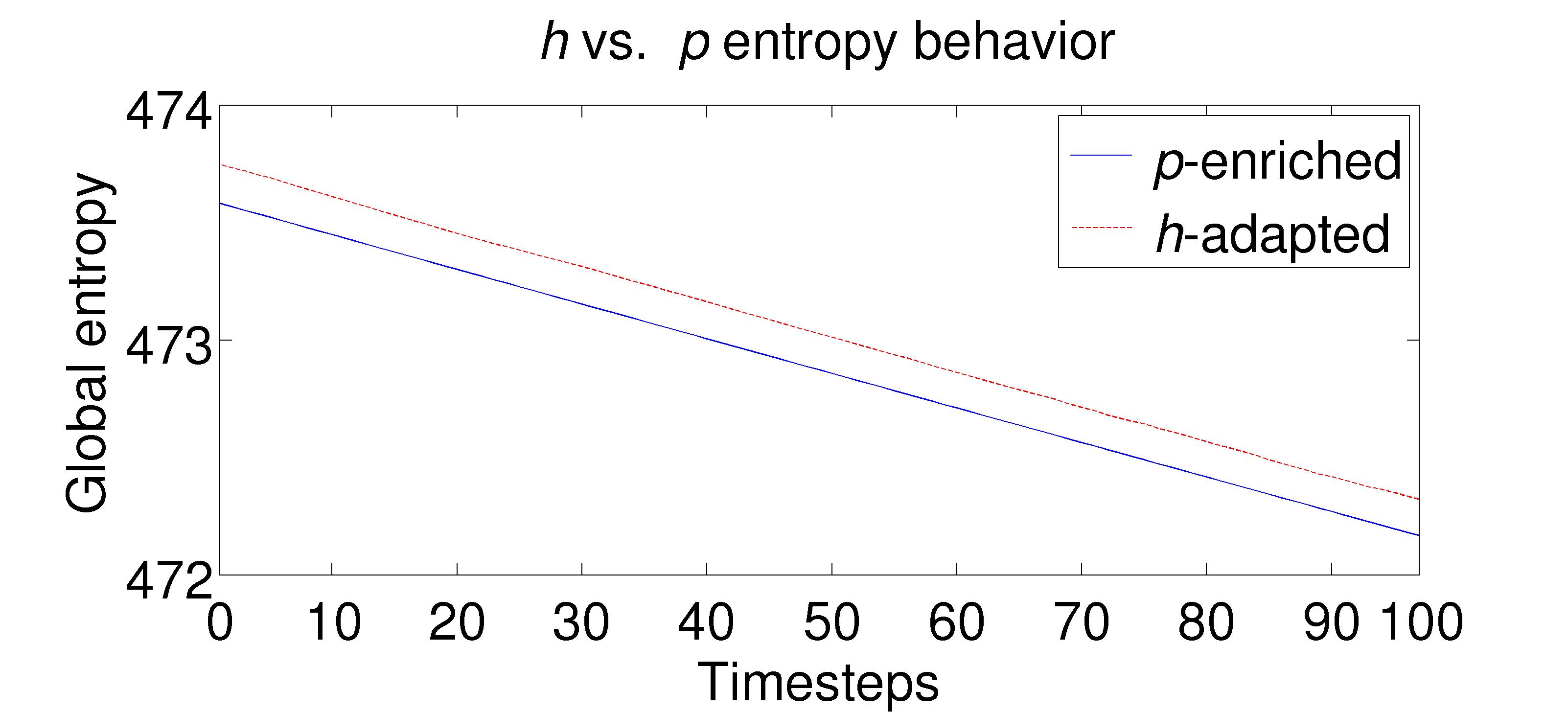}\caption{Here we show the entropy consistency of each scheme from Figure \ref{fig:Auplus}, where the uncoupled $p$-enrichment scheme shows a slightly offset decay in $\mathscr{S}_{\mathfrak{R}}^{k+1}$ than that of the uncoupled $h$-adaptive scheme.}
\label{fig:AuEntropy}  
\end{figure}

We also note that both solutions shown in Figure \ref{fig:Auplus} --- the uncoupled $h$-adaptive and the uncoupled $p$-enrichment schemes --- are entropy consistent as displayed in Figure \ref{fig:AuEntropy} up to $T\sim 17$ minutes using timesteps of $\Delta t = 10$ s.  However, this example is quite simple, and we have not shown the full $hp$-adaptive results.  Let us now address a more complicated system: the Belousov-Zhabotinskii reaction.

\subsection{\texorpdfstring{\S 4.3 Autocatalysis: the BZ reaction with $\mathscr{D}_{ij}(\alpha)$}{\S 4.3 Autocatalysis: the BZ reaction with functional diffusion}}

Here we consider a mixed regime, where the reactivity and diffusivity demonstrate a complicated and subtle interplay between the reaction and diffusion limited and dominated regimes, respectively.  That is, the classical reaction-diffusion known as the Belousov-Zhabotinskii (BZ) reaction is known to demonstrate reactive oscillations in time, and comprises what is known as an excitable medium; which is simply a medium whose propagation is nonlinearly constrained by a dispersion limited (visible to the eye) refractory period caused by local gradients in the rate limiting reagent.  In this sense, the non-equilibrium thermodynamics of the BZ reaction illustrates a nice example of a reaction regime that oscillates between a diffusion dominated process and a reaction dominated process.  

The generalized chemical kinetics of the system are characterized by the following coupled set of chemical reactions:\  \begin{subequations}\begin{align}\label{BZ1} \mathrm{BrO}_{3}^{-} + \mathrm{Br}^{-} \  &  \ce{->T[{$ \ \ \ k_{f_{1}} \ \ $}] } \ \mathrm{HBrO}_{2} + \mathrm{P} \\  \label{BZ2}\mathrm{HBrO}_{2} +  \mathrm{Br}^{-}  \ &  \ce{->T[{$ \ \ \ k_{f_{2}} \ \ $}] } \ 2\mathrm{P} \\ \label{BZ3}\mathrm{BrO}_{3}^{-} + \mathrm{HBrO}_{2} \ &  \ce{->T[{$ \ \ \ k_{f_{3}} \ \ $}] } \ 2\mathrm{HBrO}_{2} + 2\mathrm{M} \\  \label{BZ4} 2\mathrm{HBrO}_{2} \ &   \ce{->T[{$ \ \ \ k_{f_{4}} \ \ $}] }  \ \mathrm{BrO}_{3}^{-} + \mathrm{P} \\ \label{BZ5}\mathrm{B} + \mathrm{M}  \ &  \ce{->T[{$ \ \ \ k_{f_{5}} \ \ $}] } \ c\mathrm{Br}^{-}\end{align}\end{subequations} such that $\alpha_{1}= [\mathrm{HBrO}_{2}]$, $\alpha_{2}= [\mathrm{Br}^{-}]$, $\alpha_{3}=[M]$, $\alpha_{4}=[\mathrm{BrO}_{3}^{-}]$, $\alpha_{5}=[\mathrm{P}]$ and $\alpha_{6}= [\mathrm{B}]$ where in addition we take a positive constant $c\in\mathbb{R}^{+}$ depending on the initial concentration of $\mathrm{B}$ (\emph{i.e.} $c=c(\alpha_{6}^{0})$ where $\alpha_{6|t=0}=\alpha_{6}^{0}$) and contained in the interval $c\in\Big[\frac{1}{4},\frac{1+\sqrt{2}}{2}\Big]$ whenever dynamic oscillations are present in the reaction space $\mathfrak{R}$.  That is, $c$ is considered here as an adjustable stoichiometric coefficient, up to an adjustable rate constant $k_{f_{5}}$, which depends on the choice of catalyst $\mathrm{M}$ and the oxidized (generally organic) species $\mathrm{B}$, where $\mathrm{P}$ and $\mathrm{BrO}_{3}^{-}$ are the generalized reaction products.  As is the standard convention, we neglect the hydrogen cations in (\ref{BZ1})-(\ref{BZ5}).

The classical approach to modeling the BZ reactive system is to employ the Oregonator model \cite{Merkin,Dockery}, which is a simplification of the FKN (Fields, K\"{o}ros and Noyes) chemical mechanism \cite{Field,Field2} that relies strongly on the underlying assumption that the products of the reaction are decoupled from the dynamics of the oscillation mechanism \cite{Taylor}.  It should also be noted that alternative model systems do exist, such as the radicalator and GTF models \cite{Taylor}, which generally attempt to formulate systems reliant on many more degrees of freedom in $\boldsymbol{\alpha}$, and to eliminate, for example, the adjustable parameter $c$.  These extended models introduce their own problems however (\emph{e.g.} $n=26$ and $r_{max}=80$), and it remains unclear, at present, whether they model the experimental behavior better than the classical Oregonator.

\begin{figure}[!t]
\centering
\includegraphics[width=8.cm]{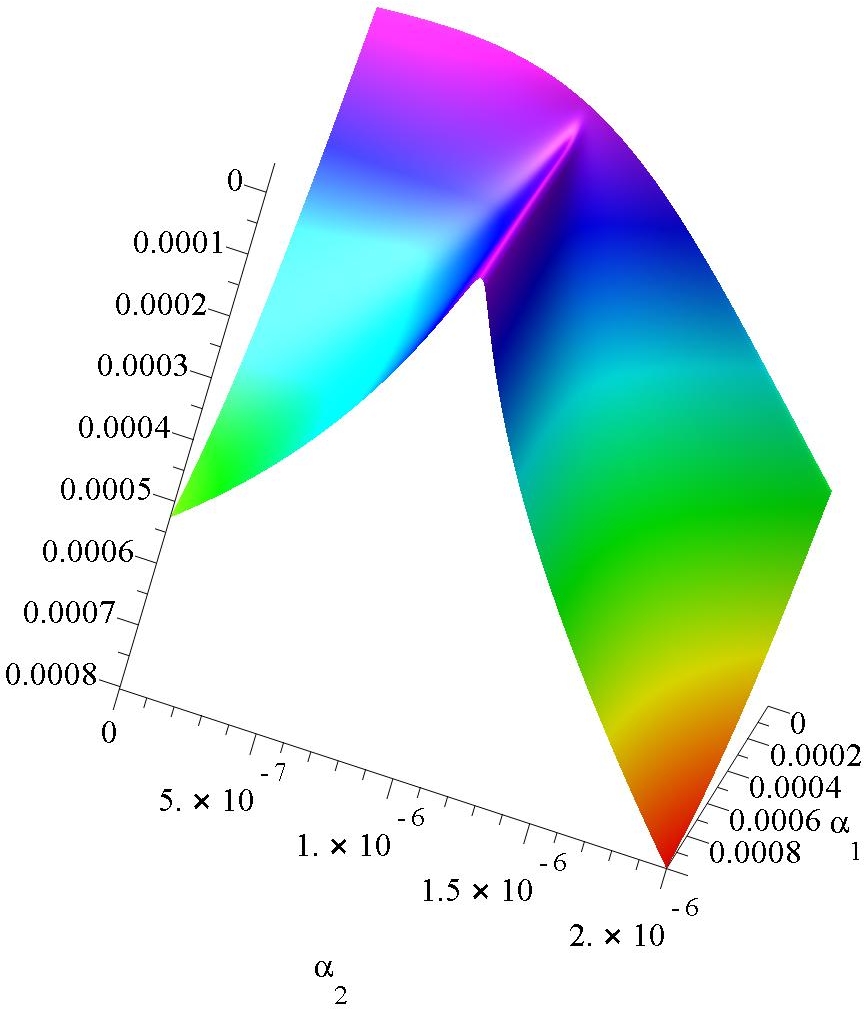} \includegraphics[width=8cm]{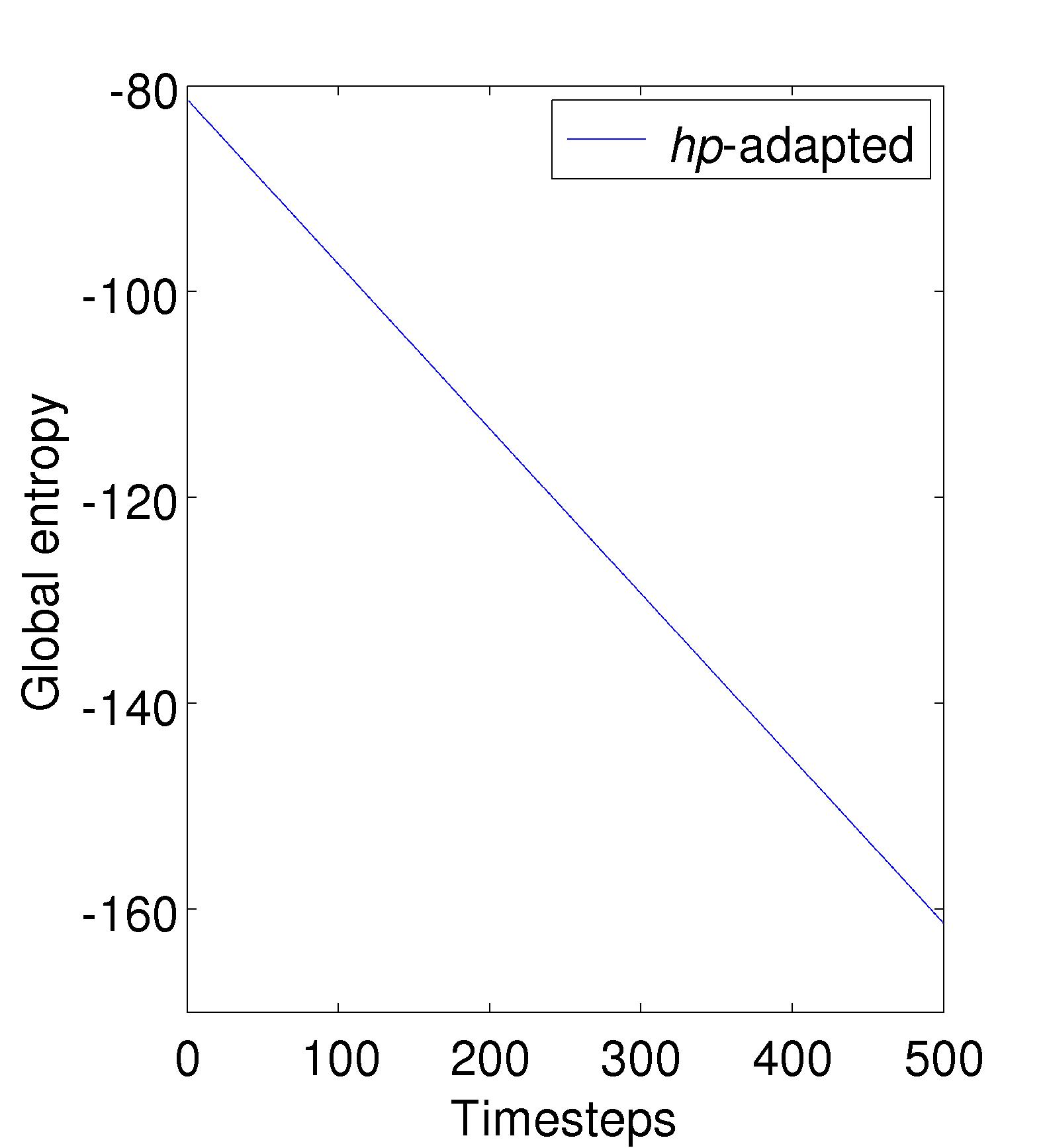}\caption{We plot (\ref{solRicBZ}) near a limit cycle when $\zeta_{n}>0$ on the left, and the on the right is the entropy consistency of the solution from Figure after $T=25$ seconds.}
\label{fig:limitcyc}  
\end{figure}

Below, we expand on the classical Oregonator model to couple all six unknowns of the system (including the product well constituent $[\mathrm{P}]$) and derive a new formalism for treating this system which utilizes the methods implicit in the scheme presented in \textsection{2} and \textsection{3}.  However, expanding the Oregonator to dynamically include the bath components leads to a number of additional complications in the solution space.   For example, it is well known that the Oregonator model admits bistable nonequilibrium solutions \cite{Dockery}, while here, this bistability is observed, and moreover --- as we predict from studying the surface of the reaction coordinate (\emph{e.g.} see Figure \ref{fig:limitcyc}) --- the solution in this extended treatment may develop local multistability behavior due to the highly irregular reactive surface.  However, we also observe a dampening effect noticeable near the limit cycle, which is not usually present in the standard Oregonator model, though very nicely agrees with the experimental observation in oscillating reaction experiments.  This is not surprising since local fluctuations in the bath concentration can have large effects on the local magnitude of the functional parameters (\emph{e.g.} $\mathscr{D}_{ij}(\alpha)$ and $K_{eq}(\alpha)$).  These observations require closer analysis, and generally lie beyond the scope of this paper.  Here, we restrict ourselves primarily to introducing the model.  

Starting with the quiescent reactor scheme from \textsection{2} we partially decouple our kinetic equations for $t\in(0,T)$ from (\ref{quiescent}) by solving the following fully explicit integrated rate law: \begin{subequations}\begin{align}\label{2BZ1} \partial_{t}\alpha_{1} = &k_{f_{1}}\alpha_{4}\alpha_{2}  + k_{f_{3}}\alpha_{4}\alpha_{1} - k_{f_{2}} \alpha_{1}\alpha_{2} - 2k_{f_{4}}\alpha_{1}^{2} \\ \label{2BZ2}\partial_{t}& \alpha_{2}= ck_{f_{5}}\alpha_{6}\alpha_{3} -k_{f_{1}}\alpha_{4}\alpha_{2} -   k_{f_{2}} \alpha_{1}\alpha_{2}  \\ \label{2BZ3} & \quad \partial_{t}\alpha_{3} = 2k_{f_{3}}\alpha_{1}\alpha_{4}- k_{f_{5}}\alpha_{6}\alpha_{3} \\ \label{2BZ4}& \partial_{t}\alpha_{4} =  k_{f_{4}}\alpha_{1}^{2}-k_{f_{1}}\alpha_{2}\alpha_{4} - k_{f_{3}}\alpha_{1}\alpha_{4} \\ \label{2BZ5} & \partial_{t}\alpha_{5} =  k_{f_{4}}\alpha_{1}^{2}+ 2k_{f_{2}}\alpha_{1}\alpha_{2} + k_{f_{1}}\alpha_{4}\alpha_{2} \\ \label{2BZ6} &\qquad\qquad\partial_{t}\alpha_{6} = -k_{f_{5}}\alpha_{3}\alpha_{6}.\end{align}\end{subequations}  We also note that we explicitly solve here for the reaction products $\mathrm{B},\mathrm{P}$ and $\mathrm{BrO}_{3}^{-}$, which are conventionally neglected due to the fact that they are each present in excess throughout the medium and are known to have negligible effect on the dynamics. However, in our system they play a rather central role, as we will see below.

Now, it is immediately clear that all of the corresponding differential equations (\ref{2BZ2})-(\ref{2BZ6}) are linear with respect to time, with the exception being (\ref{2BZ1}).  That is, the linear equations, as in \textsection{4}, may be solved using (\ref{firord}), which yields in the $\alpha$ notation that: \begin{equation}\begin{aligned}\label{BZdis}& \alpha_{2}^{n+1}= \left(\frac{ ck_{f_{5}}\alpha^{n}_{6}\alpha^{n}_{3}}{k_{f_{1}}\alpha_{4}^{n}+k_{f_{2}}\alpha_{1}^{n}}\right) + e^{-(k_{f_{1}}\alpha_{4}^{n}+k_{f_{2}}\alpha_{1}^{n})\Delta t}\left(\alpha_{2}^{n} - \frac{ ck_{f_{5}}\alpha^{n}_{6}\alpha^{n}_{3}}{k_{f_{1}}\alpha_{4}^{n}+k_{f_{2}}\alpha_{1}^{n}} \right), \\ & \quad\qquad \alpha_{3}^{n+1}= \left(\frac{  2k_{f_{3}}\alpha^{n}_{1}\alpha^{n}_{4}}{k_{f_{5}}\alpha_{6}^{n}}\right) + e^{-k_{f_{5}}\alpha_{6}^{n}\Delta t}\left(\alpha_{3}^{n} - \frac{2k_{f_{3}}\alpha^{n}_{1}\alpha^{n}_{4}}{k_{f_{5}}\alpha_{6}^{n}} \right) \\  & \alpha_{4}^{n+1}= \left(\frac{ k_{f_{4}}(\alpha^{n}_{1})^{2}}{k_{f_{1}}\alpha_{2}^{n}+k_{f_{3}}\alpha_{1}^{n}}\right) + e^{-(k_{f_{1}}\alpha_{2}^{n}+k_{f_{3}}\alpha_{1}^{n})\Delta t}\left(\alpha_{4}^{n} - \frac{ k_{f_{4}}(\alpha^{n}_{1})^{2}}{k_{f_{1}}\alpha_{2}^{n}+k_{f_{3}}\alpha_{1}^{n}} \right),   \\  & \alpha_{5}^{n+1}= \Delta t (k_{f_{4}}\alpha_{1}^{2}+ 2k_{f_{2}}\alpha_{1}\alpha_{2} + k_{f_{1}}\alpha_{4}\alpha_{2}) + \alpha_{5}^{n}, \ \ \mathrm{and} \ \ \alpha_{6}^{n+1}= \alpha_{6}^{n}  e^{-k_{f_{5}}\alpha_{3}^{n}\Delta t} .\end{aligned}\end{equation}  

The first equation, on the other hand, requires a solution to the classical nonlinear Riccati equation, which means solving for \begin{equation}\label{Ricatti}\varpi' = C_{1}\varpi^{2}+C_{2}\varpi + C_{3},\end{equation} where here we may treat as constants $C_{1},C_{2},C_{3}\in\mathbb{R}$.  Generally we may solve (\ref{Ricatti}) by way of the fundamental theorem of calculus, such that we first obtain \[\int^{\varpi}\left(\frac{1}{C_{1}s^{2}+C_{2}s+C_{3}}\right)ds-t=0, \] for $C_{4}\in\mathbb{R}$.  Then the general solution is simply determined by the sign of the discriminant $\zeta = C_{2}^{2}-4C_{1}C_{3}$ of the polynomial in $s$ from the denominator, such that upon integrating we arrive with the general solution: \begin{equation}\label{solRic}  \varpi = \left\{\begin{matrix}  -1/C_{1}t - \sqrt{C_{3}/C_{1}} & \mathrm{for} \ C_{2}^{2}=4C_{1}C_{3} \\  \frac{1}{2}C_{1}^{-1}\left(\tan(\frac{1}{2} t \xi^{1/2})\xi^{1/2} - C_{2}\right) & \mathrm{for} \ C_{2}^{2}<4C_{1}C_{3} \\  -\frac{1}{2}C_{1}^{-1}\left(\tanh(\frac{1}{2} t\zeta^{1/2})\zeta^{1/2}+C_{2}\right)  & \mathrm{for} \ C_{2}^{2}>4C_{1}C_{3}\end{matrix}\right.\end{equation} 

Now, in order to find the well-posed discretized version of (\ref{Ricatti}) over $\Delta t^{n}$, we note that the differential equation in $\alpha_{1}$ similarly yields, \[\int^{\alpha_{1}^{n+1}}_{\alpha_{1}^{n}}\left(k_{f_{1}}\alpha^{n}_{4}\alpha^{n}_{2}+(k_{f_{3}}\alpha^{n}_{4}-k_{f_{2}}\alpha^{n}_{2})s-2k_{f_{4}}s^{2}\right)^{-1}ds-\Delta t^{n}=0,\] while the general solution to the differential equation yields 
\begin{equation}
\label{ricsoln2}
\alpha_{1}^{n} = -\frac{1}{2C_1}\left(\tanh\left(\frac{t+\bar{C}}{2}\zeta^{1/2}\right)\zeta^{1/2}+C_2\right)
\end{equation}
where $\bar{C}$ is a constant that depends on $\alpha(0)$ and $\zeta =C_2^2-4C_1C_3>0$.  Now we will be concerned with determining $\alpha_{1}^{n+1}$ given $\alpha_{1}^n$ for $t^{n+1}=t^n+\Delta t$.  Then setting $\zeta_{n}=(C_2^{n})^2-4C_1C_3^{n}>0$, and letting $C_{1}=-2k_{f_{4}}$, $C_{2}^{n}=(k_{f_{3}}\alpha^{n}_{4}-k_{f_{2}}\alpha_{2}^{n})$ and $C^{n}_{3}=k_{f_{1}}\alpha^{n}_{4}\alpha_{2}^{n}$, we have
\[
\alpha_{1}^{n+1} = -\frac{1}{2C_1}\left(\tanh\left(\frac{t^n+\Delta t+\bar{C}}{2}\zeta_{n}^{1/2}\right)\zeta_{n}^{1/2}+C^{n}_2\right)
\]
but this involves the unknown $\bar{C}$ and not $\alpha_{1}^n$. Then by rewriting the argument of the $\tanh$ into two groups:
\[
\alpha_{1}^{n+1} = -\frac{1}{2C_1}\left(\tanh\left(\frac{t^n+\bar{C}}{2}\zeta_{n}^{1/2}+\frac{\Delta t}{2}\zeta_{n}^{1/2}\right)\zeta_{n}^{1/2}+C^{n}_2\right)
\]
and expanding using 
\[
\tanh(x+y) = \left(\frac{ \tanh(x)+\tanh(y) }{1+\tanh(x)\tanh(y)}\right),
\]
we see that
\begin{equation}\label{back}
\alpha_{1}^{n+1} = -\frac{1}{2C_1}\left(\frac{\tanh\left(\frac{t^n+\bar{C}}{2}\zeta_{n}^{1/2}\right)+\tanh\left(\frac{\Delta t}{2}\zeta_{n}^{1/2}\right)}{1+\tanh\left(\frac{t^n+\bar{C}}{2}\zeta_{n}^{1/2}\right)\tanh\left(\frac{\Delta t}{2}\zeta_{n}^{1/2}\right)}\zeta_{n}^{1/2}+C^{n}_2\right).
\end{equation}
Then recognizing that the term from (\ref{ricsoln2}),
\[
\tanh\left(\frac{t^n+\bar{C}}{2}\zeta_{n}^{1/2}\right)\quad\mathrm{rearranges \ to,}\quad -(2C_1\alpha_{1}^n+C^{n}_2)\zeta_{n}^{-1/2} = \tanh\left(\frac{t^n+\bar{C}}{2}\zeta_{n}^{1/2}\right),
\] such that substituting back into (\ref{back}) eliminates the term with the unknown $\bar{C}$ in terms of the known $\alpha_{1}^{n}$, whereby we arrive with
\[
\alpha_{1}^{n+1} = -\frac{1}{2C_1}\left(\frac{ -(2C_1\alpha_{1}^{n}+C_2)\zeta_{n}^{-1/2}+\tanh\left(\frac{\Delta t}{2}\zeta_{n}^{1/2}\right)}{1-(2C_1\alpha_{1}^{n}+C_2)\zeta_{n}^{-1/2}\tanh\left(\frac{\Delta t}{2}\alpha_{1}^{1/2}\right)}\zeta_{n}^{1/2}+C_2\right),
\] so that letting $\iota^{n} = \sqrt{C_{1}C_{3}^{n}}$ and noticing that $C_{1}\leq 0$ we recover the full solution:  \begin{equation}\label{solRicBZ}  \alpha_{1}^{n+1} = \left\{\begin{matrix} \left(\Delta t (\iota^{n}\alpha_{1}^{n} + C_{3}^{n}) - C_{1}\alpha_{1}^{n}\right)/\left( \Delta t (\iota^{n} + C_{1}\alpha_{1}^{n}) -C_{1}\right)  & \mathrm{for} \ \zeta_{n} = 0, \\  -\frac{1}{2C_1}\left(\frac{ -(2C_1\alpha_{1}^{n}+C_2)\zeta_{n}^{-1/2}+\tanh\left(\frac{\Delta t}{2}\zeta_{n}^{1/2}\right)}{1-(2C_1\alpha_{1}^{n}+C_2)\zeta_{n}^{-1/2}\tanh\left(\frac{\Delta t}{2}\alpha_{1}^{1/2}\right)}\zeta_{n}^{1/2}+C_2\right)  & \mathrm{for} \ \zeta_{n} > 0.\end{matrix}\right.\end{equation}



It remains to address our variable in $\alpha$ diffusion tensor.  That is, in a slightly more general setting than in \textsection{4.1} and \textsection{4.2}, it is the case that the diffusion tensor is known to obey functional dependencies such that, for example $\mathscr{D}_{ij}=\mathscr{D}_{ij}(\alpha,\vartheta)$ for $\vartheta$ the temperature at which the reaction occurs (thus constant in the isothermal approximation).  These dependencies may be determined in a number of different ways, namely: $(a)$ they may be determined empirically, $(b)$ they may be determined using basic collisional theory arguments, $(c)$ they may be determined using the Stokes-Einstein relation, $(d)$ they may be determined by applying the fluctuation dissipation theory, and so forth.  In any case, the coefficients $\mathscr{D}_{i}$ may be simply taken as the row sum over the matrices $\mathscr{D}_{ij}$, such that $\mathscr{D}_{i}=\sum_{j=1}^{n}\mathscr{D}_{ij}$.

We present a simple functional form for the mass diffusion (neglecting concentration gradient dependencies (\emph{e.g.} $\mathscr{D}_{ij}=\mathscr{D}_{ij}(\nabla_{x}\alpha,\alpha,\vartheta)$ in \cite{Bardow}) as discussed and derived in Ref.~\cite{Hirschfelder,Chapman,Giovangigli3} which arises as a natural result of the Chapman-Enskog theory.  That is, recall that the $j$-th constituent $\mathfrak{M}_{j}$ is written as a concentration with respect to the specific volume $\rho^{-1}$.  Fixing our $\alpha_{j}$'s as the molar fraction of species $\mathfrak{M}_{j}$, where $M_{j}$ corresponds to its specific molecular mass, we may write the standard multicomponent diffusion tensor as satisfying: \begin{equation}\label{massdiffusion}\mathscr{D}_{ij} = \left(\frac{1}{M_{j}}\sum_{k}\alpha_{k}M_{k}\right)\frac{K^{ji}-K^{ii}}{|K|},\end{equation} where the cofactor matrices are given by: \begin{equation}K^{ji}= (-1)^{i+j} \left|\begin{array}{cccccc}
     0 & \hdots & K_{1,i-1} &K_{1,i+1} & \ldots & K_{1,n}\\
    \vdots &  & \vdots & \vdots &  &\vdots  \\
 K_{j-1,1} & \ldots & K_{j-1,i-1} & K_{j-1,i+1} & \hdots & K_{j-1,n} \\
    K_{j+1,1} & \hdots & K_{j+1,i-1} & K_{j+1,i+1} & \hdots  & K_{j+1,n} \\
      \vdots &   & \vdots & \vdots & & \vdots \\ 
     K_{n,1} & \hdots  & K_{n,i-1} & K_{n,i+1} & \hdots & 0 \end{array}\right|\end{equation} with entries defined by, \[ K_{ij}= \frac{\alpha_{i}}{[\mathscr{D}_{ij}]}+\frac{M_{j}}{M_{i}}\sum_{k\neq i}\frac{\alpha_{k}}{[\mathscr{D}_{ik}]} \ \mathrm{if } \ i \ne j,  \quad  \mathrm{and} \ \mathrm{zero} \ \mathrm{when} \  i=j,\] such that the binary mixtures are set componentwise by way of the reduced molecular mass via, \[[\mathscr{D}_{ij}]=\mathscr{C}_{ij} p^{-1}\sqrt{\vartheta^{3}(M_{i}+M_{j})/2M_{i}M_{j}}.\]  The Chapman-Enskog prefactors $\mathscr{C}_{ij}=\mathscr{C}_{ij}(\alpha_{i},\alpha_{j})$  (see Ref.~\cite{Hirschfelder,Chapman} for derivation and details) are defined in terms of the reduced temperature of the mixture $\vartheta^{*}_{ij}=\vartheta^{*}_{ij}(\alpha_{i},\alpha_{j})$, a unitless function of the first order deviation from the idealized rigid sphere model denoted $\Omega_{ij}^{(1,1)}$, and the cross sectional radius $\sigma_{ij}=\sigma_{ij}(\mathfrak{M}_{i},\mathfrak{M}_{j})$ in \AA \ such that in the first approximation, $\mathscr{C}_{ij}=2.628\times10^{-3}(\sigma_{ij}^{2}\Omega_{ij}^{(1,1)}\vartheta_{ij}^{*})^{-1}$ (\emph{viz.} equation 7.4--4 in \cite{Hirschfelder}), where $\vartheta$ is in K, $p$ is in atm, the $M_{k}$ are in $\mathrm{g}\cdot\mathrm{mol}^{-1}$ and the $[\mathscr{D}_{ij}]$ are in $\mathrm{cm}^{2}\cdot\mathrm{s}^{-1}$.   

For simplicity, we take the deviation parameter $\Omega_{ij}^{(1,1)}$ to unity to restrict to the rigid sphere approximation. Then setting $\sigma_{ij}=\pi b_{ij,max}^{2}$, where $b_{ij,max} = (r_{i}+r_{j})$ such that the $r_{i}$ correspond to the approximate maximum radius of a single molecule of $\mathfrak{M}_{i}$ in the rigid sphere approximation to the reactive cross section as shown in \cite{Houston} (where molecular geometries and bond lengths are approximated using the ghemical/Mopac suite), and approximating the critical relations in the binary mixtures via the weighted sum $\vartheta_{ij}^{*}=\sum_{k\in\{i,j\}}\alpha_{k}\vartheta^{*}_{k}$, as seen in \cite{Akasaka} for example, where the approximate pure critical temperatures are given using the technique developed in \cite{Stein}.

Further note that we may rewrite (\ref{massdiffusion}) using the classical adjoint matrix such that \begin{equation}\label{massdiffusion2}\mathscr{D}_{ij} = \left(\frac{1}{M_{j}}\sum_{k}\alpha_{k}M_{k}\right)(K^{-1})_{ij}-(K^{-1})_{ii},\end{equation} where $(K^{-1})_{ij}$ represents the $ij$-th entry of the full rank inverse matrix $K^{-1}$.

\begin{table}[t]
\centering
{
\renewcommand{\arraystretch}{1.3}
\begin{tabular}{|c|c|c|c|c|c|c|}
\hline
\bf{Species} & $\alpha_{1}$ &  $\alpha_{2}$ &  $\alpha_{3}$ &  $\alpha_{4}$ &  $\alpha_{5}$ &  $\alpha_{6}$ \\ 
\hline\hline
$\vartheta_{ii}^{*}$  & $.343$ &  $.364$ & $.428$ &  $.345$ & $.341$ & $.428$  \\
\hline
  $r_{i}/$\AA &  $3.35$ & $1.2$ & $6.4$ &  $2.8$ &  $2.3$ & $6.4$ \\ 
\hline
\end{tabular}
}
\label{table:diffconst2}
\caption{We show the results of the crude approximate values for the reduced temperature $\vartheta_{ii}^{*}$ and the effective maximum cross sectional radius $r_{i}$ of each pure species $\mathfrak{M}_{i}$. }
\end{table}

\begin{figure}[!t]
\centering
 \includegraphics[width=8cm]{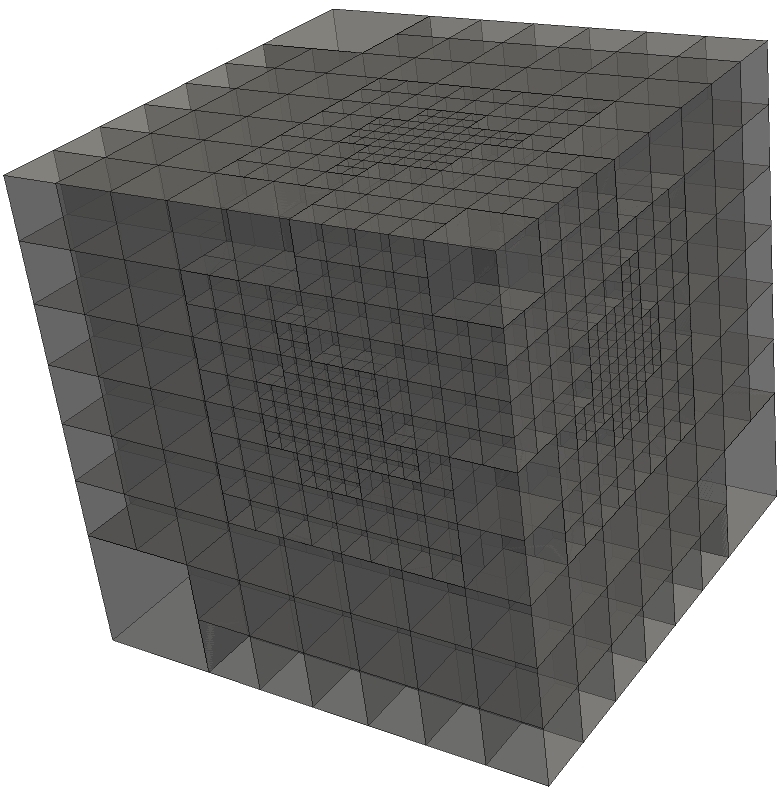}\includegraphics[width=8.cm]{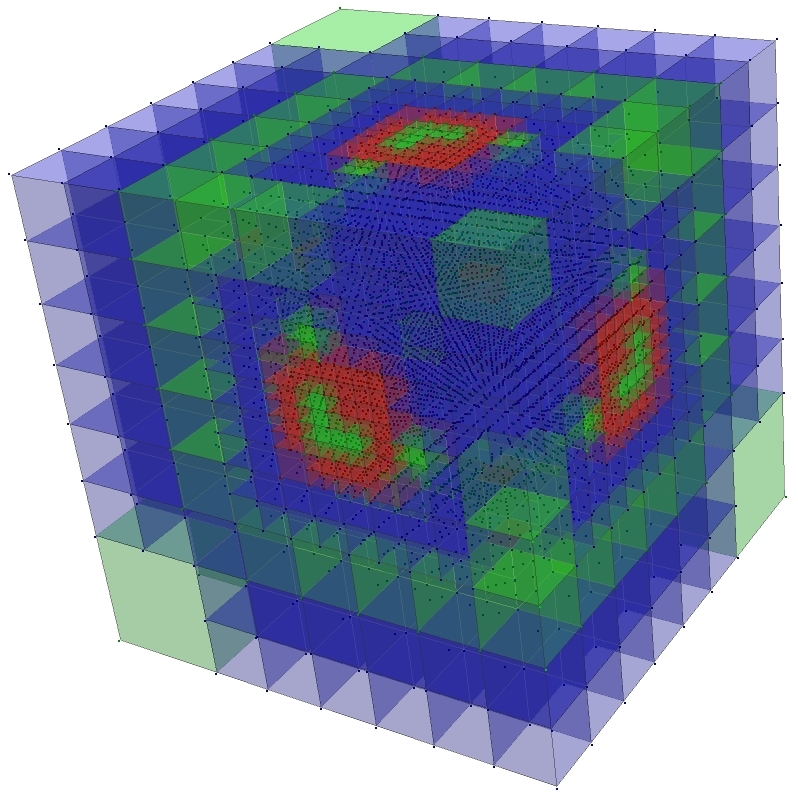} \caption{On the left is the $h$-adaptation at $T=3$ seconds from initial as a wireframe hexahedral mesh, where $h\in \{1/4,1/8,1/16,1/32\}$.  On the right we show the $p$-enrichment at $T=3$ seconds with $p\in\{1,2,3\}$, where $p=1$ is red, $p=2$ is green, and $p=3$ is blue.  The domain is $\Omega = [-5,5]^{3}$}
\label{fig:hBZ}  
\end{figure}

We take as our example case the system studied in \cite{Dockery}, modified to include the functional dependencies of $\mathscr{D}_{ij}$.  That is we use the ferroin system such that $\mathrm{P}=\mathrm{HOBr}$, $\mathrm{M}=\mathrm{Fe}(\mathrm{Phen})_{3}^{3+}$ and $\mathrm{B}=\mathrm{M}$.  For full consistency we recast (\ref{BZ1})--(\ref{BZ5}) in the form: \begin{subequations}\begin{align}\label{3BZ1} \mathrm{BrO}_{3}^{-} + \mathrm{Br}^{-} \  &  \ce{->T[{$ \ \ \ k_{f_{1}} \ \ $}] } \ \mathrm{HBrO}_{2} + \mathrm{P} \\  \label{3BZ2}\mathrm{HBrO}_{2} +  \mathrm{Br}^{-}  \ &  \ce{->T[{$ \ \ \ k_{f_{2}} \ \ $}] } \ 2\mathrm{P} \\ \label{3BZ3} \mathrm{BrO}_{3}^{-} + \mathrm{HBrO}_{2} \ &  \ce{->T[{$ \ \ \ k_{f_{3}} \ \ $}] } \ 2\mathrm{HBrO}_{2} + 2\mathrm{M} \\ \label{3BZ4} 2\mathrm{HBrO}_{2} \ &   \ce{->T[{$ \ \ \ k_{f_{4}} \ \ $}] }  \ \mathrm{BrO}_{3}^{-} + \mathrm{P} \\ \label{3BZ5} 2\mathrm{M}  \ &  \ce{->T[{$ \ \ \ k_{f_{5}} \ \ $}] } \ c\mathrm{Br}^{-}\end{align}\end{subequations} which means that (\ref{BZdis})  must be augmented in the second, third and sixth constituents, such that:  \begin{equation}\begin{aligned} \partial_{t}& \alpha_{2}= ck_{f_{5}}\alpha_{3}^{2} -k_{f_{1}}\alpha_{4}\alpha_{2} -   k_{f_{2}} \alpha_{1}\alpha_{2} \\  & \quad \partial_{t}\alpha_{3} = 2k_{f_{3}}\alpha_{1}\alpha_{4}- 2 k_{f_{5}}\alpha_{3}^{2}.\end{aligned}\end{equation}  The sixth constituent vanishes here while the second yields: \begin{equation}\begin{aligned}\label{BZdis2 }& \alpha_{2}^{n+1}= \left(\frac{ ck_{f_{5}}(\alpha^{n}_{3})^{2}}{k_{f_{1}}\alpha_{4}^{n}+k_{f_{2}}\alpha_{1}^{n}}\right) + e^{-(k_{f_{1}}\alpha_{4}^{n}+k_{f_{2}}\alpha_{1}^{n})\Delta t}\left(\alpha_{2}^{n} - \frac{ ck_{f_{5}}(\alpha^{n}_{3})^{2}}{k_{f_{1}}\alpha_{4}^{n}+k_{f_{2}}\alpha_{1}^{n}} \right).\end{aligned}\end{equation}

\begin{table}[t!]
\centering
{
\renewcommand{\arraystretch}{1.3}
\begin{tabular}{|c|c|c|c|c|}
\hline
$k_{f_{1}}$ &  $k_{f_{2}}$ &  $k_{f_{3}}$ &  $k_{f_{4}}$ &  $k_{f_{5}}$  \\ 
\hline\hline
 $2.5 \ [\mathrm{H}^{+}]^{2} \ \mathrm{M}^{-3}\mathrm{s}^{-1}$ &  $3\times10^{6} \ [\mathrm{H}^{+}] \ \mathrm{M}^{-2}\mathrm{s}^{-1}$ &  $40 \ [\mathrm{H}^{+}] \ \mathrm{M}^{-2}\mathrm{s}^{-1}$ &  $3\times10^{3} \ \mathrm{M}^{-1}\mathrm{s}^{-1}$ & $0.1 \ \mathrm{s}^{-1}$ \\
\hline
\end{tabular}
}
\label{table:BZkinetics}
\caption{The reaction rates used for the BZ reaction.}
\end{table}

The third requires solving another Ricatti equation (albeit a simpler one), such that we consider the equation $\alpha_{3}' = C_3 - C_1\alpha_{3}^2$, where similar to before, we begin with the general solution
\begin{equation}\label{back2}
\alpha_{3} = \frac{\sqrt{C_1C_3}}{C_1}\tanh\left( (t+\bar{C})\sqrt{C_1C_3}\right)
\end{equation}
and perform the expansion for $\alpha^{n+1}=\alpha(t^n+\Delta t)$, such that setting $C^{n}_{3}= 2k_{f_{3}}\alpha_{1}^{n}\alpha_{4}^{n}$ and $C_{1}=2k_{f_{5}}$ we acquire
\[
\alpha_{3}^{n+1} = \left(\frac{\sqrt{C^{n}_1C_3}}{C^{n}_1}\frac{ \tanh\left( (t^{n}+\bar{C})\sqrt{C^{n}_1C_3}\right) + \tanh\left(\Delta t \sqrt{C^{n}_1C_3}\right) }{ 1+  \tanh\left( (t^{n}+\bar{C})\sqrt{C_1^{n}C_3}\right)\tanh\left(\Delta t^{n} \sqrt{C_1^{n}C_3}\right) } \right).
\]
Then rearranging (\ref{back2}) such that
\[
\frac{C_1^{n}}{\sqrt{C_1^{n}C_3}}\alpha_{3}^n = \tanh\left( (t^n+\bar{C})\sqrt{C^{n}_1C_3} \right),
\]
we again may eliminate the $\bar{C}$, obtaining 
\begin{eqnarray*}
\alpha_{3}^{n+1} &=& \left(\frac{\sqrt{C^{n}_1C_3}}{C^{n}_1}\frac{ \frac{C^{n}_1}{\sqrt{C^{n}_1C_3}}\alpha_{3}^n + \tanh\left(\Delta t \sqrt{C^{n}_1C_3}\right) }{ 1 +  \frac{C^{n}_1}{\sqrt{C^{n}_1C_3}}\alpha_{3}^n\tanh\left(\Delta t \sqrt{C^{n}_1C_3}\right) }\right) \\
                &=& \left(\frac{ \alpha_{3}^n + \frac{\sqrt{C^{n}_1C_3}}{C^{n}_1}\tanh\left(\Delta t \sqrt{C^{n}_1C_3}\right) }{ 1 +  \frac{C^{n}_1}{\sqrt{C^{n}_1C_3}}\alpha_{3}^n\tanh\left(\Delta t \sqrt{C^{n}_1C_3}\right) }\right);
\end{eqnarray*} and thus ultimately recovering the full solution:\begin{equation}\begin{aligned}\label{new3}\alpha_{3}^{n+1} = & \left\{\begin{matrix} \alpha_{3}^{n}/(2k_{f_{5}} \alpha_{3}^{n}\Delta t  +1) & \mathrm{for} \ \alpha_{4}^{n}\alpha_{1}^{n} = 0,  \\  \left(\frac{ \alpha_{3}^n + \frac{\sqrt{C^{n}_1C_3}}{C^{n}_1}\tanh\left(\Delta t \sqrt{C^{n}_1C_3}\right) }{ 1 +  \frac{C^{n}_1}{\sqrt{C^{n}_1C_3}}\alpha_{3}^n\tanh\left(\Delta t \sqrt{C^{n}_1C_3}\right) }\right)  & \mathrm{for} \ C_{1}^{n}C_{3} > 0.\end{matrix}\right. \end{aligned}\end{equation} 

\begin{figure}[!t]
\centering
 \includegraphics[width=8cm]{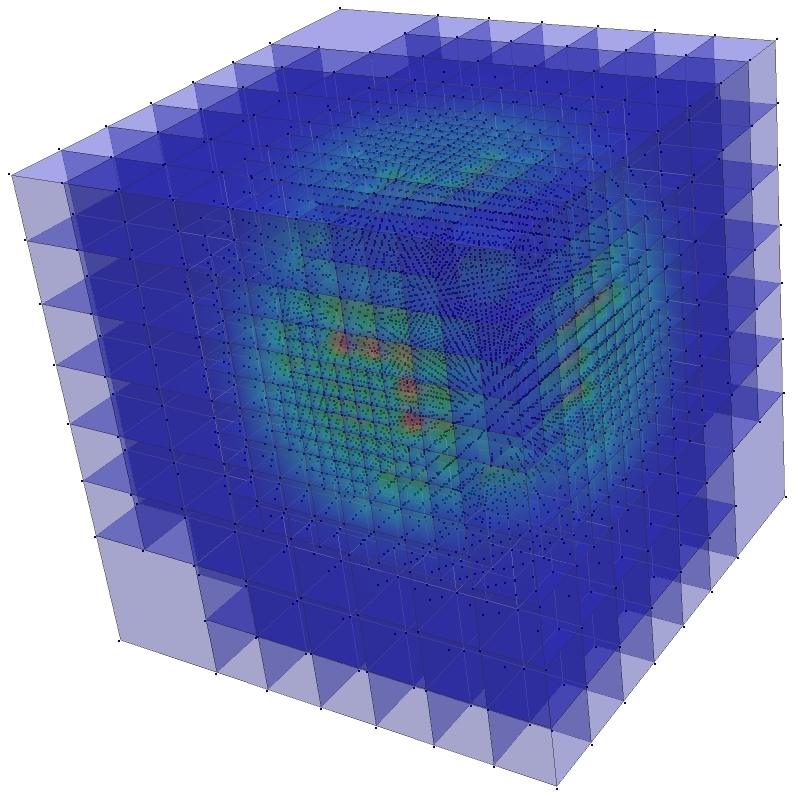}\includegraphics[width=8.cm]{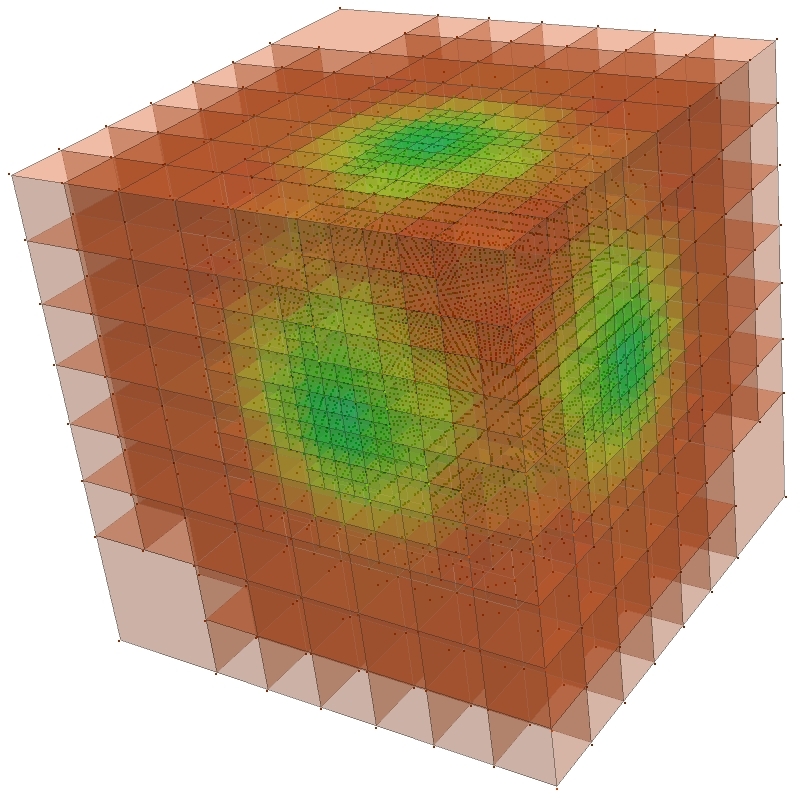} \caption{On the left is the gradient of $\alpha_{1}$ (\emph{i.e.} $\boldsymbol{\sigma}_{1}$) at $T=3$ seconds, while the right shows $\delta$ from \textsection{3} at $T=3$ seconds on $\Omega = [-5,5]^{3}$.  We emphasize relative magnitudes here, where the gradient $\boldsymbol{\sigma}_{1}$ goes from highest (red) to lowest (blue), and $\delta$ is lowest (red) to highest (blue).}
\label{fig:sig}  
\end{figure}

The malonic acid concentration $[\mathrm{CH}_{2}(\mathrm{COOH})_{2}]$ is consider ``absorbed'' into the kinetics of $k_{f_{5}}$, and the reaction rates are given in Table \ref{table:BZkinetics}, where $[\mathrm{H}^{+}]=0.8 \ \mathrm{M}$, $[\mathrm{Fe}(\mathrm{Phen})_{3}^{3+}]=2.3\times 10^{-3} \ \mathrm{M}$, and  the reaction term $c=0.43$ from (\ref{BZ1})--(\ref{BZ5}).  Note that the entropy terms $\mathscr{S}_{\mathfrak{R}}$ from \textsection{3} present a very delicate problem in this nonequilibrium setting.  Most explicitly, the equilibrium constant itself $K_{eq}$ is not well-defined here, and determining the effective activities $\tilde{a}_{i}$ of the reactions is a difficult problem.  In our results we merely assume that $K_{eq}$ scales with $\max_{i}k_{f_{i}}$, which though crude, is enough to assure that the entropy conditions from \textsection{3} are preserved.  Also notice that using only a very slightly more complicated approximate form, such as $K_{eq}\sim \alpha_{5}^{4}/\alpha_{1}^{2-c}\alpha_{2}\alpha_{4}$ immediately leads to numerical instabilities due to the concentration scalings in the BZ reaction.   

Let us further comment that it seems that in fact this subtle equilibrium behavior at $t^{n+1}$ must be determined by relying upon a bath assumption at $t^{n}$ that leads to a second order condition on the first order assumption employed in the classical Oregonator model.  More clearly, in the Oregonator model the bath concentrations are assumed large and constant to a first order approximation, while here, we make no such assumption \emph{a priori} but must rely upon an equilibrium condition that suggests that $\alpha_{5}^{4}\gg\alpha_{1}^{2-c}\alpha_{2}\alpha_{4}$ or at the very least that $\tilde{a}_{5}^{4}\gg\tilde{a}_{1}\tilde{a}_{2}\tilde{a}_{4}$, which we consider a second order bath assumption because it plays no direct role in the system dynamics, but only on the \emph{a posteriori} entropy consistency and $hp$-adaptivity of the solution.  

Here we present some numerical results from the BZ solution, using the initial conditions given by: \[\begin{aligned} & \alpha_{1,0} = 4\times10^{-5}+1\times10^{-6}e^{-((x-1.6)^{2}+(y-1.6)^{2}+(z-1.6)^{2})/2.5}, \\ & \alpha_{2,0} =  1\times10^{-7}+1\times10^{-6}e^{-((x-1.5)^{2}+(y-1.5)^{2}+(z-1.5)^{2})/2.5}, \\ & \alpha_{3,0} =  2.3\times10^{-3}+1\times10^{-4}e^{-((x-1.4)^{2}+(y-1.4)^{2}+(z-1.4)^{2})/2.5}, \\ & \qquad\qquad \alpha_{4,0}=1\times10^{-3},\quad\mathrm{and}\quad\alpha_{5,0}=1\times 10^{-3}.\end{aligned}\]  As we see in Figures \ref{fig:hBZ} and \ref{fig:sig}, the $hp$-refinement and coarsening is driven by the structure of the initial-boundary conditions.  We get oscillation behavior of our solutions, and as seen in \ref{fig:limitcyc}, the entropy consistency is preserved and drives the $hp$-refinement regime.  

\subsection{\S 4.4 Error behavior at equilibrium}

Finally we consider a simple equilibrium problem comprised of two constituents and constructed in such a way as to allow for complete decoupling between the constituents in the mass action, and thus obtain an exact analytic solution that may be easily employed for error analysis.  We assume for this case that $\mathscr{D}_{i}(\alpha)=0$ since the error behavior of the very same LDG method employed here has been previously analyzed by the authors in \cite{MESV}.

That is, consider the elementary equilibrium reaction satisfying: \begin{equation}\label{equilibrium} \nu_{1}^{f}\alpha_{1}\ce{<=>T[{$ \ \ k_{f} \ \ $}][{$ \ \ k_{b} \ \ $}] } \ \nu_{2}^{b}\alpha_{2}  \end{equation} such that the coupled system of differential equations is comprised of, \begin{equation}\label{simple2}\alpha_{1}'= \nu_{1}^{f}(k_{b}\alpha_{2}^{\nu_{1}^{b}}-k_{f}\alpha_{1}^{\nu_{1}^{f}}),\quad \alpha_{2}'= \nu_{2}^{b}(k_{f}\alpha_{1}^{\nu_{1}^{f}}-k_{b}\alpha_{2}^{\nu_{1}^{b}}),\quad\mathrm{so \ that}\quad \nu_{2}^{b}\alpha_{1}'=-\nu_{1}^{f}\alpha_{2}'.\end{equation}  Integrating for any $t\in[0,T_{eq})$ with $T_{eq}$ the equilibrium time (which exists \emph{a priori} for $\min\{k_{b},k_f\}\neq 0$) and letting the initial concentration $\alpha_{2,0}=0$, then we further notice that at each $t$ we have \begin{equation}\label{one}\alpha_{1}(t)=\alpha_{1,0}-\frac{\nu_{1}^{f}}{\nu_{2}^{b}}\alpha_{2}(t),\quad\mathrm{and \ at \ equilibrium \ that} \ \ \alpha_{1}(T_{eq})=\alpha_{1,0}-\frac{\nu_{1}^{f}}{\nu_{2}^{b}}\alpha_{2}(T_{eq}).\end{equation} 

\begin{figure}[!t]
\centering
\includegraphics[width=8.5cm]{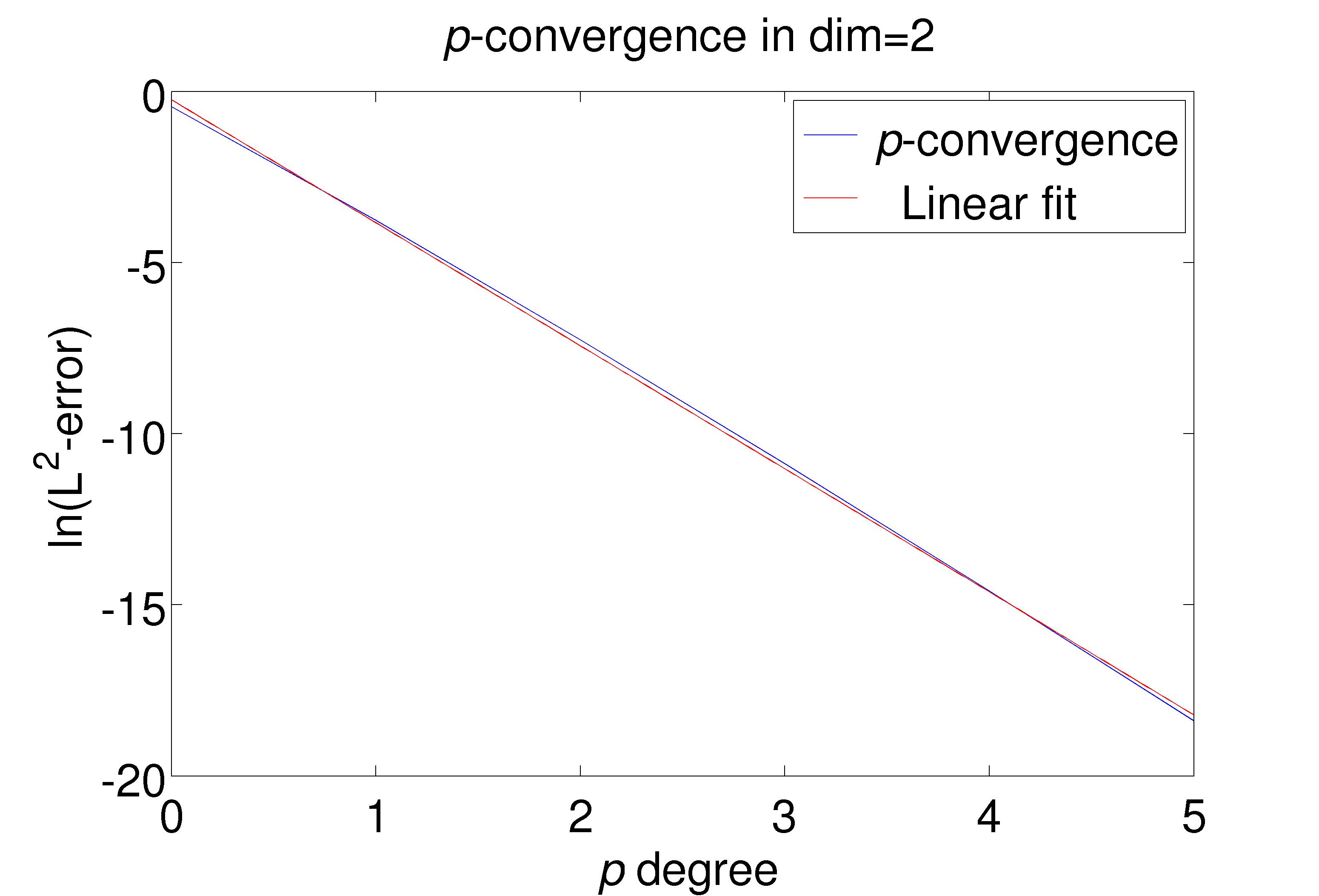} \includegraphics[width=8.5cm]{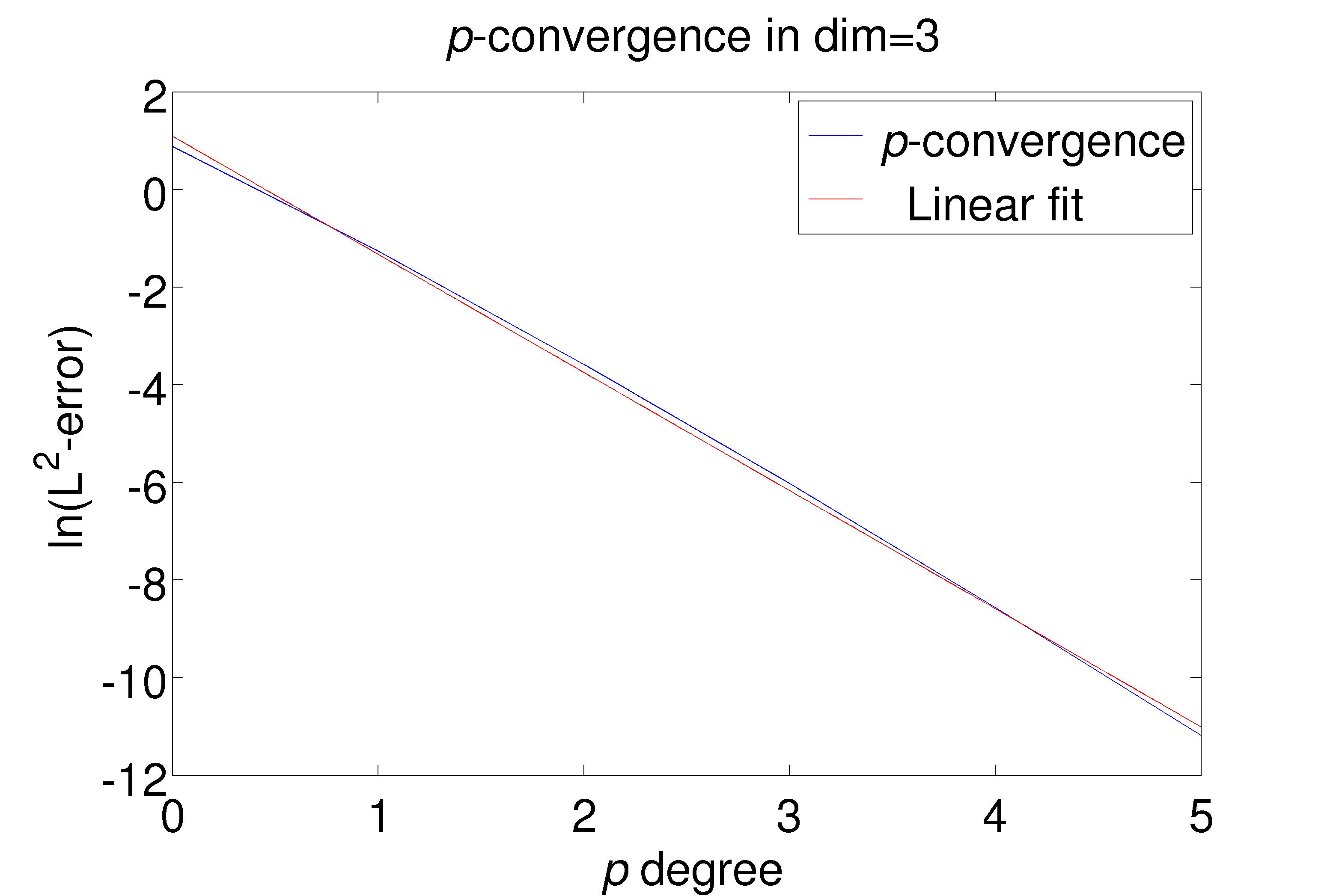} \caption{We plot the $p$-convergence of the equilibrium solution, where in $N=2$ we set $h=1/32$ and for $N=3$ we have $h=1/16$.}
\label{fig:pconv}  
\end{figure} 

\begin{table}[!t]
\centering
\begin{tabular}{|c | c | c |}
\hline
$p$ & $L^{2}$-error for $N=2$, $h=32$  & $L^{2}$-error for $N=3$, $h=16$  \rule{0pt}{3ex} \rule[0ex]{0pt}{0pt} \\ 
\hline\hline
0 &  $0.638527$ & $2.41962$  \rule{0pt}{3ex} \rule[0ex]{0pt}{0pt}\\
\hline 
1 &  $0.0230384$ & $0.283632$  \rule{0pt}{3ex} \rule[0ex]{0pt}{0pt}  \\
\hline
2 &  $0.000702172$ &  $0.027854$  \rule{0pt}{3ex} \rule[0ex]{0pt}{0pt} \\
\hline
3 &  $1.88756\times 10^{-5}$ & $0.00241581$ \rule{0pt}{3ex} \rule[0ex]{0pt}{0pt} \\
\hline 
4 & $4.59182\times 10^{-7}$ & $0.000190342$  \rule{0pt}{3ex} \rule[0ex]{0pt}{0pt} \\
\hline 
5 & $1.02788\times 10^{-8}$  & $1.38649\times 10^{-5}$ \rule{0pt}{3ex} \rule[0ex]{0pt}{0pt} \\
\hline 
\end{tabular}
\caption{We give the $L^{2}$-errors shown in Figure \ref{fig:pconv}.}
\label{table:ptab}
\end{table}

\begin{figure}[!t]
\centering
\includegraphics[width=8.5cm]{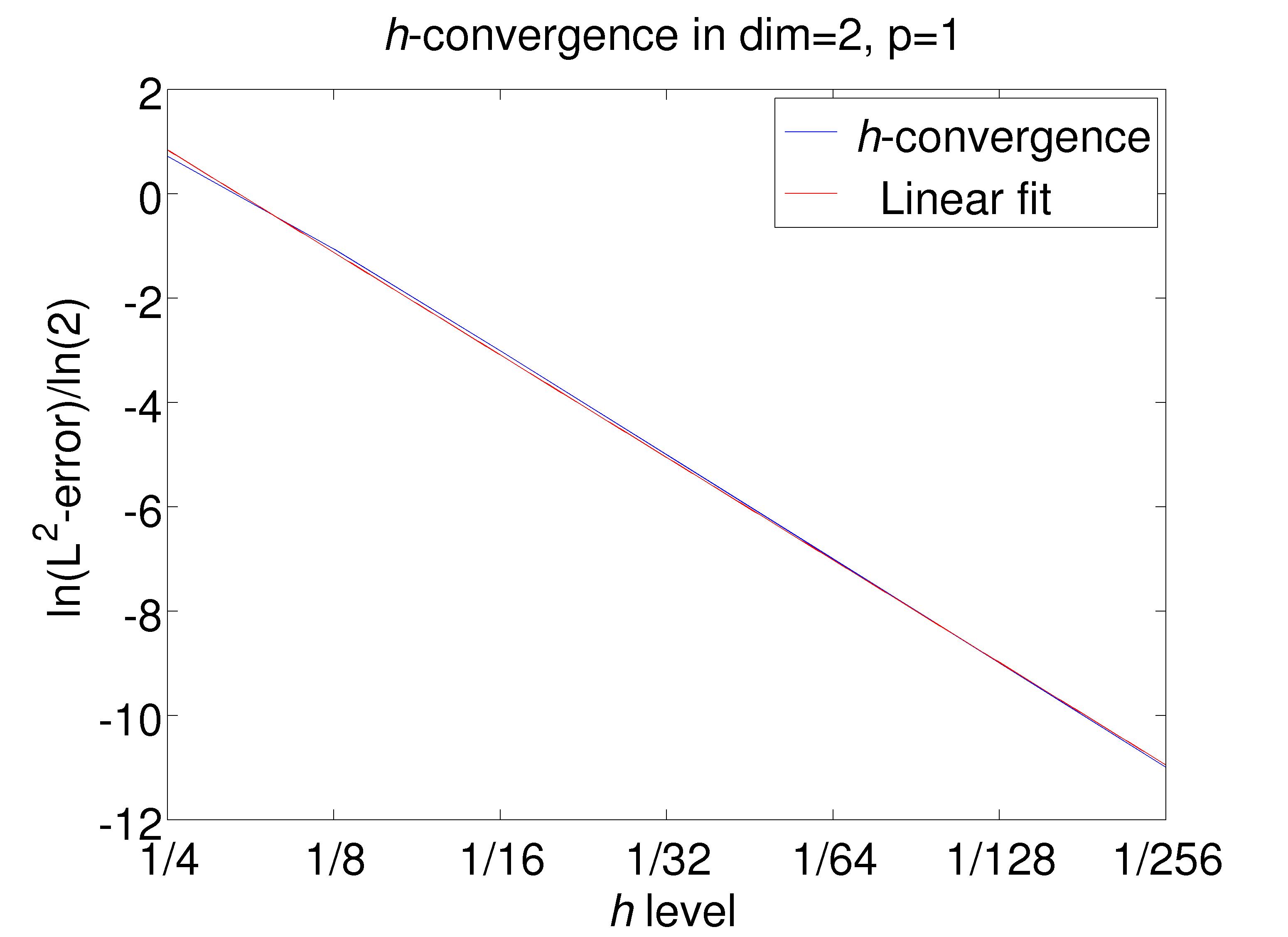} \includegraphics[width=8.5cm]{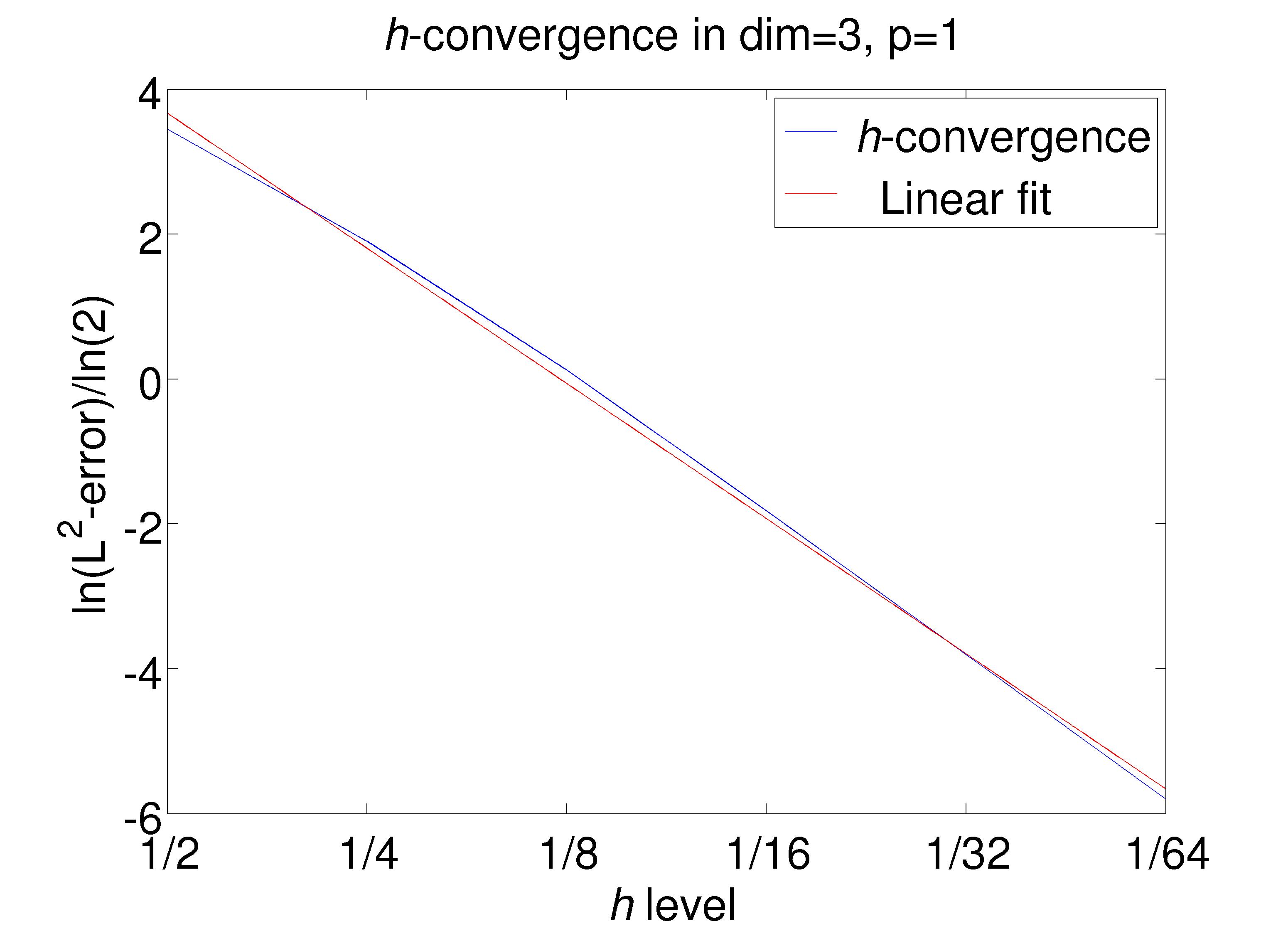} \caption{Here we show the $h$-convergence of the equilibrium solution, where $p=1$ in both cases.  For $N=2$ the best fit rate of convergence is $\sim 2$, and in $N=3$ is $\sim 1.9$.}
\label{fig:hconv}  
\end{figure}

\begin{table}[!t]
\centering
\begin{tabular}{|c | c | c || c | c |}
\hline
     & \multicolumn{2}{c||}{$p=1$} & \multicolumn{2}{c|}{$p=2$} \\
\cline{2-5}
$h$  & $L^{2}$-error & Convergence Rate & $L^{2}$-error & Convergence Rate \rule{0pt}{3ex} \rule[0ex]{0pt}{0pt} \\ 
\hline\hline
1/4   & 1.64322     & ---  & 0.535138                & ---  \rule{0pt}{3ex} \rule[0ex]{0pt}{0pt} \\
\hline
1/8   & 0.479583    & 1.78 & 0.0779288               & 2.78 \rule{0pt}{3ex} \rule[0ex]{0pt}{0pt} \\
\hline 
1/16  & 0.123667    & 1.96 & 0.0101416               & 2.94 \rule{0pt}{3ex} \rule[0ex]{0pt}{0pt} \\
\hline
1/32  & 0.0311632   & 1.99 & 0.00128043              & 2.99 \rule{0pt}{3ex} \rule[0ex]{0pt}{0pt} \\
\hline
1/64  & 0.00780632  & 2    & 0.000160455             & 3    \rule{0pt}{3ex} \rule[0ex]{0pt}{0pt} \\
\hline 
1/128 & 0.00195255  & 2    & $2.00695\times 10^{-5}$ & 3    \rule{0pt}{3ex} \rule[0ex]{0pt}{0pt} \\
\hline 
1/256 & 0.000488199 & 2    & $2.50908\times 10^{-6}$ & 3    \rule{0pt}{3ex} \rule[0ex]{0pt}{0pt} \\
\hline 
\end{tabular}
\caption{We give the $L^{2}$-errors and convergence rates shown in Figures \ref{fig:hconv}--\ref{fig:hconv2} for $N=2$.}
\label{table:htab}
\end{table}

\begin{figure}[!t]
\centering
\includegraphics[width=8.5cm]{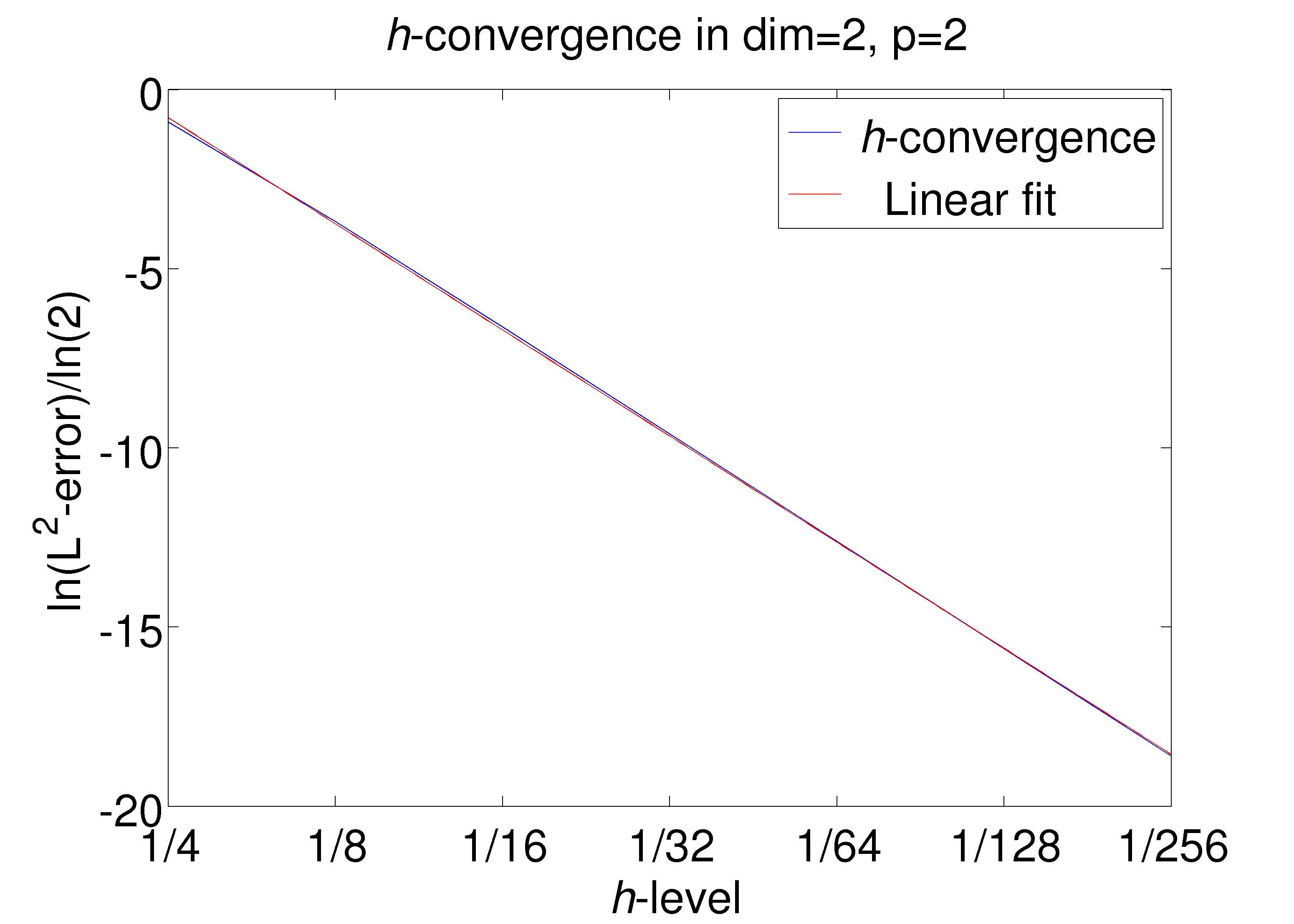} \includegraphics[width=8.5cm]{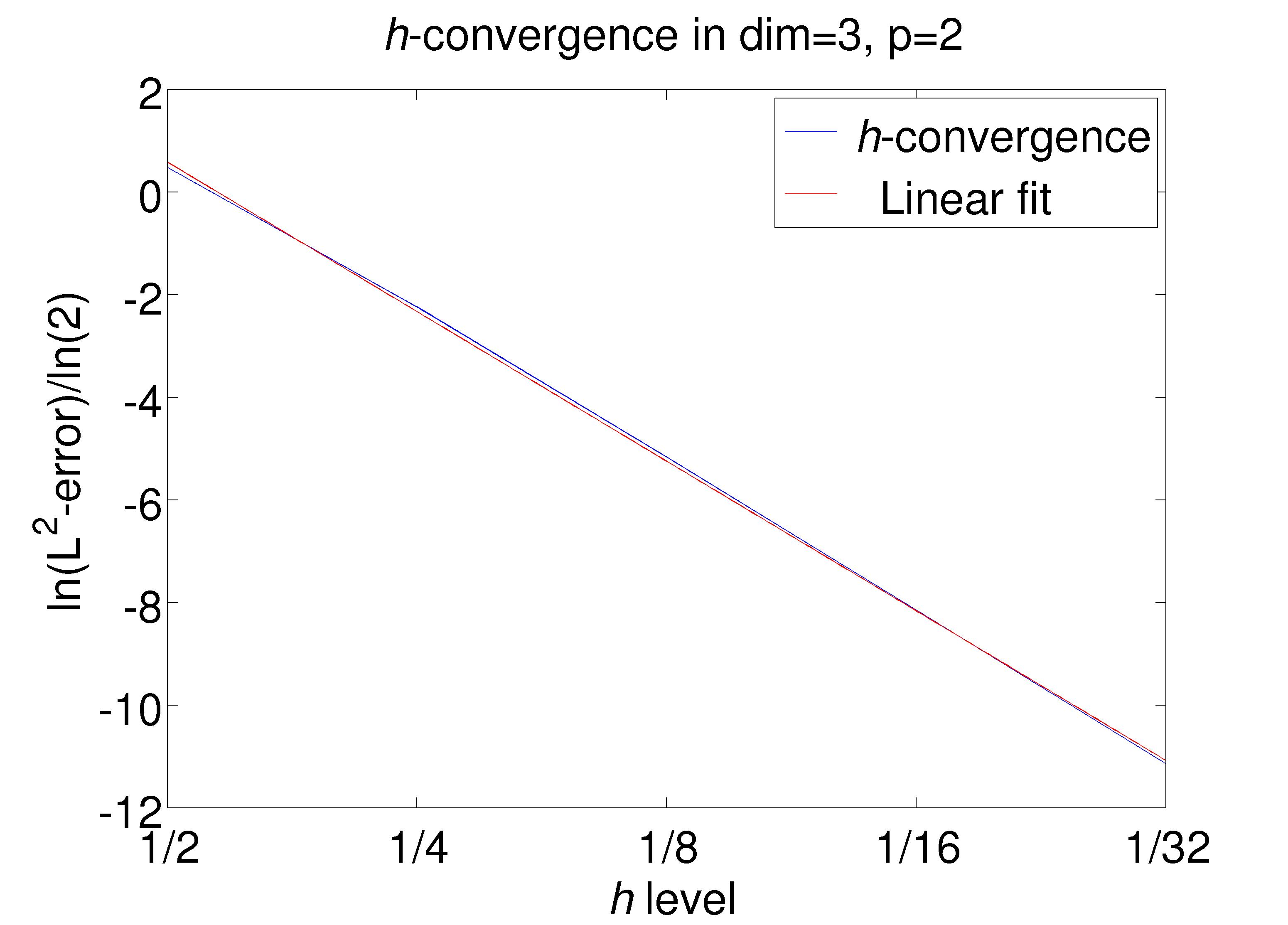} \caption{The $h$-convergence of the equilibrium solution with $p=2$.   For $N=2$ the best fit rate of convergence is $\sim 3$, and in $N=3$ is $\sim 2.9$.}
\label{fig:hconv2}  
\end{figure}

\begin{table}[!t]
\centering
\begin{tabular}{|c | c | c || c | c |}
\hline
     & \multicolumn{2}{c||}{$p=1$} & \multicolumn{2}{c|}{$p=2$} \\
\cline{2-5}
$h$  & $L^{2}$-error & Convergence Rate & $L^{2}$-error & Convergence Rate \rule{0pt}{3ex} \rule[0ex]{0pt}{0pt} \\ 
\hline\hline
1/4  & 3.73802   & ---  & 0.535138    & ---  \rule{0pt}{3ex} \rule[0ex]{0pt}{0pt}  \\
\hline
1/8  & 1.08911   & 1.78 & 0.0779288   & 2.72 \rule{0pt}{3ex} \rule[0ex]{0pt}{0pt}  \\
\hline 
1/16 & 0.283632  & 1.94 & 0.0101416   & 2.93 \rule{0pt}{3ex} \rule[0ex]{0pt}{0pt}  \\
\hline
1/32 & 0.0716523 & 1.98 & 0.00128043  & 2.98 \rule{0pt}{3ex} \rule[0ex]{0pt}{0pt}  \\
\hline
1/64 & 0.0179601 & 2    & 0.000160455 & 3    \rule{0pt}{3ex} \rule[0ex]{0pt}{0pt}  \\
\hline 
\end{tabular}
\caption{We give the $L^{2}$-errors and convergence rates shown in Figure \ref{fig:hconv}--\ref{fig:hconv2} for $N=3$.}
\label{table:htab2}
\end{table}

But then, for the special case of $\nu_{1}^{b}=\nu_{1}^{f}=1$ assuming ideal behavior where we may take that $K_{eq}=k_{f}/k_{b}=\alpha_{2}(T_{eq})/\alpha_{1}(T_{eq})$, where the equations of (\ref{one}) provide that $\alpha_{2}=\alpha_{1,0}-\alpha_{1}$, and also yields for the equilibrium constant that $K_{eq}=(\alpha_{1,0}-\alpha_{1}(T_{eq}))/\alpha_{1}(T_{eq})$.  Using these relations, we then rewrite $\alpha_{1}'$ in the first equation of (\ref{simple2}) as \begin{equation}\begin{aligned}\alpha_{1}' & = k_{b}\alpha_{2}-k_{f}\alpha_{1} \\ &  =  k_{b}\alpha_{1,0}-(k_{b} + k_{f})\alpha_{1} \\ & = (k_{f}+k_{b}) (\alpha_{1}(T_{eq})- \alpha_{1}),\end{aligned}\end{equation} which has a solution of the form of (\ref{firord}), giving that \begin{equation}\label{exeq}\alpha_{1} = \exp^{-\int_{\mathfrak{X}} (k_{f}+k_{b} )ds}\left(\alpha_{1,0} - \alpha_{1}(T_{eq})\right) + \alpha_{1}(T_{eq}).\end{equation}  for any $\mathfrak{X}\subset [0,T_{eq})$ containing the initial state and any $t\geq T_{eq}$, which is just to say the solution only depends only upon the initial and equilibrium concentration of $\alpha_{1}$ --- hence fully independent of $\alpha_{2}$.

By contrast we implement our predictor multi-corrector in the naive way to recover $\mathscr{A}(\boldsymbol{\alpha})$ by simply solving the discrete form of (\ref{simple2}) with $\nu_{1}^{b}=\nu_{1}^{f}=1$, such that by the usual procedure we arrive with the solutions \begin{equation} \begin{aligned}\label{naive}\alpha_{1}^{n+1} = \exp^{-\int_{\Delta t}  k_{f} dt}\left(\alpha_{1}^{n} - \frac{\alpha_{2}^{n}}{K_{eq}}\right) +   \frac{\alpha_{2}^{n}}{K_{eq}}, \\  \alpha_{2}^{n+1} = \exp^{-\int_{\Delta t}  k_{b} dt}\left(\alpha_{2}^{n} - K_{eq}\alpha_{1}^{n}\right) +  K_{eq}\alpha_{1}^{n},\end{aligned}\end{equation} where at equilibrium the constant terms balance to unity.

Then to test our method we compare the error behavior of (\ref{exeq}) to (\ref{naive}) where we take an end time for our simulation $T$ which is appropriately set to $T\geq T_{eq}$.  Here we have a stable equilibrium solution (See definition 11.21 in \cite{Smoller}), such that we expect the solution to rapidly converge to the equilibrium point in time to machine precision, where the only error remaining should be that taken with respect to the standard $L^{2}$-projection.  We use the following initial conditions: \[\alpha_{1,0}=1+4e^{(\boldsymbol{x}+\frac{1}{2})^{2}/3.75},\quad\mathrm{and}\quad\alpha_{2,0}=0.\]   All the solutions were run at SSP$(5,3)$ using $\Delta t = 1$ s with $T=200$ s. The $p$-convergence results are shown in Figure \ref{fig:pconv} and Table \ref{table:ptab} in both dimension two and three for regular meshes.  The $h$ convergence results are shown Figures \ref{fig:hconv} and \ref{fig:hconv2} as well as in Tables \ref{table:htab}--\ref{table:htab2} in both dimension two and three.  Here we use spatially homogeneous refinements of integral value in each direction to obtain the expected results.

\section{\texorpdfstring{\protect\centering $\S 5$ Conclusion}{\S 5 Conclusion}}

We have developed a predictor multi-corrector time-operator splitting RKLDG SSP scheme that utilizes a stability preserving $hp$-adaptive entropy consistency scheme for its coarsening and refinement methodology.  The scheme is presented and implemented for arbitrary spatial (\emph{i.e.} $N\leq 3$) and component (\emph{i.e.} $n$ computable) dimension, and includes methods which adopt varying functional parameters (\emph{e.g.} $\mathscr{D}(\alpha)$ and $K_{eq}(\alpha)$) as well as arbitrary extended Robin boundary data as so applied to a generalized subset of reaction-diffusion equations which we denote: \emph{quiescent reactors}.

The entropy methodology that serves as the fabric of the $hp$-adaptive scheme, is extended from the regularity analysis of \cite{MCV1}, and provides for a sharp stability condition on the computational well-posedness of the reaction-diffusion system.  

In addition we have presented solutions to a number of application models, most notably we have derived a novel solution to the classical BZ reaction.  This is a difficult and complicated reaction regime which is often used to underscore the nuance involved in rate-coupled reaction mechanisms.  

Our future directions are to extend the quiescent reactors to include fluid reactors where density $\rho=\rho(t,\boldsymbol{x})$ and temperature (energy) $\mathfrak{E}=\mathfrak{E}(\vartheta)=\mathfrak{E}(\vartheta(t,\boldsymbol{x}))$ are fully coupled, in addition to adding turbulence models and electromagnetic fields for weakly ionized and plasma reactors.

{\setlength\parskip{0pt} 
\bibliographystyle{abbrv}
\def\cprime{$'$} \def\cprime{$'$}
  \def\polhk#1{\setbox0=\hbox{#1}{\ooalign{\hidewidth
  \lower1.5ex\hbox{`}\hidewidth\crcr\unhbox0}}}
  \def\polhk#1{\setbox0=\hbox{#1}{\ooalign{\hidewidth
  \lower1.5ex\hbox{`}\hidewidth\crcr\unhbox0}}}
  \def\polhk#1{\setbox0=\hbox{#1}{\ooalign{\hidewidth
  \lower1.5ex\hbox{`}\hidewidth\crcr\unhbox0}}}

}

\end{document}